\documentclass[epsfig]{article}
\usepackage{makeidx}
\usepackage{graphicx}
\newcommand{\bfb}{\mathbf{b}}
\newcommand{\bfn}{\mathbf{n}}
\newcommand{\bfx}{\mathbf{x}}

\newcommand{\refginput}{3}
\newcommand{\refnew}{4}
\newcommand{\refnewat}{4.2}
\newcommand{\refnewion}{4.3}

\newcommand{\refexampostwo}{6.3}

\newcommand{\refnonstfreq}{7.4.2}
\newcommand{\refnonstrte}{7.4.3}
\newcommand{\refnonstdiscr}{7.4.6}
\newcommand{\refnonstacc}{7.4.8}
\newcommand{\refnonstblank}{7.7}
\newcommand{\refnonstconv}{7.8}

\newcommand{\refions}{11}
\newcommand{\refionbf}{11.2}
\newcommand{\refionbb}{11.3}

\newcommand{\refxfiles}{11.7}
\newcommand{\refnsttwophys}{12}
\newcommand{\refnsttwohyd}{12.1.1}

\newcommand{\refnsttwocolh}{12.3.2}
\newcommand{\refnsttwodiel}{12.3.3}

\newcommand{\refnsttwocompt}{12.5}
\newcommand{\refnsttwoconv}{12.6}
\newcommand{\refnsttwoeos}{12.7}
\newcommand{\refnsttwoprd}{12.9.1}
\newcommand{\refnsttwodisk}{12.8}

\newcommand{\refnsttwooptab}{13.1}
\newcommand{\refnsttwortetwo}{13.1}

\makeindex

\begin{document}

\title{\bf {\sc tlusty}  User's Guide II: \\Reference Manual}  
\author{I. Hubeny\footnote{University of Arizona, 
Tucson; USA; hubeny{\tt @}as.arizona.edu}~ and 
T. Lanz\footnote{Observatoire de C\^{o}te d'Azur, France; {\tt lanz{\tt @}oca.eu}}}
\date{\today}
\maketitle

\begin{abstract}
This is the second part of a three-volume guide to {\sc tlusty} and {\sc synspec}.
It presents a detailed reference manual for {\sc tlusty}, which contains
a detailed description of basic physical assumptions and equations used to
model an atmosphere, together with an overview of the numerical
methods to solve these equations.
\end{abstract}

\tableofcontents
\newpage

%--------------------------------------------------------------------------

\section{Introduction}

This paper in the second part of a series of three papers that provide
a detailed guide to model stellar atmosphere and accretion disk program
{\sc tlusty}, and associated spectrum synthesis program {\sc synspec}.
The first paper (Hubeny \& Lanz 2017a; hereafter referred to as Paper~I)
provides a brief introductory guide to these programs, without much
description of the underlying physical, mathematical, and numerical
background.  

The aim of this paper is to provide just that. We shall summarize the basic
structural equations of the problem, and outline the adopted numerical methods 
to solve the resulting equations. An outline of the actual operation of
{\sc tlusty}, and a detailed explanation and description of the input parameters,
the individual computational strategies, and various tricks to solve potential
numerical problems, is covered in the subsequent Paper~III (Hubeny \& Lanz 2017c). 

 A detailed description of the original version of {\sc tlusty} is given 
by Hubeny (1988). That paper
describes the basic concepts, equations, and numerical methods used.
However, because the program has evolved considerably since
1988, the description presented in that paper has become in many places obsolete. 
The major new developments are described in several papers:
Hubeny \& Lanz (1992) presented the Ng and Kantorovich accelerations;
Hubeny \& Lanz (1995) developed the hybrid CL/ALI method,
and the concept of superlevels and superlines treated by means of
an Opacity Distribution Functions (ODF) or Opacity Sampling (OS). 
A treatment of level dissolution, occupation probabilities,
merged levels, and corresponding pseudocontinuum is described by 
Hubeny, Hummer, \& Lanz (1994). An extension to high-temperature conditions,
including Compton scattering, X-ray opacities with inner-shell (Auger) ionization,
is described in Hubeny et al. (2001).
A general and comprehensive overview of the physical and mathematical 
formulation of the problem is presented in Hubeny \& Mihalas (2014;  
Chaps.12, 13, 14, 17, 18).

%-------------------------------------------------------------------------------------------------

\section{Physical background}
\label{phys}

{\sc tlusty} is designed to compute the so-called classical model atmospheres;
that is, plane-parallel, horizontally homogeneous atmospheres in hydrostatic 
and radiative (or radiative+convective) equilibrium. For a comprehensive
discussion and detailed description of the basic physics and numerics of the 
problem, refer to Hubeny \& Mihalas (2014; Chap. 18).

In the next section, we describe the basic assumptions and structural 
equations specific to stellar atmospheres. The equations are generally non-local and
therefore depend on the geometry of the problem. Analogous assumptions and
equations for accretion disks will be described in the subsequent section, while the
local physics and corresponding equations, valid for both atmospheres 
and disks, will be covered in the rest of the chapter.

\subsection{Basic equations of stellar atmospheres}
\label{phys_bas}

\noindent
$\bullet$\,\textsf{Radiative transfer equation} 
\index{Radiative transfer equation}

 It is convenient to use the second-order form
\begin{equation}
\label{rte}
\frac{d^2 (f_\nu J_\nu)}{d\tau_\nu^2} = J_\nu - S_\nu,
\end{equation}
where  $f_\nu$ is the variable Eddington factor, $J_\nu$ the mean intensity of
radiation at frequency $\nu$,  $\tau_\nu$ the monochromatic optical depth and $S_\nu$ the source function, defined by
\index{Source function}
\begin{equation}
\label{sf}
S_\nu = \eta_\nu^{\rm tot}/\chi_\nu,
\end{equation}
where $\chi_\nu$ is the total absorption coefficient, and $\eta_\nu^{\rm tot}$ the total
emission coefficient. The Eddington factor is defined by
\index{Eddington factor}
\begin{equation}
\label{vef}
f_\nu\equiv K_\nu/J_\nu=\int_{-1}^1 I_\nu(\mu)\mu^2\,d\mu \bigg/ 
\int_{-1}^1 I_\nu(\mu)\,d\mu,
\end{equation}
where $\mu=\cos\theta$, with $\theta$ being the angle between the direction of
propagation of the radiation and the normal to the surface.
The optical depth is defined by
\index{Optical depth}
\begin{equation}
d\tau_\nu \equiv -\chi_\nu dz = (\chi_\nu/\rho)\, dm,
\end{equation}
where $z$ is a geometrical distance measured along the normal to the surface 
from the bottom of the atmosphere to the top,  $m$ the column mass, and $\rho$
the mass density -- see equation (\ref{dm}). The column mass is taken as the basic geometrical coordinate.

The upper boundary condition is written as
\begin{equation}
\label{rte_ubc}
\left[\frac{\partial(f_\nu J_\nu)}{\partial\tau_\nu}\right]_0 = g_\nu J_\nu(0)
- H_\nu^{\rm{ext}},
\end{equation}
where $g_\nu$ is the surface Eddington factor defined by
\index{Eddington factor}
\begin{equation}
\label{vefg}
g_\nu\equiv \frac{1}{2}\int_{-1}^1 I_\nu(\mu,0)\mu\,d\mu \big/ J_\nu(0),
\end{equation}
and
\begin{equation}
\label{hext}\,
H_\nu^{\rm{ext}}\equiv\frac{1}{2}\int_{0}^1 I_\nu^{\rm{ext}}(\mu)\mu\,d\mu
\end{equation}
where $I_\nu^{\rm{ext}} (\mu)$ is an external incoming intensity at the top of
\index{External irradiation}
the atmosphere. In most cases one assumes no incoming radiation,
$I_\nu^{\mathrm{ext}} (\mu)=0$, but can be taken as a non-zero input quantity
 if needed.

The lower boundary condition is written in a similar way
\begin{equation}
\label{rte_lbc}
\left[\frac{\partial(f_\nu J_\nu)}{\partial\tau_\nu}\right]_{\tau_{\rm{max}}}
= H_\nu^+ - \frac{1}{2} J_\nu,
\end{equation}
where $H_\nu^+ = \frac{1}{2}\int I_\nu^+(\mu,\tau_{\mathrm{max}})\mu\, d\mu$.
One typically assumes the diffusion approximation at
the lower boundary, in which case $I_\nu^+(\mu)=B_\nu+\mu(dB_\nu/d\tau_\nu)$, 
hence equation (\ref{rte_lbc}) is written as
\begin{equation}
\label{rte_lbc2}
\left[\frac{\partial(f_\nu J_\nu)}{\partial\tau_\nu}\right]_{\tau_{\mathrm{max}}}
=\left[\frac{1}{2}(B_\nu - J_\nu)+
\frac{1}{3}\frac{\partial B_\nu}{\partial\tau_\nu}\right]_{\tau_{\mathrm{max}}},
\end{equation}
where $B_\nu$ is the Planck function,
\begin{equation}
\label{planck}
B_\nu = \frac{2h\nu^3}{c^2} \frac{1}{\exp(h\nu/kT)-1},
\end{equation}
where $T$ is the temperature, and $h$, $k$, $c$ are the Planck constant, Boltzmann constant,  and the speed of light, respectively.

Equations (\ref{rte}), (\ref{rte_ubc}), and (\ref{rte_lbc2}) contain
only the mean intensity of radiation, $J_\nu$, a function of frequency and
depth, but not the specific intensity, $I_{\mu\nu}$, which is also a function of the
polar angle $\theta$ ($\mu =\cos\theta$). 
This is made possible by introducing the Eddington 
factor, which is computed in the formal solution of the transfer equation, and 
is held fixed during the subsequent iteration of the linearization process.
By the term ``formal solution'' we mean a solution of the transfer equation with 
{\em known} source function. It is done between two consecutive iterations of the iterative scheme, with the current values of the state parameters
-- see \S\,\ref{formal}.\\

\noindent
$\bullet\,$\textsf{Hydrostatic equilibrium equation}
\index{Hydrostatic equilibrium}

The equation is conveniently written as
\begin{equation}
\label{he}
\frac{dP}{dm} = g,
\end{equation}
where $P$ is the total (gas plus radiation) pressure, and $m$ the Lagrangian
mass, or column mass,
\index{Column mass}
\begin{equation}
\label{dm}
dm = -\rho\,dz,
\end{equation}
$g$ is the surface gravity, which is assumed constant throughout the atmosphere,
and given by $g=G M_\ast/R_\ast^2$, where $M_\ast$ and $R_\ast$ are the stellar mass and radius, respectively; $G$ is the gravitational constant. 
The surface gravity $g$  is one of the basic parameters of the problem.
\index{Surface gravity}

The total pressure is generally composed of three parts, the gas pressure,
$P_{\mathrm{gas}}$, the radiation pressure, $P_{\mathrm{rad}}$, and a ``turbulent
pressure'', $P_{\mathrm{turb}}$. The gas pressure is given, assuming an ideal gas
equation of state, by
\begin{equation}
\label{eos0}
P_{\mathrm{gas}}= NkT ,
\end{equation}
where $T$ is the (electron) temperature, and  $N$ is the total particle number density.  
The radiation pressure is given by
\begin{equation}
\label{radpres}
P_{\rm rad}= \frac{4\pi}{ c}\int_0^\infty K_\nu d\nu.
\end{equation}
The so-called ``turbulent pressure'' is not a 
\index{Turbulent pressure}
well-defined quantity; it is introduced to mimic a pressure
associated with a random motion of ``turbulent eddies'' as
$P_{\mathrm{turb}}\propto\rho v_{\mathrm{turb}}^2$, where 
\index{Turbulent velocity}
$v_{\mathrm{turb}}§$ is the microturbulent velocity. It can be included
in {\sc tlusty} model calculations, but it is not recommended since its physical
meaning is questionable.

The hydrostatic equilibrium equation can then be written as 
\begin{equation}
\label{he1}
\frac{d(P_{\rm gas}+P_{\mathrm{turb}})}{ dm} = g - \frac{4\pi}{ c}\int_0^\infty \frac{dK_\nu}{ dm}\, d\nu =
g-\frac{4\pi}{ c}\int_0^\infty \frac{\chi_\nu}{\rho}H_{\nu}\, d\nu,
\end{equation}
where $H_\nu$ and $K_\nu$ are the first and second angular moments of the
specific intensity. \\

\noindent
$\bullet$\,\textsf{Energy balance equation}
\index{Energy balance equation!atmospheres}

In the convectively stable layers, the energy balance is represented 
by the radiative equilibrium equation.
For the purposes of numerical stability, it is considered in {\sc tlusty} 
as a linear combination
of its two possible forms -- the terms in square brackets of 
equation (\ref{re}), that both should be identically equal to zero,
\index{Radiative equilibrium equation}
\begin{equation}
\label{re}
\alpha\bigg[\int_0^\infty\!\!\left(\chi_\nu J_\nu - \eta_\nu^{\rm tot}\right)d\nu\bigg] +
\beta \bigg[\int_0^\infty \frac{d(f_\nu J_\nu)}{ d\tau_\nu}\,d\nu -
\frac{\sigma_R }{ 4\pi}\,T_{\mathrm{eff}}^4\bigg] = 0,
\end{equation}
where $\alpha$ and $\beta$ are empirical coefficients that satisfy
$\alpha\rightarrow 1$ in the upper layers, and $\alpha\rightarrow 0$ for deep
layers, while the opposite applies for $\beta$. The division between the
``surface'' and ``deep'' layers is a free parameter.
In equation (\ref{re}), $\sigma_R$ is the Stefan-Boltzmann constant
and $T_{\mathrm{eff}}$ the effective temperature, which is a measure of
\index{Effective temperature}
the total energy flux coming from the interior. It is another basic parameter
of the problem.

The first term of equation (\ref{re}) is called the ''integral form'', while the
second the ''differential'' form. Using Eqs. (\ref{chi}) and (\ref{eta}), and
assuming coherent scattering in the scattering part of the emission
coefficient, the
integral form may be rewritten in a traditional form
\begin{equation}
\int_0^\infty\!\!\left(\chi_\nu J_\nu - \eta_\nu^{\rm tot} \right) d\nu =
\int_0^\infty\!\!\left(\kappa_\nu J_\nu - \eta_\nu \right) d\nu,
\end{equation}
where $\kappa_\nu$ and $\eta_\nu$ are the extinction and thermal emission
coefficients, respectively.\\

\noindent
$\bullet\,$\textsf{Convection}
\index{Convection}

The atmosphere is convectively unstable if the Schwarzschild criterion for
convective instability is satisfied,
\index{Schwarzschild stability criterion}
\index{Adiabatic gradient}
\begin{equation}
\label{schw}
\nabla_{\rm rad} > \nabla_{\rm ad},
\end{equation}
where $\nabla_{\rm rad}= (d \ln T/d\ln P)_{\rm rad}$ is the logarithmic 
temperature gradient in radiative equilibrium, and $\nabla_{\rm ad}$ is the 
adiabatic gradient. The latter is viewed as a function of temperature and
pressure, $\nabla_{\rm ad}= \nabla_{\rm ad}(T,P)$. 
The density $\rho$ is considered to be a function of $T$ and
$P$ through the equation of state.

If convection is present, equation (\ref{re}) is modified to read
\begin{equation}
\label{re_conv}
\alpha\bigg[\int_0^\infty\!\!\left(\kappa_\nu J_\nu - \eta_\nu\right)d\nu
+ \frac{\rho}{4\pi}\frac{dF_{\rm conv}}{dm}  \bigg] +
\beta \bigg[\int_0^\infty\! \frac{d(f_\nu J_\nu)}{ d\tau_\nu}\,d\nu -
\frac{\sigma_R }{ 4\pi}\,T_{\rm eff}^4
+ \frac{F_{\rm conv}}{ 4\pi}\bigg] = 0
\end{equation}
where $F_{\rm conv}$ is the convective flux, given by
\begin{equation}
\label{conv}
F_{\rm conv} = (gQH_P/32)^{1/2}(\rho c_P T)(\nabla-\nabla_{\rm el})^{3/2} (\ell/H_P)^2,
\end{equation}
where 
$H_P \equiv -(d\ln P/dz)^{-1} = P/(\rho g)$
is the pressure scale height, $c_P$ is the specific heat at constant pressure, 
$Q \equiv -(d\ln\rho/d\ln T)_P$, and
$\ell/H_P$ is the ratio of the
convective mixing length to the pressure scale height, taken as a free parameter
\index{Mixing length}
of the problem, $\nabla$ is the actual logarithmic temperature gradient, and
$\nabla_{\rm el}$ is the gradient of the convective elements. The latter is 
determined by considering the efficiency of the convective transport; see,
e.g.,  Hubeny and Mihalas (2014; \S\,16.5),
\begin{equation}
\label{nablae}
\nabla-\nabla_{\rm el} = (\nabla-\nabla_{\rm ad}) + B^2/2 - 
B\sqrt{B^2/2 - (\nabla-\nabla_{\rm ad})},
\end{equation}
where
\begin{equation}
\label{convb}
B= \frac{12\sqrt 2 \sigma_R T^3}{\rho c_p (gQH_P)^{1/2} (\ell/H_P) }
\, \frac{\tau_{\rm el} }{ 1+ \tau_{\rm el}^{2}/2},
\end{equation}
with $\tau_{\rm el} = \chi_R \ell$ is the optical thickness of the
characteristic element size $\ell$. 
The gradient of the convective elements is then a function of temperature,
pressure, the actual gradient,
$\nabla_{\rm el} = \nabla_{\rm el}(T,P,\nabla)$, and the convective flux
can also be regarded as a function of $T$, $P$, and $\nabla$.

For white dwarfs, Fontaine et al. (1992) suggested a slightly
different prescription for the convective flux, called ML2, which essentially
consists of replacing the factor $\sqrt{1/32}$ in Eq. (\ref{conv}) by 2. This
possibility is also offered by {\sc tlusty}.\\

\noindent$\bullet$\,\textsf{Other structural equations}

The rest of structural equations, namely the kinetic equilibrium equation,
charge and particle conservation equation, equation of state, together with auxiliary equations such as the definition of absorption and emission coefficients, are local, and therefore the same for stellar atmospheres and accretion
disks, and will be described in \S\S\,\ref{kee} - \ref{absoemisscat}.

%--------------------------------------------------------------

\subsection{Accretion disks}
\label{disks}

The basic assumptions are the following:

\begin{itemize}
\item The disk is divided into a set of concentric rings, each behaving
 as an independent 1-D plane parallel radiating slab;
no assumptions about optical thickness are made.
One run of {\sc tlusty} calculates a vertical structure of
one ring;
\item hydrostatic equilibrium in the vertical $z$-direction;
\item energy balance is considered as a balance between the net radiation
loss (calculated exactly, without invoking neither optically
thin, nor optically thick [diffusion] approximations), and
the mechanical energy generated through viscous dissipation;
\item the dissipated energy is proportional to viscosity, which is given
through the empirical viscosity parameter $\alpha$ or through a Reynolds number;
\index{Reynolds number}
\end{itemize}

One can consider accretion disks around stars -- the so-call 
{\it classical disks}, in which case the basic radial structure is given by the 
standard model (e.g., Frank, King, Raine 1992), or disks around black 
holes -- the so-called {\it relativistic disks}, in which case the radial structure is 
given essentially by Novikov \& Thorne (1973), with refinements described in
Riffert \& Herold (1995) and  Krolik (1999). 
We use a universal  formalism, in which case  the relativistic disks are
described by means of {\it relativistic corrections}. We follow the notation
of Riffert \& Herold (1995) and Hubeny \& Hubeny (1998).

\subsubsection{General structural equations}
The basic structural equations for one individual ring are the following:\\

\noindent
$\bullet$\,\textsf{Vertical hydrostatic equilibrium equation}
\index{Hydrostatic equilibrium!disks}

The atmosphere at each disk radius $R$ (specified in the disk
midplane) is in hydrostatic equilibrium, with a depth-dependent gravity
($g$) that arises from the vertical component of the central star's
gravitational force on the disk material. Neglecting the self-gravity of
the disk and assuming that $R$ is much larger than the distance from the
central plane, $z$:
\begin{equation}
\label{hedisk}
{dP\over dz} = -g(z)\rho, \quad {\rm or} \quad {dP \over dm} =  g(z)\, ,
\end{equation}
where the depth-dependent vertical gravity acceleration is given by
\begin{equation}
\label{gdef}
 g(z) = {GM \over R^3 }{C \over B} \, z \, .
\end{equation}
$G$ is the gravitational constant, $M$ is the mass of the central object, 
and $B$ and $C$  (together with $A$ used later) are the so-called relativistic 
corrections in the notation of Riffert and Harold (1995).
\index{Relativistic corrections}
For classical disks, $A=B=C=1$.\\

\noindent$\bullet$\,\textsf{Energy balance}
\index{Energy balance equation!disks}

Generally, it is written as
\begin{equation}
\label{ebal}
\frac{\partial F_z}{\partial z} = \frac{3}{2}\left(\frac{GM}{R^3}\right)^{1/2}
\!\frac{A}{B}\, t_{\phi r},
\end{equation}
where $F_z$ is the $z$-component of the energy flux and $T_{\phi r}$
is the sheer stress, also called the viscous stress.\\

\noindent$\bullet$\,\textsf{Azimuthal momentum balance}

Under the assumption of $t_{\phi r}=0$ at the innermost orbit, it is written as
\begin{equation}
\label{mbal}
\int_{-h}^{h} t_{\phi r} dz = \frac{\dot M}{2\pi}\left(\frac{GM}{R^3}\right)^{1/2}
\frac{D}{A},
\end{equation}
where $h$ is the vertical height of the given annulus, 
and $\dot M$ is the mass accretion rate.\\

\noindent$\bullet$\,\textsf{Equation describing the source of viscous stress}
\begin{equation}
\label{vbal}
t_{\phi r}  = \frac{3}{2}\eta\left(\frac{GM}{R^3}\right)^{1/2}
\frac{A}{B},
\end{equation}
where $\eta$ is the coefficient of sheer viscosity, which is expressed through
the coefficient of kinematic viscosity $w$ as $\eta\equiv\rho w$.
\medskip

\noindent$\bullet$\,\textsf{The coefficients} $A$ - $D$ are given by:

For classical disks,
\begin{eqnarray}
A&=&B=C=1,\\
D&=&1-\left({R/R_\ast}\right)^{1/2}  ,
\end{eqnarray}
where $R_\ast$ is the radius of the central star.

For relativistic disks, they are called the relativistic corrections:
\begin{eqnarray}
A&=& 1-\frac{2}{r} + \frac{a^2}{r^2},\\
B&=& 1-\frac{3}{r} + \frac{2a}{r^{3/2}},\\
C&=& 1-\frac{4a}{r^{3/2}} + \frac{3a^2}{r^2},\\
D&=& \frac{1}{\sqrt{r}} \int_{r_i}^r \frac{x^2-6x+8a\sqrt{x}-3a^2}
     {\sqrt{x}(x^2-3x+2a\sqrt{x})  } dx,
 \end{eqnarray}
where $r$ is the radius of the annulus is expressed in units of the gravitational radius,
$r=R/(GM/c^2)$, and $a$ is the specific angular momentum (spin) of the
black hole expressed in units of $G/c$ ($a=0$ for a Schwarzschild black
hole; $a=0.998$ for maximum rotating Kerr black hole).

\subsubsection{Viscosity and the total column mass}
\index{Viscosity}

To write down practical expressions of the energy balance and for the
total column mass, one has to introduce a suitable parametrization of viscosity.
First, the sheer viscosity $\eta$ is expressed through the kinematic viscosity $w$ as
\begin{equation}
\label{kinvisc}
\eta\equiv\rho w.
\end{equation}
The corresponding vertically averaged kinematic viscosity is given by
\begin{equation}
\label{avervisc}
\bar w = \frac{\int_0^h w \rho dz}{\int_0^h \rho dz} = \frac{1}{m_0} \int_0^h \eta dz =
\frac{1}{m_0} \int_0^{m_0}\!\! w\, dm.
\end{equation}
Integrating Eq. (\ref{vbal}) from 0 to $h$, and using Eq. (\ref{mbal}) together
with Eq. (\ref{avervisc}), one can express the total column mass at the midplane
through the averaged viscosity as
\begin{equation}
\label{m0}
m_0 = \frac{1}{\bar w} \frac{\dot M}{6\pi} \frac{BD}{A^2},
\end{equation}

There are two possibilities to express the viscosity. A simple one is to parametrize
it through the Reynolds number (Lynden-Bell \& Pringle 1974, K\v r\'\i\v z \&
Hubeny 1986), in which case the vertically averaged viscosity is simply given as
\begin{equation}
\label{revisc}
\bar w = \frac{(GMR)^{1/2}}{{\rm Re}},
\end{equation}
where Re is the Reynolds number, which is a free parameter of the problem,
typically chosen between 1000 and 10000 (Lynden-Bell \& Pringle 1974).
Consequently, the total column mass $m_0$ is simply
\begin{equation}
m_0= \frac{\dot M {\rm Re}}{6\pi (GMR)^{1/2}} \frac{BD}{A^2},
\end{equation}
which has a big computational advantage that $m_0$ is given as a function
of $M$, $\dot M$, and $R$ (and the spin $a$ in the case of relativistic disks),
and therefore is known a priori. However, it is not clear how to choose a proper
value of the Reynolds number and, moreover, whether the same Reynolds number
applies for all radial distances in a disk.

Therefore, a much more commonly used prescription is based on the so-called
$\alpha$-parametrization (Shakura \& Sunyaev 1973). There are several variants
of this parametrization; we use here a version in which the vertically
averaged sheer viscosity is taken proportional to the vertically averaged (total) pressure,
\begin{equation}
\bar{t_{\phi r}} \equiv \frac{1}{h} \int_0^h t_{\phi r} dz = \alpha \bar P,
\end{equation}
in which case
\begin{equation}
\label{inttfir}
\int_0^h t_{\phi r} dz = h \alpha \bar P = m_0 \alpha (\bar P/\bar\rho),
\end{equation}
where the averaged density is given by $\bar\rho = m_0/h$. The vertically
averaged kinematic viscosity is given by substituting Eqs. (\ref{vbal}) integrated over
$z$ into Eq. (\ref{inttfir}),
\begin{equation}
\bar w = \alpha\, \frac{2}{3} \left(\frac{R^3}{GM}\right)^{1/2} \frac{B}{A} 
\left(\frac{\bar P}{\bar\rho}\right).
\end{equation}
A disadvantage of the $\alpha$-prescription is that the vertically averaged kinematic
viscosity, and the total column mass are not known a priori since they depend on
$(\bar P/\bar\rho)$, which can only be accurately computed when the model is
constructed. 

In the case of dominant radiation pressure, one can, however, derive a relation
between $\bar P$ and $\bar\rho$ prior to solving for the detailed structure
(e.g., Hubeny \& Hubeny 1998). In this case,
the vertically-averaged kinematic viscosity is given through $\alpha$ as
\begin{equation}
\label{barw_alp}
\bar{w}  = 2 \dot M^2 \alpha \, \left({G M  \over R^3}\right)^{1/2}\!
\left({\sigma_{\rm e} \over 8 \pi m_H c} \right)^2 \, {D^2 \over A C} ,
\end{equation}
and, with the $\alpha$-parametrization of viscosity, the total
column mass is given by
\begin{equation}
\label{m0_alp}
m_0  = {16 \pi \over 3} \, \left( {m_H c \over \sigma_{\rm e}} \right)^2
 \left({ R^3 \over G M }\right)^{1/2}\! {1 \over \dot M \alpha }
\, {B C \over A D} \, ,
\end{equation}

In the case where the radiation pressure is not dominant, the (approximate) relation
between the total column mass $m_0$ and the viscosity parameter
$\alpha$ is given by 
\begin{equation}
\label{m0}
\alpha m_0 \left(a + \beta m_0^{1/4}\right) -\gamma =0.
\end{equation}
where
\begin{equation}
a=(\sigma_R T_{\rm eff}^4 \chi_{\rm e}/c)^2/(3Q),
\end{equation}
\begin{equation}
\beta=0.8 R_g\, \kappa_0^{1/8} (2Q/\pi R_g)^{1/16} T_{\rm eff}^{1/2},
\end{equation}
and
\begin{equation}
\gamma= (\dot M /4\pi) (GM/R^3)^{1/2} (D/A),
\end{equation}
where $R_g=1.3 \times 10^8$ is the gas constant, 
$\chi_{\rm e}=\sigma_{\rm e}/m_H=0.39$, and $\kappa_0 =6.4\times 10^{24}$ is
the coefficient in the Kramers-type expression for the Rosseland mean opacity,
\begin{equation}
\kappa_R \approx \kappa_0 \rho T^{-7/2}.
\end{equation}
Equation (\ref{m0}) follows from Eqs. (\ref{mbal}) and (\ref{inttfir}), where
one makes an approximation that
\begin{equation}
(\bar P/\bar\rho) \approx (\bar P/\bar\rho)_{\rm rad} + (\bar P/\bar\rho)_{\rm gas},
\end{equation}
where $(\bar P/\bar\rho)_{\rm rad}$ and $(\bar P/\bar\rho)_{\rm gas}$ correspond
to the radiation-pressure dominated and gas-pressure dominated situation, respectively.

The non-linear equation (\ref{m0}) for $m_0$ is solved by the Newton-Raphson
method. In the case of negligible gas pressure we have  $\beta m_0^{1/4} \ll a$,
so Eq. (\ref{m0}) becomes a simple linear equation for $m_0$. It can be easily
verified that its solution is identical to that given by Eq. (\ref{m0_alp}).\\

\noindent$\bullet$\,\textsf{Parametrization of the local viscosity}

The (depth-dependent) viscosity $w$ is allowed to vary as a step-wise 
power law of the mass column density, viz. 
\index{Viscosity!local}
\begin{equation}
\label{visc1}
w(m) = w_0 \left( {m/m_0} \right)^{\zeta_0}\, , \quad m>m_{\rm d} \, ,
\end{equation}
\begin{equation}
\label{visc2}
w(m) = w_1 \left( {m/m_0} \right)^{\zeta_1}\, , \quad m<m_{\rm d} \, ,
\end{equation}
where $m_{\rm d}$ is the division point.
In other words, we allow for a different power-law exponent for inner and
outer layers, and also for a different portion of the total energy dissipated in these
layers. This represents a generalization
of an approach we used previously, based on a single power-law representation.
For details, refer to Paper~III, \S\,\refnsttwodisk.

We stress that this parametrization is an empirical one. The most natural way
of treating local viscosity would be to keep the coefficient of kinematic viscosity
constant with depth, i.e., $\zeta_0=\zeta_1=0$. We have originally introduced
a power-law parametrization to avoid numerical problems at the surface
(K\v r\'\i\v z \& Hubeny 1986), where $\zeta_1 > 0$. One can also simulate
a dissipation that occurs preferentially at the surface layers, with viscosity
increasing with decreasing depth which  leads to the formation of a disk corona. 
In this case one choses $\zeta_1 < 0$. Parameter $\zeta_0$ is typically taken as 0.

\subsubsection{Actual form of structural equations to be solved}

As in the case of stellar atmospheres, the basic geometrical coordinate is
the column mass, $m$. The above structural equations, supplemented by 
the radiative transfer equation, are written as follows:\\

\noindent$\bullet$\,\textsf{Radiative transfer equation}
\index{Radiative transfer equation!disks}

This equation, and its upper boundary condition,  is exactly the same as 
for stellar atmospheres, equations (\ref{rte}) and (\ref{rte_lbc}); the only difference
is the lower boundary condition that represents a symmetry condition
at the midplane,
\begin{equation}
\label{rte_lbcd}
H_\nu = 0, \quad {\rm or} \quad \frac{d(f_\nu J_\nu)}{d\tau_\nu} = 0.
\end{equation}

\noindent
$\bullet$\,\textsf{Vertical hydrostatic equilibrium equation}
\index{Hydrostatic equilibrium!disks}

Equations (\ref{hedisk}), using Eq. (\ref{gdef}) is rewritten as
\begin{equation}
\label{heqz}
\frac{dP}{dm} = Q z,
\end{equation}
where
\begin{equation}
\label{qdef}
Q \equiv \frac{GM}{R^3} \frac{C}{B}.
\end{equation}

\noindent$\bullet$\,\textsf{Energy balance}
\index{Energy balance equation!disks}

Integrating Eq.\,(\ref{ebal}) over $z$, and using Eq.\,(\ref{mbal}), one obtains for the
total energy flux at the surface which is expressed, in analogy to stellar atmospheres,
through the effective temperature,
\index{Effective temperature!disks}
\begin{equation}
\label{teff_def}
F_z(h) \equiv \sigma_R T_{\rm eff}^4\  =\ 
{3\over 8\pi} {G M \dot M \over R^3}\,  {D \over B}\, ,
\end{equation}
The integral form of the energy balance equation follows directly from
Eqs. (\ref{ebal}), (\ref{vbal}), and (\ref{kinvisc}):
\begin{equation}
\label{ebal_int}
4\pi \int_0^\infty (\eta_\nu- \kappa_\nu J_\nu) d\nu = 
\frac{9}{4} \frac{GM}{R^3} \left(\frac{A}{B}\right)^2 \rho w,
\end{equation}
where the left-hand side, analogous to the case of stellar atmospheres, expresses
the net energy radiated away from the unit volume, while the right-hand side 
expresses the total energy generated by viscous dissipation in the same unit
volume.
The differential form, in analogy with the atmospheric case, is written as:
\begin{equation}
\label{ebal_dif}
4\pi\int_0^\infty \frac{d(f_\nu J_\nu)}{d\tau_\nu} d\nu = \sigma_R T_{\rm eff}^4
[1-\theta(m)],
\end{equation}
where the function $\theta$ is defined by
\begin{equation}
\label{theta}
\theta(m) \equiv \frac{1}{\bar m_0} \int_0^m w(m^\prime)\,dm^\prime,
\end{equation}
which is a monotonically increasing function of $m$ with $\theta(0)=0$ and
$\theta(m_0)=1$, for any dependence of the local viscosity on depth.

As in the case of stellar atmospheres, one uses here a linear combination of
Eqs. (\ref{ebal_int}) and (\ref{ebal_dif}). If convection is present, the convective
flux is added analogously as in Eq. (\ref{re_conv}) for stellar atmospheres.
The complete energy balance equation then reads:
\begin{eqnarray}
\label{re_con_disk}
\alpha\bigg[\int_0^\infty\!\!\left(\kappa_\nu J_\nu - \eta_\nu\right)d\nu +E_{\rm diss}
+ \frac{\rho}{4\pi}\frac{dF_{\rm conv}}{dm}  \bigg] + \nonumber \\
\beta \bigg[\int_0^\infty\! \frac{d(f_\nu J_\nu)}{ d\tau_\nu}\,d\nu -
\frac{\sigma_R }{ 4\pi}\,T_{\rm eff}^4[1-\theta(m)]
+ \frac{F_{\rm conv}}{ 4\pi}\bigg] = 0,
\end{eqnarray}
where
\begin{equation}
\label{ediss}
E_{\rm diss} = \frac{9}{16\pi} \frac{GM}{R^3} \left(\frac{A}{B}\right)^2 \rho w
\end{equation}
is the total energy generated by viscous dissipation per unit volume.\\
%--------------------------

\begin{table}
\caption{Comparison of stellar atmospheres and accretion disks}
\begin{center}
\begin{tabular}[h]{|c|c|}
\hline
\hline
Stellar atmospheres  &  Accretion disks  \\ 
\hline
\hline
radiative equilibrium:   &  no radiative equilibrium: \\
$\int (\eta_\nu - \kappa_\nu J_\nu)d\nu =0$  & 
$\int (\eta_\nu - \kappa_\nu J_\nu)d\nu =E_{\rm diss}$\\
\hline
$g$ constant    &  $g=Q z$\\ & (dependent on vertical distance) \\
\hline
$\tau_{\rm tot} \rightarrow \infty$  &  $\tau_{\rm tot}$ finite; not a priori known\\
\hline
Lower boundary: diffusion approx.   &  Lower boundary: symmetry   \\
\hline
Basic parameters: $T_{\rm eff}$, $g$, $[A]$  &  
Basic parameters: $T_{\rm eff}$, $Q$, $m_0$, $[A]$\\
\hline
\end{tabular}
\end{center}
Here, $E_{\rm diss}$ is the dissipated energy given by 
Eq. (\ref{ediss}), $Q$ is the gravity acceleration parameter given by Eq. (\ref{qdef}),
$\tau_{\rm tot}$ denotes any optical depth (monochromatic or mean) at the midplane,
and $[A]$ denotes a collection of chemical abundances of all species that are
taken into account (see below). The entry ``Lower boundary'' means the form of the
lower boundary condition for the radiative transfer equation.
Notice that in the case of disks, the parameters
$T_{\rm eff}$, $Q$, $m_0$ are not the fundamental ones, but are evaluated from
the more fundamental parameters that specify the disk and the ring within it, namely
$M$, $\dot M$, $R$, $\alpha$ (or Re), and, in the case of black-hole disks, the spin $a$.

\end{table}

%---------------------------------------------------------

\noindent$\bullet$\,$z$-$m$ \textsf{relation}

Unlike the case of stellar atmospheres, where the vertical coordinate $z$
has no special meaning, for disks it has to be calculated for given column
mass and current density because the gravitational acceleration depends
on it through equations (\ref{hedisk}) and (\ref{gdef}). The $z$-$m$ relation,
given by Eq. (\ref{dm}) has to be considered as one of the structural equations.\\

\noindent$\bullet$\,\textsf{Other structural equations}

The rest of structural equations, namely the kinetic equilibrium equations
for explicit levels, equation of state, and charge conservation equation, together
with auxiliary equations such as the definition of absorption and emission coefficients,
are local, and therefore the same as in the case of stellar atmospheres.

% --------------------------------------------------------------------------

\subsection{Kinetic equilibrium and NLTE}
\label{kee}

\noindent
$\bullet\,$\textsf{Kinetic equilibrium equations}

Since one does not assume Local Thermodynamic Equilibrium (LTE), the
atomic level populations have to be determined by solving the corresponding 
kinetic equilibrium equation. To this end, one selects the set of chemical species,
each with a number of ionization stages, and for each stage a number of energy 
levels, for which the kinetic equilibrium equation is solved. These are called
{\em explicit atoms},
\index{Explicit atoms}
{\em explicit ions}, and {\em explicit levels}, respectively. Their choice
\index{Explicit ions}
\index{Explicit levels}
is completely driven by input data, described in Paper~III, Chaps.\,\refginput, \refnew,
and \refions.

For each explicit atomic species, $I$, the set of kinetic equilibrium equations may be
written as
\begin{equation}
\label{ke1}
\boldmath{\mathcal{A}}_I \cdot {\bf n}_I = {\bf b}_I,
\end{equation}
where $\boldmath{\mathcal{A}}$ is the rate matrix, 
${\bf n}\equiv \{n_1,n_2,\ldots,n_{N\!L_I}\}$ is a vector of populations 
(number densities) of all explicit levels
(in all ionization stages); their number
being $N\!L_I$.
In the following, we will drop subscript $I$.

We stress that the term ``level'' means here either a genuine energy eigenstate
of an atom/ion, or several states lumped together (for instance a level composed
of all components of a multiplet), or even the so-called {\em superlevel}, which
\index{Superlevels}
is an aggregate of many different energy levels; for details, see \S\,\ref{superl}.

The elements of the rate matrix $\boldmath{\mathcal{A}}$ are given by 
\begin{eqnarray}
\label{ratmat}
{\mathcal{A}}_{ii} &=&\sum_{j\not= i}(R_{ij}+C_{ij}),\\
{\mathcal{A}}_{ij} &=&-(R_{ji}+C_{ji}),\ {\mathrm{for}}\ j\neq i\ {\mathrm{and}}
\ i\neq k,\\
{\mathcal{A}}_{kj} &=&1,\hspace{5.5em} {\mathrm{for}}\ j=1,\ldots,N\!L_I,
\end{eqnarray}
where $R_{ij}$ and $C_{ij}$ are the radiative and collisional rates,
respectively, and
$k$ is the index of the \textit{characteristic level}, i.e. the level for
\index{Characteristic level}
which the rate equation is replaced by the particle conservation (abundance
definition) equation. 

Assuming $i<j$, the radiative rates are given by 
\index{Radiative rates}
\begin{eqnarray}
\label{rij}
R_{ij} &=& \frac{4\pi}{ h} \int_{0}^\infty \frac{\sigma_{ij}(\nu)}{ \nu} J_\nu d\nu, \\
\label{rji}
R_{ji} &=& \frac{4\pi}{ h} \int_{0}^\infty \frac{\sigma_{ij}(\nu)}{ \nu} G_{ij}(\nu)
\left(\frac{2h\nu^3 }{ c^2} + J_\nu \right)\! d\nu,
\end{eqnarray}
where $\sigma_{ij}(\nu)$ is the corresponding cross section, and
$G_{ij}(\nu)$ is defined by
\begin{equation}
\label{gfac}
G_{ij}(\nu)\equiv
\left\{ \begin{array}{ll}
g_i/g_j, & \mathrm{for~bound\!\!-\!\!bound}, \\ [6pt]
n_{\mathrm{e}}\Phi_i(T)\exp(-h\nu/kT), & \mathrm{for~bound\!\!-\!\!free}.
\end{array} \right.
\end{equation}
where $g_i$ is the statistical weight, $n_{\rm e}$ the electron density,
and $\Phi_i(T)$ is the Saha-Boltzmann factor,
\index{Saha-Boltzmann factor}
\begin{equation}
\label{sgb}
\Phi_{i}(T)=\frac{g_i}{ 2g_1^+}\left(\frac{h^2}{ 2\pi m_{\mathrm{e}} kT}\right)^{3/2}
e^{(E_I-E_i)/kT};
\end{equation} 
$E_I$ is the ionization potential of the ion to which level $i$ belongs; $E_i$
is the excitation energy of level $i$, $g_1^+$ is the statistical weight of
the ground state of  the next ion, and $m_{\rm e}$ the electron mass.

If one adopts an occupation probability formalism that describes bound level 
dissolution resulting from perturbations with neighboring particles
(see \S\,\ref{dissol})
the above equation remains the same, replacing
$\sigma_{ij}(\nu) \rightarrow \sigma_{ij}(\nu) w_j$, and 
\begin{equation}
\label{gfac2}
G_{ij}(\nu) =
\left\{ \begin{array}{ll}
(g_i w_i)/(g_j w_j), & {\mathrm{for~bound\!\!-\!\!bound}}, \\ [6pt]
(w_i/w_j) n_{\mathrm{e}}\Phi_i(T)\exp(-h\nu/kT), & {\mathrm{for~bound\!\!-\!\!free}}.
\end{array} \right.
\end{equation}
where $w_i$ is the occupation probability of level $i$.

In many instances one employs the concept of {\em detailed radiative balance}.
It is defined such as the transition $i\leftrightarrow j$ is in detailed radiative
balance if $n_iR_{ij} = n _j R_{ji}$. This is numerically equivalent to setting
$R_{ij} = R_{ji} = 0$. This concept is often used to compute intermediate NLTE
models because they are usually easier to converge than the full NLTE models.

The collisional rates, assuming that only collisions with free electrons are
\index{Collisional rates}
important (again, $i<j$), are given by
\begin{eqnarray}
\label{crat2}
C_{ij} &=& n_{\mathrm{e}} \Omega_{ij},\nonumber \\
C_{ji} &=& (n_i^\ast/n_j^\ast) C_{ij},
\end{eqnarray}
where $\Omega_{ij}$ is the collisional cross section, and $n_i^\ast$
the LTE population of level $i$.

The complete set of kinetic equilibrium equations is written as
\begin{equation}
\label{ke2}
\mathcal{A}\cdot {\bf n} = {\bf b},
\end{equation}
where the full rate matrix $\boldmath{\mathcal{A}}$ is a block--diagonal matrix
composed of all the individual matrices $\boldmath{\mathcal{A}}_I$, and in a
similar manner for the vector of populations $\bfn$ and the right--hand--side
vector $\bfb$ -- see below.

In LTE, the kinetic equilibrium equation in the form of Eq. (\ref{ke2}) is
not solved. Instead, the atomic level populations are determined by the
Saha-Boltzmann relation,
\begin{equation}
\label{lte}
n_i^\ast = n_{\rm e} n_1^+ \Phi_i(T),
\end{equation}
where $n_1^+$ is the population of the ground state of the next higher ion.
%As is customary, superscript ${}^\ast$ denotes LTE population. 
It is often instructive
to introduce the NLTE {\em departure coefficient}, or a $b$-factor, as
\begin{equation}
\label{bfac}
b_i \equiv n_i/n_i^\ast,
\end{equation}
that shows a magnitude of NLTE effects for a given level.\\

\noindent
$\bullet\,$\textsf{Particle conservation equation}
\index{Particle conservation equation}

The set of rate equations for all levels of an atom would form a linearly 
dependent system. Therefore, one equation of the set has to be replaced by the
\textit{number conservation}; also called the \textit{abundance definition
equation}, 
\index{Abundance definition equation}
\begin{equation}
\label{ade}
\sum_{i=1}^{N\!L_I} n_i(1+S_i) = N_I \equiv A_I N_H =
(N-n_{\mathrm{e}}) \alpha_I,
\end{equation}
where $A_I \equiv N_I/N_H$ is the
\textit{abundance} of the species $I$, defined here as a ratio of the total
\index{Abundance}
number of atoms $I$, in all degrees of ionization, to the total number of
hydrogen atoms, per unit volume.  The last equality introduces
the notion of \textit{fractional abundance}, $\alpha_I=A_I \big/\sum_J A_J$ 
of the chemical element $I$. The summation extends over all species,
including hydrogen (for which, by definition, $A_H=1$). {\sc tlusty} usually
works in terms of chemical abundances with respect to hydrogen, but it
also allows to define an abundance with respect to any other species,
called {\em reference species}, to be able to compute models of
extremely hydrogen--poor or even hydrogen--free atmospheres.

The set of abundances of the individual species forms another basic
input parameter of the problem.

The factor $S_i$
accounts for the (LTE) populations of the higher, non-explicit levels, 
and is given by
\begin{equation}
\label{upsum0}
S_i = \left\{ \begin{array}{ll}
0 & {\mathrm{if\ }} i {\mathrm{\ is\ not\ the\ ground\ state\ an\ ion}} ,\\ [5pt]
n_{\mathrm{e}} \Sigma_J  &  {\mathrm{if\ }} i {\mathrm{\ is\ the\ ground\ state\
       of\ ion\ }}J+1 .
\end{array} \right.
\end{equation}
where $\Sigma_J$ is the so-called upper sum for an ion $J$ which
\index{Upper sum}
expresses the total population of higher, non-explicit states of ion $J$.
In the previous versions of {\sc tlusty} it was expressed through
its partition function as (Hubeny 1988)
\begin{equation}
\label{upsum1}
\Sigma_J = (h^2/2\pi m_{\rm e} kT)^{3/2} e^{\chi_J/kT} (U_J/g_1^{J+1})
- \sum_{i=1}^{N\!L_J} w_i \Phi_i(T),
\end{equation}
where the first term represents the total population of ion $J$, and the
second term the LTE population of all explicit levels. Here  $\chi_J$ the 
ionization energy of  ion $J$, and $U_J$ its partition function. However, an 
evaluation of the upper sum using Eq. (\ref{upsum1}) may sometimes be inaccurate.
Moreover, when one uses a model atom with many explicit levels, the contribution
of the upper levels is very small. Therefore, a safe and still reasonably accurate
approach for models with a sufficient number of explicit levels 
is to set the upper sum to zero,
\begin{equation}
\Sigma_J=0.
\end{equation}
This is actually used in {\sc tlusty}, version 205, by default for all ions
except neutral hydrogen and hydrogenic ions. 

An accurate way of expressing the contribution of upper states is by
introducing the so-called {\em merged level}. This approach
is currently used only for hydrogen and hydrogenic ions. 
A merged level is a sort of superlevel (see \S\,\ref{superl}) representing all
\index{Merged level!definition}
levels higher than the highest explicit regular level. Its population is given 
by the sum of LTE populations of these states, computed with the 
occupation probability
(see \S\,\ref{dissol}). 
\begin{equation}
n_{\rm mer} = \sum_{j=N\!L+1}^{N^\ast}\!\!\! n_j^\ast =\!\!
\sum_{j=N\!L+1}^{N^\ast} \!\!n_{\rm e} n_1^+ w_j \Phi_j(T),
\end{equation}
where $n_1^+$ is the population of the ground state of the next ion,
which in the case of hydrogen is the proton number density, $n_1^+\!=\! n_{\rm p}$.
$N^\ast$ is a large number, taken in {\sc tlusty} as 80, but its exact value does
not matter because the occupation probability for such high states is
essentially zero. 

A merged level is treated as an explicit level with statistical weight
$g_{\rm mer} = \sum_{j=N\!L+1}^{N^\ast} w_j g_j \exp(E_j/kT)$, so that its
Saha-Boltzmann factor is simply given by
$\Phi_{\rm mer}(T) = C T^{-3/2} g_{\rm mer}/(2g_1^+)$, 
with $C=(h^2/2\pi m_{\rm e}k)^{3/2}$.
Its energy is formally set
to the ionization energy, $E_{\rm mer}= E_{\rm ion}$, and its occupation
probability $w_{\rm mer} =1$ because the occupation probabilities of the
individual components are already included in $g_{\rm mer}$.
For a treatment of
transitions involving the merged level, see \S\,\ref{trans}. If the merged
level is set for an atom/ion $I$, then the upper sum has to be set to zero, 
$\Sigma_I=0$,
and all the level parameters are computed analytically.

The choice of the 
\index{Characteristic level}
characteristic level, denoted $k$, is arbitrary. Usually, one either choses the last level
($k=N\!L_I$), or a level with the highest population. The elements of the right-hand-side vector $\bf b$ are
given by 
\begin{equation}
\label{bvec}
b_i = (N-n_{\mathrm{e}})\alpha_I\,\delta_{ki},
\end{equation}
that is, the only non-zero element of $\bfb$ is the term corresponding to level
$k$.\\

\noindent
$\bullet$\,\textsf{Charge conservation equation}

The condition of global charge neutrality is expressed as
\index{Charge conservation equation}
\begin{equation}
\label{chce}
\sum_i n_i Z_i + Q - n_{\mathrm e} = 0,
\end{equation}
where $Z_i$ is the charge associated with level $i$; that is, $Z_i=0$ for levels
of neutral atoms, $Z_i=1$ for levels for once ionized ions, etc.,
The summation in Eq. (\ref{chce}) extends over all levels of
all ions of all species. Quantity $Q$ accounts for the additional charge coming
from the ions that are not treated explicitly -- see \S\,\ref{eos}.
\\

\noindent
$\bullet$\,\textsf{Mass density, and fictitious massive particle density
equations}

The mass density is expressed in terms of atomic level populations as
\begin{equation}
\label{rho}
\rho=\sum_{i} m_i n_i,
\end{equation}
where $m_i$ is the mass of the atom to which level $i$ belongs. It can also be
expressed as
\begin{equation}
\label{rho2}
\rho=(N-n_{\mathrm{e}})\,\mu m_H,
\end{equation}
where $m_H$
is the mass of the hydrogen atom, and $\mu$
the \textit{mean molecular weight}, defined by
\index{Mean molecular weight}
\begin{equation}
\label{mu}
\mu = \sum_{I}{\alpha_I(m_I/m_H)} = \frac{\sum_{I}{A_I(m_I/m_H)}}{\sum_I A_I}
\end{equation}
where $m_I$ is the mass of an atom of species $I$.

One can also introduce a \textit{fictitious massive particle density}, defined as
\index{Fictitious massive particle density}
\begin{equation}
\label{mpd}
n_{\mathrm{m}} \equiv (N-n_{\mathrm{e}})\mu,
\end{equation}
so that the mass density can be written as
\begin{equation}
\label{rho3}
\rho=n_{\mathrm{m}} m_H.
\end{equation}
This option is offered in {\sc tlusty}, but is rarely used.\\

\subsection{Absorption, emission, and scattering coefficients}
\label{absoemisscat}

\noindent
$\bullet$\,\textsf{Absorption}

The above set of structural equations has to be complemented by equations
defining the absorption and emission coefficients.

The absorption coefficient (or opacity) is given by 
\index{Absorption coefficient}
\begin{equation}
\label{chi}
\chi_\nu = \kappa_\nu + \kappa_\nu^{\mathrm{sc}}\, ,
\end{equation}
where $\kappa_\nu$ is the extinction coefficient (sometimes called the
true absorption coefficient), and $\kappa_\nu^{\mathrm{sc}}$ is the scattering coefficient. The extinction coefficient is given by
\begin{eqnarray}
\label{kappa}
\kappa_\nu&=&\sum_i\sum_{j>i}\left[n_i-n_j G_{ij}(\nu)\right]\sigma_{ij}(\nu)
+\sum_i\left[n_i-n_k G_{ik}(\nu)e^{-h\nu/kT}\right]\sigma_{ij}(\nu)\nonumber\\
      &+&\sum_\kappa n_{\mathrm{e}}n_{\kappa}{\sigma}_{\kappa\kappa}(\nu,T)
\left(1-e^{-h\nu/kT}\right) + \kappa_\nu^{\mathrm{add}},
\end{eqnarray}
where the first term represents the contribution of the bound--bound transitions,
the second term the bound--free transitions (with $k$ being the final state of
the corresponding process), and the third term the free--free transitions. The
summations extends over all level of all species. The term
$\kappa_\nu^{\mathrm{add}}$ represents any additional opacity. It is used to
account for opacity sources that are not written in terms of explicit
bound--bound or bound--free transitions.

The scattering part of the absorption coefficient is given by
\begin{equation}
\label{kappasc}
\kappa_\nu^{\mathrm{sc}} = n_{\mathrm{e}} \sigma_{\mathrm{e}} + 
\sum_{i}n_{i} \sigma_{\mathrm{Ray},i}
\end{equation}
where the first term accounts for electron scattering, and the second
term for the Rayleigh scattering.
Here $\sigma_{\mathrm{e}}$ is the Thomson cross section,  
\index{Thomson scattering}
\index{Rayleigh scattering}
$\sigma_{\mathrm{Ray},i}$ is the Rayleigh scattering cross section of species
$i$, and $n_i$ is the number density of species $i$. The summation extends over
all species for which Rayleigh scattering gives a non--negligible contribution to
the total scattering opacity. 

For high-energy objects (extremely hot white dwarfs or hot accretion disks), one
may consider an inelastic electron scattering, called Compton scattering, 
as described in Hubeny et al. (2001). A brief outline is given below, and for more
details refer to Paper~III, \S\,\refnsttwocompt.\\

\noindent
$\bullet$\,\textsf{Emission}

The total emission coefficient is also given as a sum of thermal and scattering
contributions. The latter refers only to \textit{continuum scattering};
scattering in spectral lines is usually treated with complete frequency redistribution, 
in which case the scattering is in fact a part of the thermal emission coefficient. The continuum scattering part is usually treated separately from the thermal part,
and the ``thermal emission coefficient'' is usually called the ``emission
coefficient.'' Specifically,
\index{Emission coefficient}
\begin{equation}
\label{eta}
\eta_\nu^{\mathrm{tot}}=\eta_\nu + \eta_\nu^{\mathrm{sc}},
\end{equation}
where
\begin{eqnarray}
\label{etanu}
\eta_\nu&=&(2h\nu^3/c^2) \bigg[\sum_{j}\sum_{j>i} n_jG_{ij}(\nu)\sigma_{ij}(\nu)
+ \sum_{i} n_k G_{ik}(\nu) \sigma_{ik}(\nu)e^{-h\nu/kT} \nonumber \\
&+&\sum_{\kappa}n_{\mathrm{e}} n_{\kappa}\sigma_{\kappa\kappa}(\nu,T)e^{-h\nu/kT}
\bigg] + \eta_\nu^{\mathrm{add}}.
\end{eqnarray}
The additional emissivity, if included, is usually given by
$\eta_\nu^{\mathrm{add}}=\kappa_\nu^{\mathrm{add}} B_\nu$; with $B_\nu$
being the Planck function. The form of the
scattering part of the emission coefficient depends on additional assumptions
-- see below.
In the simple case that we assume an isotropic phase function,
and electron scattering is treated as coherent Thomson scattering, then
\begin{equation}
\label{etasc}
\eta_\nu^{\mathrm{sc}} = \kappa_\nu^{\mathrm{sc}} J_\nu.
\end{equation}

\noindent
$\bullet$\,\textsf{Scattering}

Various scattering mechanisms are treated differently, depending on their
nature. The three types of scattering {\sc tlusty} is designed to treat are
the following:\\ [3pt] 
\noindent -- {\em Thomson and Rayleigh scattering}.
\index{Rayleigh scattering}
These scattering processes are coherent (without a change in frequency
of the absorbed and scattered photon), and they are also approximated to be isotropic. Consequently, their treatment is simple.
As follows from equations (\ref{kappasc}) and (\ref{etasc}), the total source
function is given by
\begin{equation}
\label{scatsf}
S_\nu^{\rm tot} = \frac{\eta_\nu}{\chi_\nu} + 
\frac{\kappa_\nu^{\rm sc}}{\chi_\nu} J_\nu,
\end{equation}
so that the scattering part of the source function, the second term in
equation (\ref{scatsf}), is simply proportional to the mean intensity at the
same frequency, which does not cause any numerical complications.\\ [3pt]
\noindent -- {\em Scattering in spectral lines},
If one assumes complete frequency redistribution in all lines, as is
done in equation (\ref{etanu}), the emission coefficient for lines does not
contain the radiation intensity explicitly (it enters implicitly through the
solution of kinetic equilibrium equation that contains radiative rates),
so that formally there is no distinction between the thermal and scattering
processes in a line.

{\sc tlusty} also offers an approximate treatment of partial frequency redistribution,
through the so-called partial coherent scattering approximation; for a detailed
discussion, refer to Hubeny \& Mihalas (2014; \S\,15.3).
This option is rarely used, and is briefly described in \S\,\refnsttwoprd. \\ [3pt]
\noindent -- {\em Compton scattering}
is a non-coherent electron scattering, which is
particularly important for high temperature models at high photon
energies.. It is important for hot accretion disks (Seyfert galaxies,
X-ray binary disks), and also for extremely hot white dwarfs, possibly
with a hydrogen-burning layer on their surface (the so-called super-soft sources),
or pre-white dwarfs.
Its treatment in {\sc tlusty} follows from the formalism of Hubeny et al. (2001),
which is based on a Kompaneets approximation (e.g., Rybicki \& Lightman,
1979), for which the
Compton scattering part of the source function is given by
\begin{equation}
\label{compt}
S_\nu^{\rm Compt} = (1-x) J_\nu +(x-3\Theta)J_\nu^\prime 
+ \Theta J_\nu^{\prime\prime} + \frac{c^2}{2h\nu^3}J_\nu 2x 
(J_\nu^\prime - J_\nu),
\end{equation}
where 
\begin{equation}
x=\frac{h\nu}{m_{\rm e}c^2}, \quad \Theta=\frac{kT}{m_{\rm e}c^2}, 
\end{equation}
and
\begin{equation}
\label{compt_der}
J_\nu^\prime  \equiv \frac{\partial J_\nu}{\partial\ln\nu},\quad 
J_\nu^{\prime\prime} \equiv \frac{\partial^2 J_\nu}{\partial(\ln\nu)^2}.
\end{equation}
A discretization of equations (\ref{compt}) and (\ref{compt_der})
and their implementation in the linearization scheme, is described in detail in 
Hubeny et al. (2001; Appendix A); see also Paper~III, \S\,\refnsttwocompt.

%---------------------------------------------------------------------------------------------------

\subsection{Atomic transition processes}
\label{trans}

A treatment of atomic transitions, in particular the corresponding
cross sections, is an important ingredient of the modeling process,
for both stellar atmospheres and accretion disks. Although handling
of transition cross sections is essentially transparent for a casual user 
because most of the atomic data is communicated to the code through
the set of already prepared atomic data files (see Paper~III, Chap.\,\refions), there
is still a possibility for the user to choose from several options, or 
add specific processes that operate on, or are important only for certain 
classes of objects. This is done by specific keyword parameters, that are
be described in Paper~III, Chap.\,\refnsttwophys.

The two essential classes of transitions are the {\em radiative} (an interaction
of an atom/ion with a photon) and the {\em collisional} (an interaction of an
atom with another particle, typically electron) transitions. We will briefly describe
them in the following.

\subsubsection{Bound-bound radiative transitions}

{\sc tlusty} distinguishes two types of bound-bound
processes (lines):\\
-- ordinary line -- a transition between two ordinary levels;\\
-- superline -- a  transition involving a superlevel, i.e. a transition between an
\index{Superlines}
ordinary level and superlevel, or between two superlevels.\\

\noindent$\bullet\,$ {\sf Ordinary lines}

The cross section for an ordinary line that enters equations
(\ref{rij}) and (\ref{rji}) for radiative rates, as well as equations (\ref{kappa}) and
(\ref{etanu}) for the opacity and emissivity, is given by
\begin{equation}
\sigma_{ij}(\nu) = \frac{\pi e^2}{m_{\rm e} c} f_{ij} \phi_{ij}(\nu),
\end{equation}
where $f_{ij}$ is the oscillator strength, and $\phi_{ij}(\nu)$ is the
normalized absorption profile coefficient. There are several
possibilities to adopt for the profile coefficient:\\ [6pt]
(i) {\em Doppler profile}, 
\index{Doppler profile}
\begin{equation}
\phi(\nu)=\frac{1}{\sqrt{\pi}\Delta\nu_D}\exp(-x^2),
\end{equation}
where
$x=(\nu-\nu_{ij})/\Delta\nu_D$ is the frequency displacement  from 
the line center, $\nu_{ij}$,
expressed in units of the Doppler  width, $\Delta\nu_D$. The latter is
given by
\index{Doppler width}
\begin{equation}
\Delta\nu_D= \frac{\nu_{ij}}{c}\sqrt{\frac{2kT}{m_A}+ v_{\rm turb}^2},
\end{equation}
where $m_A$ is the mass of the atom, and $v_{\rm turb}$ is the turbulent
velocity. The profile depends on temperature, and therefore depends
on depth and changes from iteration to iteration. 
Experience revealed (e.g., Werner 1987) that one can safely use
a simplified approach in which a characteristic temperature
is chosen [typically $T=(3/4) T_{\rm eff}$] with which the line profile
is computed at the beginning and is stored for further use during the 
whole iteration  process.
This option is offered in {\sc tlusty}, and is often being used,
although considering depth-dependent
Doppler profile does not lead to any significant additional 
time consumption.
There is an additional subtle point. The frequency points for a line 
are chosen to provide accurate values of the integrals over frequency.
Their setting is also based on using a characteristic temperature.
When a modest number of frequency points is considered,
using depth-independent Doppler profile assures that the evaluation
of radiative rates and other integrals over frequency is accurate at all
depth points, which is not necessarily true if a depth-dependent profile
is assumed. In order to treat a depth-dependent profile accurately at
all depth points, one would need to select many more frequency points.\\ [6pt]
(ii) {\em Voigt profile}, 
\index{Voigt profile}
\begin{equation}
\label{voigt}
\phi(\nu)= \frac{1}{\sqrt{\pi}\Delta\nu_D} H(a,x),
\end{equation}
where $H$ is the Voigt
function, and $a$ is the damping parameter expressed in units of
Doppler width, $a=\Gamma/(4\pi\Delta\nu_D)$. The damping parameter
\index{Damping parameter}
is usually composed of three parts, corresponding to natural (lifetime) 
broadening, Stark broadening, and Van der Waals broadening. \\ [6pt]
(iii) {\em Special hydrogen line profiles}. For the purposes of model
\index{Hydrogen line broadening}
construction it is possible to consider hydrogen lines with a Doppler
profile, as was done in the past, see e.g. Mihalas (1978). However,
it is preferable to take the profile coefficients in a more accurate form.
{\sc tlusty} offers three possibilities, ordered here with increasing accuracy:\\ [2pt]
-- a simple approximation by Hubeny, Hummer, \& Lanz (1995; Appendix A), 
that essentially gives a Doppler profile in the line center, a Holtsmark profile 
\index{Holtsmark profile}
in the wings, and a simple bridging  law in the intermediate region.\\
-- using tables computed by Lemke (1997) for the first  members 
of the Lyman, Balmer, Paschen, and Brackett series up to the higher level
with main quantum number $n=10$ (the tables contain more lines, but the
results for lines with  $n>10$ were found to be incorrect);\\
-- using Tremblay-Bergeron (2009) tables, that take into account the
effects of level dissolution, and are therefore the most accurate ones, in particular
for high densities, and so preferable for instance for white dwarf models.
\smallskip

To describe an ordinary line, one therefore needs the {\em oscillator strength}
(or, equivalently, one of the Einstein coefficients), the line center frequency
(which is given by the difference of level energies), a switch indicating the
choice of the type of profile, and in case of a Voigt profile, parameters that
specify the damping parameters. All these parameters are communicated
to the code through the corresponding entries in the input atomic data
files -- see Paper~III, \S\,\refionbb\ and \S\,\refnsttwohyd.\\

\noindent$\bullet\ $ {\sf Superlines}
\index{Superlines}

The cross sections for superlines can be treated in two different ways:\\
-- using the Opacity Distribution Functions (ODF); or\\
-- using the Opacity Sampling (OS) approach.

The ODF approach, which consists of resampling a complicated composite
\index{Opacity Distribution Function}
cross section  for a superline to form a monotonic function of frequency, was used in the past (e.g., Hubeny \& Lanz 1995),
and although it is still offered by {\sc tlusty}, it is outdated,
and will not be described here.

The OS approach is the standard one used in {\sc tlusty} to describe a
superline cross section. It is evaluated by summing the contributions from
the individual ordinary lines that form the superline. The name Opacity
\index{Opacity Sampling}
Sampling comes from the fact that one computes the composite cross
section for a set of frequency points that are in principle randomly distributed,
and do not necessarily have to describe all the details of the exact cross
section. In other words, the composite cross section may miss cores of
some lines, and may miss some windows between lines, but in a statistical
sense provides a reasonable description. The frequency sampling used
in {\sc tlusty} is simply a set of equidistantly spaced frequencies with a
frequency step that is set up by input data. When decreasing the sampling step,
one in fact recovers an essentially exact form of the composite cross section.
This is the approach used for instance in the computed grids of NLTE metal
line-blanketed model atmospheres for O and B stars (Lanz \& Hubeny 2003, 2007).

\subsubsection{Bound-free radiative transitions}
\index{Bound-free cross sections}

Bound-free (photoionization) cross sections for transitions from ordinary
levels are evaluated using various standard formulae. For hydrogen and
hydrogenic ions one uses the standard hydrogenic cross section
(e.g., Hubeny \& Mihalas 2014, eq. 7.91), viz.
\begin{equation}
\label{bfhyd}
\sigma_{\rm bf}(\nu) = \frac{64\pi^4 Z^4 e^{10} m_{\rm e}}{3\sqrt{3}ch^6}
\frac{\bar{g}_{\rm bf}(n,\nu)}{n^5 \nu^3} 
= 2.815\times 10^{29}\, Z^4\, \frac{\bar{g}_{\rm bf}(n,\nu)}{n^5 \nu^3},
\end{equation}
where $n$ is the main quantum number of the lower bound state,
$Z$ is the charge of the ion ($Z=1$ for hydrogen), 
and $\bar{g}_{\rm bf}(n,\nu)$ is the bound-free Gaunt factor. By default,
it is evaluated by an analytical approximation as in Mihalas et al.(1975), 
but {\sc tlusty} offers some alternative expressions. For details, refer to 
Paper~III, \S\,\refionbf.

For other species, the cross sections may be evaluated by various 
approximate fitting formulae (which is an outdated
option, although still available in {\sc tlusty}), or, for most transitions, by
tabular values based on the Opacity Project (OP -- Seaton 1995) calculations. 
They are described  in detail in Paper~III, \S\,\refionbf. 
The values are already included in the atomic data
files, so that the user does not have to provide any additional information, unless
they want to change or improve the default cross sections. For levels for which
no  cross sections are available, hydrogenic form is usually assumed.

The OP cross sections contain 
a number of photoionization resonances.
Before transporting them to the {\sc tlusty} input atomic data files, they
are usually smoothed, forming the so-called {\em resonance-averaged profile} (RAP),
see, e.g., Allende Prieto et al. (2003). 
The smoothing serves two purposes. Firstly, it decreases the number of frequency
points needed to represent the cross section. However, since the number of
frequencies is large anyway, this is not critical. Secondly, and more importantly,
it avoids spurious peaks in the cross sections that may arise due to an insufficient
frequency resolution adopted in the original atomic structure calculations. 
The cross sections are then represented by
several tens up to several hundred values. For details, refer to  Paper~III, 
\S\,\refnonstblank.

\index{Superlevels}
For bound-free transitions from superlevels, the photoionization cross sections
are pre-calculated by summing the cross sections of the individual components,
and stored in special input files -- see Paper~III, \S\,\refionbf.

The inverse process, the radiative recombination, is described by the
same cross section as for the photoionization. Because the cross sections
include resonances, there is no need to consider the dielectric
recombination as a separate process, or requiring separate data for dielectronic
recombination rate (although such on option is also included in {\sc tlusty}
for historical reasons). For a discussion of the physics, see Hubeny \& Mihalas
(2014, \S\,9.3), and for details on the numerical implementation in {\sc tlusty},
see Paper~III, \S\,\refnsttwodiel.

\subsubsection{Inner-shell photoionization}
\index{Inner-shell photoionization}

In order to better describe high temperature structures where the
X-ray flux is dominant, a simple treatment of the inner-shell photoionization,
or Auger process, has been implemented in {\sc tlusty} by Hubeny et al. (2001).
\index{Auger process}
It employs a simplifying assumption
that if an Auger electron is energetically possible, than it is in fact produced
and the photoionization results in a jump by two stages of ionization to
a ground state configuration. Therefore, both fluorescence and multiple 
Auger electron ejection arising from inner shell photoionization are
neglected. The necessary data were taken from the X-ray photoionization 
code XSTAR (Kallman 2000)
\index{XSTAR code}

In practice, a different structure of input atomic data is employed
for ions for which one allows inner-shell photoionization processes.
They are described in more detail in Paper~III, \S\,\refxfiles.

\subsubsection{Free-free radiative transitions}

The cross section is given by the standard formulae [e.g. Hubeny \&
Mihalas (2014, eq. 7.100)]
\index{Free-free cross section}
\begin{equation}
\label{ff}
\sigma_{\!f\!f}(\nu,T) = \frac{\sqrt{32\pi} Z^2 e^6 \bar{g}_{\!f\!f}(\nu,T) }
{3\sqrt{3}ch (km_{\rm e}^3)^{1/2}} \frac{1}{T^{1/2}\nu^3} =
3.69\times 10^8 Z^2 \bar{g}_{\!f\!f}(\nu,T) \frac{1}{T^{1/2}\nu^3},
\end{equation}
where $\bar{g}_{f\!f}(\nu,T)$ is the free-free Gaunt factor. It is evaluated 
by the approximate expression as in Mihalas et al. (1975). For non-hydrogenic
ions, one still uses formula (\ref{ff}) with the Gaunt factor set to unity,
$\bar{g}_{f\!f}(\nu,T)=1$. In the past, one usually employed a {\em modified
free-free} cross section (Mihalas et al. 1975) that contained a contribution
of photoionization from higher, non-explicit levels taken in a hydrogen
approximation (since all these have a common $\nu^{-3}$ frequency
dependence). This option is still being offered in {\sc tlusty}, but is essentially
obsolete. It is briefly described in Paper~III, \S\,\refnewion2.

\subsubsection{Particle-induced transitions (collisions)}

\noindent $\bullet\ ${\sf Collisions with electrons}

By default, one considers collisions with electrons, for both bound-bound
and bound-free transitions. The code contains a number of hardwired
expressions for evaluating the electron collision rates. The user can
either employ a default expression, or choose some other form of
expression. For details, refer to Paper~III, \S\,\refionbf\  
and \S\,\refionbb.\\

\noindent $\bullet\ ${\sf Charge transfer reactions}
\index{Charge transfer}

{\sc tlusty} offers an option to include one such particular reaction, namely the
single electron exchange with neutral hydrogen,
\begin{equation}
{\rm X}^Z + {\rm H} \leftrightarrow {\rm X}^{Z-1} + {\rm H}^{+},
\end{equation}
in which species X in ionization stage $Z$ exchanges an electron
with hydrogen, becoming an ion with charge $Z-1$ (charge-exchange
recombination), or an inverse process (charge-exchange ionization).
The reaction rates are taken from Kingdon \& Ferland (1996) who provide
useful analytical fits to theoretical as well as experimental results
in the form
\begin{equation}
\sigma_{\rm rec}(T) = aT_4^b(1+e^{dT_4}),
\end{equation}
with $T_4=T/10^4$, and $a,b,d$ being the fitting parameters.
For more details, refer to Paper~III, \S\,\refnsttwocolh.

%------------------------------------------------------------------------------------

\subsection{Level dissolution; occupation probabilities, and pseudocontinua}
\label{dissol}

An improved description of the atomic level populations and related
transition rates, also implemented in {\sc tlusty}, adopts the concept of
\index{Occupation probability}
occupation probabilities and level dissolution (Hummer \& Mihalas
1988; Hubeny, Hummer \& Lanz 1994; Hubeny \& Mihalas 2014, \S\, 9.4).
It is based on introducing the {\em occupation probability} of a level.
In LTE, it is given through the generalized  Boltzmann formula,
\begin{equation}
(n_i/n_I)^\ast = w_i (g_i/U_I) e^{-E_i/kT},
\end{equation}
where the occupation probability $w_i$ is the probability that an atom
is in state $i$ relative to that in a similar ensemble of non-interacting atoms.
The same general definition applies also in the case of NLTE.
Correspondingly, $(1-w_i)$ is the probability that the state $i$ is
dissolved, i.e., it lies in the continuum, or, in other words, the corresponding
electron instead of being in the bound state $j$ is in fact free.

The transition from a lower bound state $i$ to the state $j$ is therefore
split into two parts: (i) a transition to a non-dissolved fraction of the level
forms an ordinary line, while (ii) a transition to the dissolved fraction of
the level is in fact a bound-free process, called {\em pseudocontinuum}.
The total pseudocontinuum cross section is a sum of the contributions
from the dissolved fractions of all the states $j$ higher than $i$. It is very
difficult to compute it exactly; the current expression is based on the
approximation devised by Daeppen et al. (1987) and Hubeny, Hummer,
\& Lanz (1994), viz.
\begin{equation}
\label{sigmat}
\sigma_{ik}^{\rm tot}(\nu) = D_i(\nu)\sigma_{ik}^{\rm ext}(\nu),
\end{equation}
where $\sigma_{ik}^{\rm ext}(\nu)=\sigma_{ik}(\nu)$ for $\nu\geq \nu_{ik}$ is 
a usual cross section for the bound-free transition from level $i$, while 
for $\nu< \nu_{ik}(\nu)$ represents an extrapolated cross section;
$\nu_{ik}$ is the ionization frequency from level $i$, and $D_i$ is
the so-called dissolved fraction, approximated by
\index{Dissolved fraction}
\begin{equation}
\label{dissolf}
D_i(\nu) =  \left\{ \begin{array}{lll}
1 & {\rm if} & \nu\geq \nu_{ik},\\ [6pt]
1-w_{m_{i^\ast}}\!(\nu) & {\rm if} & \nu < \nu_{ik}.
\end{array} \right. ,
\end{equation}
where  
$m_{i^\ast}\!(\nu)=[i^{-2}-(\nu/\nu_{ik})]^{-1/2}$ is an effective quantum
number of the highest state that can be reached from state $i$ by
the absorption of a photon with frequency $\nu$. It does not have to
be an integer. Its occupation probability is computed by the same
analytic expression as for integer values, Eqs. (\ref{occup}) - (\ref{betac}).

Equation (\ref{sigmat}) was derived, and is reasonably accurate, for 
frequencies close to the ionization threshold. Since for decreasing
frequency $w_{m_{i^\ast}} \rightarrow 1$ as $m_{i^\ast} \rightarrow 1$, 
it was originally believed that Eq. (\ref{dissolf}) can be used even far 
from the threshold, but it turned out that its contribution can be numerically 
non-negligible or even in some cases dominant very far from the 
threshold where its application is completely unphysical.  Therefore, we
have to introduce an empirical cutoff, so equation (\ref{sigmat}) is
applied only for $\nu > \nu^{\rm cutoff}$. This quantity is a parameter
in {\sc tlusty} transported through the atomic data file. Typically it is taken as
$\nu^{\rm cutoff} \approx 3\times 10^{15}$ for the Lyman pseudocontinuum,
and $\nu^{\rm cutoff} \approx 7\times 10^{14}$ for the Balmer pseudocontinuum.

After Hubeny, Hummer, \& Lanz (1994), the occupation probability is given by
\begin{equation}
\label{occup}
w_i = f/(1+f), \quad {\rm with} \quad
f= \frac{0.1402(x+4Z_r a^3)\beta_c^3}{1+0.1285 x \beta_c^{3/2}},
\end{equation}
where $a=0.09\times n_{\rm e}^{1/6}T^{-1/2}$, $x=(1+a)^{3.15}$
and $Z_r$ is the radiator charge, and
the critical field strength $\beta_c$ is given by
\begin{equation}
\label{betac}
\beta_c=8.59\times 10^{14} Z^3 n_{\rm e}^{-2/3} k_i i^{-4},
\end{equation}
where $k_i=1$ for  $i\leq 3$, and $k_i=(16/3)i/(i+1)^{-2}$ for $i>3$.

When this formalism [based on an earlier version of the theory directly
from Hummer \& Mihalas (1988)] was applied in actual models, it was 
found (Bergeron, Wesemael, \& Fontaine 1991) that the
\index{Occupation probability!Bergeron factor}
agreement between predicted and observed hydrogen line profiles of
white  dwarfs was significantly improved if the critical field strength
$\beta_c$ is multiplied by an empirical factor of 2. This correction was
originally applied in {\sc tlusty} as well; however, later it tuned out that the
Bergeron correction was relevant only to the original Hummer-Mihalas
expressions for occupation probabilities. When using an improved version
(first presented in Hubeny, Hummer, \& Lanz 1994; Appendix A), such a
correction is unnecessary or even incorrect. This option is still included
in {\sc tlusty} for testing purposes, but should not be used for actual models.

Eqs. (\ref{occup}) - (\ref{betac}) apply for  perturbations by
charged particles. For cooler models one should also include
occupation probabilities due to perturbations with neutral perturbers.
This is not yet implemented in the current version of {\sc tlusty}.

%-----------------------------------------------------------------------------------

\subsection{Equation of state and molecules}
\label{eos}

\subsubsection{Atoms and ions only}
In the original implementation of {\sc tlusty}, the equation of state was not
formulated explicitly; it has only been used as a relation between the
gas pressure $P$ and the total particle number density, $N$ -- see equation 
(\ref{eos0}). At the same time, the mass density is given by equation
(\ref{rho2}). The total particle number density is given by the sum of all
level populations of all individual explicit species in all explicit ionization
stages. However, such a simple definition would miss the contribution of 
species that are not treated explicitly. To include them, one selects a set
of implicit species, which are usually all the remaining species that are
not treated explicitly (but their choice may be set differently by input data).
Their ionization balance is treated in LTE, that is, it is assumed to obey the 
Saha equation, 
\index{Saha equation}
\begin{equation}
\label{saha}
(N_{I,J} / N_{I+1,J}) = n_{\mathrm{e}} \, (U_{I,J}/ U_{I+1,J})\, 
(h^2/2\pi m_{\rm e}kT)^{3/2}\, e^{\chi_{I,J}/kT}  ,
\end{equation}
supplemented by the condition for the total abundance of the element,
\begin{equation}
\sum_I N_{I,J} = N_J = (N-n_{\rm e}) \alpha_J = (N-n_{\rm e}) A_J\Big/
\sum_{J^\prime} A_{J^\prime},
\end{equation}
where $J$ is the index of the atom to which the ions $I$ belong, $A_J$
is the abundance of species $J$ with respect to hydrogen, 
$A_J \equiv N_J/N_H$, and $\alpha_J$ is the fractional abundance.
The additional charge $Q$ in Eq. (\ref{chce}) is given by
\begin{equation}
Q = \sum_J \sum_I N_{I,J} Z_I,
\end{equation}
where $Z_I$ is the charge of ion $I$.

The adopted chemical abundances may be set by the input data
(see Paper~III, \S\,\refnewat), but if nothing is specified, the default abundances 
are given by the solar abundances. 
The adopted solar  abundances of the first 
30 elements (by number,  with respect to hydrogen), essentially after 
Grevesse \& Sauval (1998),  are listed in Table 2.
\index{Abundance!solar - adopted}

%%%%%%%%%%%%%%%%%%%%%%%%%
%
\begin{table}
\caption{Adopted solar abundances from Grevesse \& Sauval (1998).}
\begin{center}
\begin{tabular}[t]{|l|c|}
\hline
 Element & $N_{\rm el}/N_{\rm H}$ \rule{0in}{3ex}\\[1ex]
\hline
H    &  1.0  \\
He   & 1.00$\times 10^{-1}$ \\
Li     &1.26$\times 10^{-11}$ \\
Be   & 2.51$\times 10^{-11}$ \\
B     & 5.00$\times 10^{-10}$ \\
C     & 3.31$\times 10^{-4}$ \\
N    &  8.32$\times 10^{-5}$ \\
O    &  6.76$\times 10^{-4}$ \\
F     & 3.16$\times 10^{-8}$ \\
Ne  &  1.20$\times 10^{-4}$ \\
Na  &   2.14$\times 10^{-6}$ \\
Mg  &   3.80$\times 10^{-5}$ \\
Al    &  2.95$\times 10^{-6}$ \\
Si    &  3.55$\times 10^{-5}$ \\
P     &  2.82$\times 10^{-7}$ \\
S     &  2.14$\times 10^{-5}$ \\
Cl    &   3.16$\times 10^{-7}$ \\
Ar    &   2.52$\times 10^{-6}$ \\
K      &  1.32$\times 10^{-7}$ \\
Ca    &  2.29$\times 10^{-6}$ \\
Sc    &  1.48$\times 10^{-9}$  \\
Ti      &  1.05$\times 10^{-7}$ \\
V       &  1.00$\times 10^{-8}$ \\
Cr      &  4.68$\times 10^{-7}$ \\
Mn     &  2.45$\times 10^{-7}$ \\
Fe      &  3.16$\times 10^{-5}$ \\ 
Co     &   8.32$\times 10^{-8}$ \\
Ni       &  1.78$\times 10^{-6}$ \\
Cu      &  1.62$\times 10^{-8}$ \\
Zn      &  3.98$\times 10^{-8}$ \\
\hline
\end{tabular}
\end{center} 
\end{table}  
%%%%%%%%%%%%%%%%%%%%%%%%%%%%%%%%%%%

\subsubsection{Standard evaluation of the partition functions}
\index{Partition functions!evaluation}

Let $A$ be the atomic number, and $Z$ the charge of the ion ($Z=0$
for neutrals). The following references are used for the individual
groups of species. In some cases, there is a choice of several source,
driven by additional input -- see Paper~III, \S\,\refnsttwoeos. 
\begin{itemize}
\item for $A\leq 30$ and $Z\leq 4$, unless specified otherwise 
(see Paper~III, \S\,\refnsttwoeos),
an evaluation after Traving, Baschek, \& Holweger (1966);
\item for $A\leq 30$ and $Z\leq 2$, and for the local temperature 
$T \leq 16,000$ K,
the default partition function is evaluated after Irwin (1980);
\item for $A>30$ and $Z \leq 2$, after Kurucz (1970);
 \item for $A=26$ (Fe) or $A=28$ (Ni), and $3 \leq Z \leq 8$, after
 Sparks \& Fischel (1971);
\item for $6 \leq A \leq 8$ (CNO) and $Z > 4$, after Sparks \& Fischel (1971).
 \item for $A\leq 30$ (except C,N,O) and $Z > 4$, approximate partition
function given by the statistical weight of the ground state;
\end{itemize}
 
\subsubsection{Molecules}
The situation is more complicated 
when the formation of molecules begins to contribute. In this case, one assumes
that free atoms exist only in the neutral and once ionized states, and that there
is a number of molecular species. In this case, we assume LTE. 
Their number densities are governed by a general Saha equation
\index{Molecules!equation of state}
\begin{equation}
\label{gsaha}
N^{+Z}_{1,\ldots,m} \equiv N^{+Z}_{\{1,m\}} =
\frac{\prod_{i=1}^m N_i}{n_{\rm e}^Z }\, \Phi
\end{equation}
where
\begin{equation}
\label{dissoc}
\Phi=\frac {U^{+Z}_{\{1,m\}} (2\pi M_{\{1,m\}} kT/h^2)^{3/2} 
\left[ 2(2\pi m_{\rm e}kT/h^2)^{3/2}  \right]^Z}
{\prod_{i=1}^m U_i (2\pi M_i kT/h^2)^{3/2} }\,
e^{-\Delta E/kT},
\end{equation}
with
\begin{equation}
\Delta E = E^{+Z}_{\{1,m\}}-\sum_{i=1}^m E_i ,
\end{equation}
where $X^{+Z}_{\{1,m\}}$, with $X\equiv N,U,M,E$, denotes a quantity 
associated with composite  particle  with charge $Z$ composed
of $m$ atomic species $1,\ldots,m$; some components may be identical.
Here $N,U,M,E$ stands for number density, partition function, mass, and
energy, respectively. Here we consider only $Z=0$ or $Z=1$; and
for negative ions of atoms or molecules, electrons are considered as
a separate atomic species. The
critical quantity of $\Phi$ is given either by its definition, Eq. (\ref{dissoc}), if all
the partition functions are available, 
or is represented by a fitting formula, e.g., the one  used by Kurucz (1970;
eq. 4.35)
\begin{equation}
\Phi(T) = \exp \left[ \frac{\Delta E}{kT} - b +cT - dT^2 +eT^3-fT^4
-\frac{3}{2}(m-1-Z) \ln T \right],
\end{equation}
where $\Delta E$, $b$, $c$, $d$, $e$, and $f$ are the fitting parameters.
For some species one uses a different fitting formula by Tsuji (with data
kindly supplied by Uffe Jorgenssen, priv. comm.).

Equation (\ref{gsaha}) is supplemented by the set of particle conservation
equations for all considered chemical elements, $A$,
\begin{equation}
\label{pcmol}
\sum_{k=1}^{N\!S} N_k^{+Z} c_k^A = N_A = (N - n_{\rm e})\alpha_A,
\end{equation}
and the charge conservation equation
\begin{equation}
\label{chcmol}
\sum_{k=1}^{N\!S} N_k^{+Z} Z c_k^A = n_{\rm e},
\end{equation}
where $N\!S$ is the total number of species, atomic or molecular,
$N_k$ is the number density of species $k$, where $k$ numbers
all composited species denoted with subscript  $\{1,m\}$, and
$c_k^A$ is the number of atoms $A$ in the species $k$ (for instance,
if species $k$ is water H${}_2$O, then $c_k^{\rm H}=2$,
$c_k^{\rm O}=1$, and $c_k^{\rm X}=0$ for all other elements X).

%------------------------------------------------------------------------------------------------------
%------------------------------------------------------------------------------------------------------
  %------------------------------------------------------------------------------------------------------

\section{Numerical procedure}
\label{numer}

The set of structural equations (\ref{rte}) with boundary conditions
(\ref{rte_ubc}) and (\ref{rte_lbc2}), (\ref{he1}), (\ref{re}) or (\ref{re_conv}), 
(\ref{ke2}), (\ref{chce}), and necessary auxiliary expressions, are discretized 
in depth and frequency, replacing derivatives by differences and integrals 
by quadrature sums. This yields a set of non-linear algebraic equations. 
Detailed forms of the discretized equations are summarized in Hubeny 
\& Mihalas (2014; \S\,18.1).

Upon discretization,
the physical state of an atmosphere is fully described by the set of vectors
$\psi_d$ for every depth point $d$, $(d=1, \ldots, N\!D)$, $N\!D$ being the total
number of discretized depth points. The full state vector ${\bf \psi}_d$ is given by
\index{State vector}
\begin{equation}
\label{cl1}
{\bf \psi}_d = \{ J_1, \ldots, J_{N\!F}, N, T, n_{\rm e}, n_1, 
\ldots, n_{N\!L},[n_m],[\nabla],[z]\},
\end{equation}
where $J_i$ is the mean intensity of radiation in the $i$-th frequency 
point; we have omitted the depth subscript $d$. The quantities in the square
brackets are optional, and are considered to be components of vector
$\psi$ only in specific cases: $n_m$ if specifically selected; $\nabla$ if convection
is taken into account, and $z$ in the case of disks.
The dimension of the vector $\psi_d$ is $N\!N$, $N\!N=N\!F+N\!L+N\!C$, where
$N\!F$ is the number of frequency points, 
$N\!L$ the number of atomic energy levels for which the rate equations 
are solved (i.e., explicit levels), and $N\!C$ is the number of constraint 
equations ($N\!C=3$ in the
standard atmospheric case, but it can be as large as 6).

\subsection{Linearization}
\label{cl}

Although the individual methods of the solution differ, the resulting
non-linear algebraic equations are solved by some kind of linearization.
In the general case, the solution proceeds as a direct application 
of the Newton-Raphson method.
Suppose the required solution ${\bf\psi}_d$ can be written in terms of the current,
but imperfect, solution ${\bf\psi}_d^0$ as 
${\bf\psi}_d={\bf\psi}_d^0+{\bf\delta\psi}_d$.
The entire set of structural equations can be formally written as an operator
$P$ acting on the state vector ${\bf\psi}_d$ as 
\begin{equation}
\label{cl2}
P_d({\bf\psi}_d)=0.
\end{equation}  
To obtain the solution, we express 
$P_d({\bf\psi}_d^0+{\bf\delta\psi}_d)= 0$,
and assuming that ${\bf\delta\psi}_d$ is ``small'' compared to ${\bf\psi}_d$ we 
use a Taylor expansion of $P$:
\begin{equation}
\label{cl3}
P_d({\bf\psi}_d^0)
+\sum_j\frac{\partial P_d}{\partial\psi_{d,j}}{\bf\delta\psi}_{d,j} = 0.
\end{equation}
to solve for ${\bf\delta\psi}_d$. Because only a first--order (i.e., linear) term
of the expansion is taken into account, this approach is called a
\textit{linearization}. To obtain the corrections ${\bf \delta\psi}_d$, one has to
form a matrix of partial derivatives of all the equations with respect to all 
the unknowns at all depths---the \textit{Jacobi matrix}, or \textit{Jacobian}---and
\index{Jacobian}
to solve equation (\ref{cl3}). The kinetic equilibrium and charge
conservation equations are \textit{local}, that is, for depth point $d$ they
contain the unknown quantities $\psi_{d}$ only at depth $d$. The
radiative equilibrium equation (in the differential form), and the hydrostatic
equilibrium equation couple two neighboring depth points $d\!-\!1$ and $d$. The
radiative transfer equations couple depth point $d$ to two neighboring depths
$d\!-\!1$ and $d\!+\!1$; see equations (\ref{rte}) -- (\ref{rte_lbc}).
Consequently, the system of linearized equations can be written as
\begin{equation}
\label{cl4}
-{\bf A}_d {\bf \delta\psi}_{d-1} +{\bf B}_d {\bf \delta\psi}_{d} 
-{\bf C}_d {\bf \delta\psi}_{d+1} = {\bf L}_d,
\end{equation}
where $A$, $B$, and $C$ are $N\!N \times N\!N$ matrices, and $L$ is
a residual error vector, given by 
\begin{equation}
\label{cl4a}
{\bf L}_d=-P_d({\bf \psi}_d^0). 
\end{equation}
At the convergence limit, $L\rightarrow 0$ and thus $\delta\psi_d\rightarrow 0$.

Equation (\ref{cl4}) is solved by a standard Gauss-Jordan elimination 
\index{Gauss-Jordan elimination}
that consists of a forward elimination
\begin{equation}
\label{elim1}
{\bf D}_1 = {\bf B}_1^{-1} {\bf C}_1,\quad {\rm and}\quad
{\bf D}_d = ({\bf B}_d - {\bf A}_d {\bf D}_{d-1})^{-1} {\bf C}_d,\quad d=2,\ldots,N\!D,
\end{equation}
and
\begin{equation}
\label{elim2}
{\bf Z}_1 = {\bf B}_1^{-1} {\bf L}_1,\quad {\rm and}\quad
{\bf Z}_d = ({\bf B}_d - {\bf A}_d {\bf D}_{d-1})^{-1} 
({\bf L}_d + {\bf A}_d {\bf Z}_{d-1}) ,\quad d=2,\ldots,N.
\end{equation}
followed by a back-substitution
\begin{equation}
\label{elim3}
{\bf \delta\psi}_{\!N\!D} = {\bf Z}_{N\!D},\quad {\rm and}\quad
{\bf \delta\psi}_d = {\bf D}_d  {\bf \delta\psi}_{d+1} +  {\bf Z}_{d} ,
\quad d=N\!D-1,\ldots,1.
\end{equation}

This procedure, known as {\it complete linearization}, was  developed in the seminal 
\index{Complete linearization}
paper by Auer \& Mihalas (1969). However, since the dimension 
of the state vector
$\psi$, that is the total number of structural parameters $N\!N$ can be extremely
large; for instance, in modern metal line blanketed model atmospheres 
$N\!F$ has to be taken few times $10^5$, a direct application of the original
complete linearization is not practical. Various possibilities to improve the
performance of the method are discussed in detail in Hubeny \& Mihalas (2014,
\S\,18.3). There are several variants of such improvements which are
implemented in {\sc tlusty}.

%------------------------------------------------------

\subsection{Hybrid CL/ALI method}
\label{hybrid}

The method, developed by Hubeny \& Lanz (1995), combines the basic
advantages of the complete linearization (CL) and the accelerated lambda
iteration method (ALI). For a general description of ALI, refer to Hubeny \&
Mihalas (2014; Chap. 13). The hybrid CL/ALI scheme is essentially a linearization 
method, except that the mean intensity in some (most) frequency points is not treated 
as an independent state parameter; instead it is expressed as
\begin{equation}
\label{clali}
J_{di} = \Lambda^\ast_{di} (\eta_{di}/\kappa_{di}) + \Delta J_{di},
\end{equation} 
where $d$ and $i$ represent indices of the discretized depth and frequency
points, respectively, $\Lambda^\ast$ is the so-called approximate Lambda
operator, and $\Delta J$ is a correction to the mean intensity. The approximate
operator, in most cases taken as diagonal (local), so that its action is just
an algebraic multiplication, is evaluated in the formal solution of the transfer
equation, and is held fixed in the next iteration of the linearization procedure,
and so is the correction $\Delta J$. Since the absorption and emission
coefficients $\kappa$ and $\eta$ are known functions of temperature,
electron density, and atomic level populations, one may express the 
linearization correction to mean intensity $J_{di}$ as
\begin{equation}
\label{clali2}
\delta J_{di} = \sum_x \Lambda^\ast_{di} 
\frac{\partial (\eta_{di}/\kappa_{di} )}{\partial x_{di} }\delta x_{di},
\end{equation}
where $x = (T, n_{\rm e}, n_i)$, i.e., $x$ stands for other state parameters --temperature, electron density, and level populations.

Equation (\ref{clali2}) shows that $J_{di}$ is effectively eliminated, thus reducing
the size of vector $\psi$ to $N\!N = N\!F_{C\!L}+N\!L+N\!C$, where
$N\!F_{C\!L}$ is the number of frequency points for which the mean intensity
is kept to be linearized, called explicit frequencies. As shown in
Hubeny \& Lanz (1995), such a number can be very small, of the order
of $O(10^0)$ to a few times $10^1$. Typically, only frequencies in the
centers of the strongest lines, and a few frequencies just shortward of the 
edges of the strongest continua, are kept to be linearized.

The choice of which frequencies are linearized is driven by input data.
{\sc tlusty} can thus cover the whole range of options, from the pure
complete linearization (no frequencies treated with ALI), to full ALI, in
which no frequency points are linearized. 

% -------------------------------------------------------------------------------

\subsection{Rybicki scheme}
\label{ryb}

An alternative scheme, which can be used in conjunction with either
original complete linearization, or a hybrid CL/ALI scheme, is a generalization
of the method developed originally by Rybicki (1969) for solving a line transfer
problem. It starts with the same set of linearized state equations, and consists 
in a reorganization of the state vector and resulting Jacobi matrix in a
different form. Instead of forming a vector of all state parameters in a given
depth point, it constructs a set of vectors of mean intensity, each containing
mean intensities in one frequency point for all depths,
\begin{equation}
\delta {\bf J}_i \equiv \{\delta J_{1i}, \delta J_{2i}, \ldots,\delta J_{N\!D,i} \},\quad
i=1,\ldots,N\!F,
\end{equation}
and analogously for the vector of temperatures
\begin{equation}
\delta {\bf T} \equiv \{\delta T_1, \delta T_2, \ldots,\delta T_{N\!D} \}.
\end{equation}
In the description of the method presented in Hubeny \& Mihalas (2014; \S\,17.3),
an analogous vector $\delta{\bf N}$ for the particle number density
was introduced, but this is not done in {\sc tlusty}.

The method is designed for LTE models, although it can in principle
be used for NLTE models as well, although in that case the convergence
of the scheme is usually quite slow. Nevertheless, in some cases it
may provide a stable scheme to obtain an intermediate NLTE model,
from which one can then more easily converge a NLTE model using the
CL or hybrid CL/ALI scheme.

The point is to express all the material state parameters as functions of
temperature and density, and then to express the density
as a function of temperature, using the equation of state relating density
to gas pressure, and keeping the gas pressure fixed in a given iteration
step. This procedure is impractical for hot models where the radiation pressure
represents a significant fraction of the total pressure, but is quite reasonable 
for cool models where the radiation pressure is small or negligible. In that
case, $P \approx m g + P_0$ [see equation (\ref{he})], and since $m$ is used as
the basic depth coordinate, the gas pressure is essentially a known quantity.

The linearized radiative transfer equation (for inner depth points) can be written as
\begin{equation}
\label{ryb1}
\sum_{d^\prime=d-1}^{d+1} U_{dd^\prime, i} \delta{\bf J}_{d^\prime i} +
\sum_{d^\prime=d-1}^{d+1} R_{dd^\prime, i} \delta{\bf T}_{d^\prime} = {E}_{di},
\end{equation}
for $i=1,\ldots,N\!F$.  In the matrix notation
\begin{equation}
\label{ryb1a}
{\bf U}_i \delta{\bf J}_i + {\bf R}_i \delta{\bf T} = {\bf E}_i ,
\end{equation}
where  ${\bf U}_i$ and ${\bf R}_i$ are $N\!D \times N\!D$
tridiagonal matrices that account for coupling of the corrections to the
radiation field at frequency $\nu_i$ and material properties as functions of
$T$, at the three adjacent depth points $(d-1, d, d+1)$.
Analogously, the contribution to the linearized energy equation from each
frequency $i$ at depth point $d$ is of the form
\begin{equation}
\label{ryb2}
\sum_{i=1}^{N\!F}{\bf V}_i \delta{\bf J}_i + {\bf W} \delta{\bf T} = {\bf F} ,
\end{equation}
where ${\bf V}_i$ and ${\bf W}$ are generally bi-diagonal matrices
(in the differential form of the radiative equilibrium; for the integral
form they would be diagonal).

The overall structure here is reversed from the original
variant, in the sense that the role of frequencies and depths is reversed.
The global system is a block-diagonal (since the frequency points are
not coupled), with an additional block (``row'') with the internal matrices
being tridiagonal. Corrections of mean intensities are found from
Eq. (\ref{ryb1a}),
\begin{equation}
\label{ryb2a}
\delta{\bf J}_i = {\bf U}_i^{-1} {\bf E}_i - ({\bf U}_i^{-1} {\bf R}_i) \delta {\bf T}
\end{equation}
substituting Eq. (\ref{ryb2a}) into (\ref{ryb2}), one obtains for the correction 
of temperature
\begin{equation}
\label{ryb4}
\left( {\bf W} - \sum_{i=1}^{N\!F} {\bf V}_i {\bf U}_i^{-1} {\bf R}_i \right) 
\delta{\bf T} = 
\left( {\bf F} - \sum_{i=1}^{N\!F} {\bf V}_i {\bf U}_i^{-1} {\bf E}_i \right),
\end{equation}
which is solved for $\delta{\bf T}$., and then $\delta{\bf J}_i$ are obtained 
from Eq. (\ref{ryb2a}).

In this scheme, one has to invert $N\!F$ tridiagonal; matrices ${\bf U}_i$,
which is very fast, plus one inversion of the
$N\!D \times N\!D$ grand matrix in Eq. (\ref{ryb4}), which is also fast.
Since the computer time scales linearly with the number of frequency
points, the method can be used even for line-blanketed models.

As stated above, the method is designed for LTE models, but experience 
shows that the Rybicki scheme can also be applied to  NLTE models. The scheme 
is in this case equivalent to the Lambda iteration (because the corrections
of level populations and electron density are not solved simultaneously
with the corrections of temperature and radiations field), so it converges very 
slowly. However, the method is quite stable and it may help avoid divergences 
that sometimes plague traditional linearization methods. Therefore,
although the Rybicki scheme is not a viable option to construct well 
converged NLTE models,
it can be used for getting an intermediate model from which the full
NLTE can be converged more easily.

%------------------------------------------------------------------------------------------------

\subsection{Acceleration methods}
\label{accel}

There are three numerical schemes offered by {\sc tlusty} that belong to the
category of mathematical {\em acceleration of convergence}. They are
discussed in detail in Hubeny \& Mihalas (2014; \S\,18.4).\\ [3pt]
(1) The simplest possibility is the so-called {\em Kantorovich method} 
\index{Kantorovich method}
(Hubeny \& Lanz 1992). The scheme keeps the Jacobian fixed after 
a certain iteration, so the subsequent iterations of complete linearization
use the same Jacobian (more accurately the inverse of the Jacobian is kept fixed for
future use); only the right--hand--side vectors ${\bf L}$ are re-evaluated after
each iteration. Experience with the method shows that it is surprisingly robust. Usually, one needs to perform a few (3-5)
iterations of the full linearization scheme, depending on the problem at hand
and the quality of the initial estimate. Also, it is sometimes very advantageous
to ``refresh'' the Jacobian (i.e., set it using the current solution and invert
it) after a certain number of Kantorovich iterations. The detailed setup is controlled
by input data.\\ [3pt]
(2) {\em Ng acceleration} is a very popular acceleration scheme used in conjunction
\index{Ng acceleration}
of an ALI scheme for solving a radiative transfer problem (Auer 1984; Olson,
Auer, \& Buchler 1986). 
In the context of accelerating a complete--linearization--based scheme to calculate 
model stellar atmospheres was applied in Hubeny \& Lanz (1992).

The general idea is to construct a new iterate of the state vector based on the 
information not only from the previous iteration step, as in a standard version
of an iterative linearization procedure, but also from still earlier steps. While
the number of earlier steps may vary,
essentially all astrophysical applications use the three--point version.
Denoting $\bfx$ the collection of all state vectors $\psi_d$ at all depth
points $d$, the ``accelerated'' iterate is written as a linear combination of the three 
previous iterates, 
\begin{equation}
\label{ng}
{\bf x}^\ast = (1-a-b) {\bf x}^{(n-1)} + a {\bf x}^{(n-2)} + b {\bf x}^{(n-3)},
\end{equation}
where coefficients $a$ and $b$ are given by
\begin{eqnarray}
a &=& \left(\delta_{01}\delta_{22} - \delta_{02}\delta_{21}\right)
/\left(\delta_{11}\delta_{22} - \delta_{12}\delta_{21}\right), \\
b &=& \left(\delta_{02}\delta_{11} - \delta_{01}\delta_{21}\right)
/\left(\delta_{11}\delta_{22} - \delta_{12}\delta_{21}\right),
\end{eqnarray}
where
\begin{equation}
\label{ng2}
\delta_{ij}\equiv\Big(\Delta {\bf x}^{(n)}-\Delta {\bf x}^{(n-i)}\Big)\cdot 
                 \Big(\Delta {\bf x}^{(n)}-\Delta {\bf x}^{(n-j)}\Big), 
\end{equation}
for $i = 0,1,2,$ and $j=1,2$; and
\begin{equation}
\label{ng3}
\Delta {\bf x}^{(n)}\equiv {\bf x}^{(n)}-{\bf x}^{(n-1)}.
\end{equation}
The scalar product in equation (\ref{ng2}) is defined as
\begin{equation}
\label{ng4}
{\bf x} \cdot {\bf y} \equiv \sum_{d=1}^{N\!D} \sum_{i=1}^{N\!N} W_{di} x_{di} y_{di},
\end{equation}
where $W_{di}$ is a weighting factor, taken in {\sc tlusty} as
$W_{di}=1/\psi_{di}$.
Experience has showed that in a large majority of cases the Ng 
acceleration improves
convergence significantly; the acceleration is usually performed for the first
time at or around the 7-th iteration of the linearization scheme, and is done 
typically every 4 iterations afterwards. Again, the detailed setup is driven by
input data -- see Paper~III, \S\,\refnonstacc. In some cases, like in
models with convection, or in specific models with sharp ionization fronts, the Ng
acceleration does not help, and may even lead to numerical problems and
divergence. Therefore, one should apply the Ng acceleration judiciously.\\ [3pt]
(3) Successive over--relaxation (SOR).   It consists in multiplying the corrections
\index{Successive over-relaxation}
$\delta\psi$ by a certain coefficient $\alpha$. This coefficient can be either
set by an educated guess, or one can use the procedure suggested by
Trujillo-Bueno \& Fabiani-Bendicho (1995), namely to express $\alpha$ in terms of
the spectral radius of the appropriate iteration operator, which in turn may be
approximated by a ratio of maximum relative changes of the source function in 
the two subsequent previous iterations.

% --------------------------------------------------------------------------

\subsection{Treatment of opacities}
\label{opac}

The opacities are treated in {\sc tlusty} in several different ways:

(1) The current opacities and emissivities are evaluated on the fly for the
current structural parameters (temperature, density, atomic level populations).
This is a traditional approach, which is moreover mandatory for NLTE models. 
In this approach, the absorption and emission coefficients are evaluated essentially 
\index{Absorption coefficient}
\index{Emission coefficient}
by using their definition equations (\ref{kappa}) and (\ref{etanu}).
The actual transitions that contribute
to the total opacity are specified through selecting atomic species, ions, and
levels which are called ``explicit''. Therefore, the transitions between explicit
levels of explicit ions of explicit atoms do contribute to the total opacity. The
opacity (emissivity) of an individual explicit transition is a function of level populations 
of the lower and upper states, and the corresponding cross section which in
turn is a function of structural parameters such as temperature, electron density,
and possibly others. Since all these quantities are being updated during the
iteration process, the opacity/emissivity is recalculated again and again.

\index{Superlevels}
\index{Superlines}
(2) A variant of this approach is the opacity and emissivity of ``superlines'',
that is, transitions between ``superlevels''. A superlevel is a set of individual
energy levels that are assumed to have the same NLTE departure coefficient or,
in other words, are in LTE within each other. For details, refer to \S\,\ref{superl}.
As explained there, relevant cross sections are evaluated at the beginning 
and are held fixed afterward. Nevertheless, the 
corresponding superlevel populations are being updated, so the corresponding 
opacities are still being re-evaluated during the iteration process.

(3) As indicated in equations (\ref{chi}) and (\ref{etanu}),
{\sc tlusty} also allows for the so-called additional opacities.
These are typically transitions between upper levels of explicit species
that are not treated explicitly, and also H${}_2^+$ 
or H${}^-$ opacity (if the H${}^-$ ion is not treated explicitly)
The user has an option to include more additional opacities by means 
of adding corresponding expressions to a previously provided subroutine. 
In all these cases, the corresponding cross sections are taken as analytic 
expressions, or from simple tables.

(4) While the above procedure  is unavoidable for NLTE models, it is not efficient
for LTE models, where the opacity depends only on temperature and density, 
and emissivity is given through the Kirchhoff-Planck relation, 
$\eta_\nu=\kappa_\nu B_\nu$,
In the past, LTE models were considered, within the context of
{\sc tlusty} applications, just as intermediate models which provide suitable
starting models for the NLTE ones. However, once the range of applications
of {\sc tlusty} extended to cool and very cool objects, such as the lower end
of the main sequence, brown dwarfs, and planetary atmospheres, LTE models,
with possible additional complexities such as molecular and cloud opacities,
become important on their own merit.

In this case, it is much more efficient to pre-calculate extensive tables of opacity 
as a function of frequency, temperature, and density (or electron density).
In an actual run of {\sc tlusty}, one simply interpolates from the tabular values of
opacity to the current values of structural parameters and frequency. Such an approach has been used in a previously separate
variant of {\sc tlusty} called Cool{\sc tlusty} (e.g. Hubeny, Burrows, \& Sudarsky 2003),
\index{Cool{\sc tlusty} program}
and was extensively used for computing model atmospheres of brown
dwarfs and extrasolar planets.

Starting with version 204, such an approach has been adopted in the mainstream
{\sc tlusty} as well. We do not use tables constructed specifically for substellar
mass objects (Sharp and Burrows 2007) because they are not released for
public distribution. However, we have constructed opacity tables for temperatures
between 3,000 and 10,000 K, so that one can use these tables to compute LTE model
atmospheres for F and G type stars. The current tables do not contain molecular
opacities. Currently, we are working on preparing more extensive opacity
tables, including molecular and cloud opacities,  applicable for the whole range 
of parameters.

In this approach one does not need to select explicit atoms, ions, levels,
and transitions, because the opacities are given and the kinetic equilibrium
equations are not solved. There is a choice how to treat the equation of state
and to evaluate the necessary thermodynamic parameters needed to describe convection.
This can be done either on the fly, in which case one has to supply data for implicit
species (abundances, a mode of evaluation of the partition functions), 
or by interpolating in additional pre-calculated tables that
specify the equation of state (essentially a relation between density and pressure),
and also other thermodynamic variables needed to evaluate the convective flux --
see Paper~III, \S\,\refnsttwooptab.

(5) Finally, there is a hybrid approach, in which some opacities (or bulk of 
opacities) are treated explicitly (options 1 -- 3 above), while the remaining
opacity sources, presumably less important ones, or those for which LTE is
a good approximation (such as molecular opacities) are treated by means of an
opacity table. Such an approach is already coded in {\sc tlusty}, Version 205, but
an evaluation of the corresponding partial opacity tables was not yet fully
streamlined.

%---------------------------------------------------------------------------

\subsection{Superlevels and superlines}
\label{superl}
\index{Superlevels}
\index{Superlines}

Using the hybrid CL/ALI method, the number of frequency points can be
reduced dramatically. However, the number of explicit levels needed for
NLTE metal line-blanketed model atmospheres may still be enormous.
For instance, each ion of Fe has of the order $10^4$ energy levels.

To deal with this problem, one introduces the concept of a superlevel. 
The idea consists of grouping several, possibly many, individual energy levels
together, forming a \textit{superlevel}. The basic physical assumption is that
all genuine levels $j$ within a superlevel $J$ are in Boltzmann equilibrium with
respect to each other,
\begin{equation}
\label{superl1}
n_j/n_{j^\prime} = g_j/g_{j^\prime}\exp[-(E_j - E_{j^\prime})/kT].
\end{equation}
Therefore, the whole superlevel 
can be treated as one level for solving the kinetic equilibrium equation. 
There is a certain flexibility in choosing a partitioning of levels into
superlevels. However, in order to provide  a realistic description, the levels
forming a superlevel have to possess close energies, and to have similar
properties, for instance belonging to the same multiplet, the same spin system,
or having the same parity. The requirement of close energies is needed because
collisional rates between levels with a small energy difference tend to be large
and hence dominate over the radiative rates. With dominant collisional rates,
one indeed recovers LTE.

As mentioned in \S\,\ref{phys_bas}, we have introduced also a special kind of 
superlevels, called {\em merged levels}. They behave like normal superlevels,
the only differences are (i) they are introduced only for the higher states of
hydrogen and hydrogenic ions, and (ii)  therefore, their parameters, and the 
necessary cross sections for transition involving them are computed 
analytically, without a need of additional data.

Bound-bound transition involving at least one superlevel are called superlines.
The absorption
coefficient for a transition $I\rightarrow J$, not corrected for stimulated
emission, is given by
\begin{equation}
\label{superl2}
\kappa_{IJ}(\nu) = \sum_i \sum_j n_i w_j \sigma_{ij}(\nu),
\end{equation}
where 
$\sigma_{ij}(\nu) = (\pi e^2/m_{\mathrm{e}} c) f_{ij} \phi_{ij}(\nu)$,
is the cross section for the transition
$i\! \rightarrow\! j$, 
$f_{ij}$ is the oscillator strength, and $\phi_{ij}(\nu)$ the (normalized)
absorption profile coefficient, $n_i$ and $w_j$ are the population of the
lower level, and the occupation probability of the upper level, respectively.

Within the superlevel formalism, the absorption coefficient for transition
$I \rightarrow J$ has to be given by [for details, refer to Hubeny \& 
Mihalas (2014, \S\,18.5)]
\begin{equation}
\label{superl3}
\kappa_{I\!J}(\nu) = n_I w_J \sigma_{I\!J}(\nu);
\end{equation}
therefore the cross section is given by
\begin{equation}
\label{superl4}
\sigma_{I\!J}(\nu)=\frac{g_I \exp(-E_J/kT)\sum_i \sum_j g_i w_i w_j \sigma_{ij}(\nu)\exp(-E_i/kT) }
{\big[\sum_i g_i w_i\exp(-E_i/kT)\big]\big[\sum_j g_j w_j\exp(-E_j/kT)\big]},
\end{equation}
and 
\begin{equation}
w_J = \frac{\exp(E_J/kT)}{g_J} \sum_j g_j w_j \exp(-E_j/kT).
\end{equation}
which has the meaning of generalized occupation probability for 
superlevel $J$. It should be noted that in general $w_J \not= 1$ even if one 
sets all the occupation probabilities of the components $j$ to $w_j=1$.

Since the number of individuals levels forming a superlevel may be quite
large (of the order of several tens to several hundreds), and so the number of
individual lines forming a superline may be huge (say, of the order of 
$10^4$ to $10^5$,
or perhaps even more), the resulting superline cross section
\index{Superlines}
is a rather complicated function of frequency. It would be impractical to compute
it for every depth point independently; moreover the amount of necessary atomic data
(the individual oscillator strengths, the line broadening parameters) will also be
impractically large. Therefore, the superline cross sections are evaluated at the 
beginning of a given {\sc tlusty} run, and subsequently held fixed,  for
several depth points (typically 3, but this number is a free parameter and
can be changed if needed). For other depth points, they are interpolated. 
As stated in \S\,.\ref{trans},
the cross sections to be used in {\sc tlusty} are evaluated by means of the so-called Opacity Sampling (OS), that is at frequency points that are set by the code
to cover the whole range of a superline, and with a frequency spacing that is given
by a certain multiple of a fiducial Doppler width\footnote{Fiducial Doppler width is
defined as a Doppler width for Fe corresponding to the effective temperature}. 
This spacing is also a free parameter. For most accurate models (such as the OSTAR2003 and BSTAR2007 grids of NLTE line-blanketed model atmospheres 
of O and B stars -- Lanz \& Hubeny 2003; 2007), the spacing was taken as 0.75, 
but it can be taken much higher; as shown in Lanz \& Hubeny (2003), even vales such as 40-50 produce reasonably accurate model atmospheres.

In the past, the superline cross sections were treated by means of the Opacity
Distribution Function (ODF) -- see Hubeny \& Lanz (1995),  which uses a 
\index{Opacity Distribution Function}
resampled cross section that is
represented by a low number of frequency points. However, this option is
not accurate enough because it does not treat the overlaps of superlines properly.
Also, it requires additional input files that contain the tabulated ODF values.
Although it is still offered in {\sc tlusty}, it is not recommended.

\index{Superlines}
The superline cross sections are computed at the beginning of the calculation
based on data from Kurucz files, for instance, for Fe II the files
{\tt gf2601.gam} and {\tt gf2601.lin}. The former file contains parameters
for the individual energy levels (energy, statistical weight, quantum numbers),
while the latter contains the data for individual lines (oscillator strengths, broadening
parameters). 

The concept of an ODF is also used for treating the opacity in a transition
form a regular level to a merged level in the case of hydrogen and hydrogenic 
ions. In this case, however, the ODF is constructed on the fly because the cross
sections for the individual lines forming such a superline can be computed analytically. 
Some additional input parameters to construct such an ODF are described in 
Paper~III, \S\,\refionbb.  However, even this option is somewhat obsolete, as it
is possible to treat the hydrogen atom with many explicit levels (say, 16 or
more), in which case the higher levels are mostly dissolved, and the opacity
in the transitions from lower levels into the merged level are described through 
the pseudocontinuum opacity.

\subsection{Level grouping and zeroing}
\label{group}
Another procedure, which may significantly reduce the number of level
populations to be linearized is the idea of {\em level grouping}. A level group
is a set of several levels whose populations are assumed to vary in a
coordinated way in the linearization. More precisely, instead of linearizing
individual level populations, one linearizes the total population of the group,
assuming that the ratios of the individual level populations within the group to
the total population of the group are unchanged during a current linearization step. 
In the formal solution step, one solves the kinetic equilibrium equations 
for all the individual level populations. 

\index{Superlevels}
The concept of level groups should not be confused with the concept of superlevels;
in the former case, the level groups are only a numerical trick to make the
complete linearization matrices smaller, while the level populations are
determined exactly; the latter case---superlevels---approximates the
individual populations of the components of the superlevel by assuming that they
are in Boltzmann equilibrium with respect to each other. In fact, one may group
the individual superlevels into level groups as well.

Another numerical trick used in {\sc tlusty} is a {\it level zeroing}.  For each 
depth point, the code examines a ratio of level populations to the total 
population of an ion. If this value decreases below a certain value (a free 
parameter, typically taken $10^{-20}$), the level population of such a level is 
set to zero, and the kinetic equilibrium equation for such particular depth point 
is written as $n_i=0$ (that is, in a matrix form, $A_{ii}=1$, $A_{ij}=0$ for
$i\not= j$, and $b_i=0$).  The level still remains in the set of explicit level 
since its population might be non-negligible or even significant in other depth 
points. If, however, the population would be zeroed in all depth points, the level 
is completely removed from the set of explicit levels. This procedure helps to
improve a numerical stability of the system, avoids solving for level populations 
which are of no practical interest, and also allows the user to use relatively 
general atomic data sets with many ions.

%----------------------------------------------------------

\subsection{Formal solution}
\label{formal}

The term ``formal solution'' is used in two different meanings:

(i) In a limited context, the ``formal solution of the transfer equation'' 
means a solution of this equation for one frequency at a time, with the 
{\em specified source function}. At this step the Eddington factors are
being evaluated.

(ii) The set of all calculations between two iterations of the global iteration
(i.e., linearization) scheme. In the standard case, it means a simultaneous
solution of the radiative transfer and kinetic equilibrium equations, keeping the
values of other state parameters (temperature, density, electron
density) fixed at the current values.  Depending
on the setup of the run, it may also include the recalculation of other state
parameters, such as electron density, or mass density, or even
temperature. The point of this procedure is to help the global iteration
scheme to converge faster by improving the values of the state vector
as much as possible before entering the next iteration step. It will be called 
here {\em global formal solution}.

\subsubsection{Formal solution of the transfer equation}
{\sc tlusty} offers several types of the formal solution which are selected
by means of input data. A detailed description of the schemes is presented
in Hubeny \& Mihalas (2014; \S\,12.4).

Briefly, discretizing in frequency and angle, the transfer equation
is written as
\begin{equation}
\label{rte0}
\mu_m \frac{dI_{nm}}{d\tau_n} = I_{nm} - S_n
\end{equation}
where $I_{nm}$ is the specific intensity at frequency point $n$ and
angle point $m$; $\mu_m$ is the cosine of the polar angle. 
$S_n$ is the source function, given by
\begin{equation}
\label{sffea}
S_n = \frac{\kappa_n}{\chi_n} + \frac{\sigma_n}{\chi_n} 
\sum_{m^\prime} w_{m^\prime} I_{nm^\prime}.
\end{equation}
The last term
represents the scattering part of the source function, assuming isotropic
and coherent scattering (e.g. Thomson electron scattering; the case
of non-coherent Compton scattering is described in Paper~III, \S\,\refnsttwocompt);
$w_m$ are the angular quadrature weights.

The default method of solution is the Feautrier method that introduces
\index{Feautrier method}
symmetric and antisymmetric averages of the specific intensity, more specifically
$j_{nm} \equiv [I_n(\mu_m) + I_n(-\mu_m)]/2$, and
$h_{nm} \equiv [I_n(\mu_m) - I_n(-\mu_m)]/2$,
and rewrites the transfer equation (\ref{rte0}) in a  second-order form,
\begin{equation}
\label{rtef}
\mu_m^2\,\frac{d^2 j_{nm}}{d\tau_n^2} = j_{nm} - \frac{\kappa_n}{\chi_n} 
-\frac{\sigma_n}{\chi_n} \sum_{m^\prime=1}^{N\!A} w_{m^\prime} I_{nm^\prime},
\end{equation}
where $N\!A$ is the number of angle points in one hemisphere
($\mu > 0$). This equation is supplemented by the boundary conditions
\begin{equation}
\label{rtebc1}
\mu_m \left.\frac{dj_{nm}}{d\tau_n}\right|_0 = j_{nm}(0) - I_{nm}^{\rm ext},
\end{equation}
where $I_{nm}^{\rm ext}$ is the incoming specific intensity
$I(\nu_n, -\mu_m)$.  Equation (\ref{rtebc1}) is only first-order accurate.
Auer (1967) suggested a convenient second-order from which is based on
a Taylor expansion
\begin{equation}
\label{eq:12.2-21}
j_{nm}(\tau_{2,n})=j_{nm}(\tau_{1,n})+\Delta\tau_{{3/2},n}\frac{dj_{nm}}{d\tau_{n}}\bigg|_1
    +\frac{1}{2}(\Delta\tau_{{3/2},n})^2\,\frac{d^2j_{nm}}{d\tau_n^2}\bigg|_1,
\end{equation}
which is used to achieve a second-order accuracy,
\begin{equation}
\label{eq:12.2-22}
\mu_m\frac{j_{nm}(\tau_{2n})-j_{nm}(\tau_{1n})}{\Delta\tau_{3/2,n}} = 
j_{nm}(\tau_{1n})+\Delta\tau_{3/2,n} \frac{j_{nm}(\tau_{1n})-S_{n}(\tau_{1n})}{2\mu_m}.
\end{equation}
and for the lower boundary
\begin{equation}
\label{rtebc2}
\mu_m \left.\frac{dj_{nm}}{d\tau_n}\right|_{\tau_{\rm max}}\!\! = I^+_{nm}(\tau_{\rm max}),
\end{equation}
where $I^+_{nm}(\tau_{\rm max})$ is the outward-defected specific intensity at
the deepest point, given either by the diffusion approximation (for stellar
atmospheres)
\begin{equation}
\label{rtebc3}
I^+_{nm}(\tau_{\rm max}) = B(\nu_n,\tau_{\rm max}) +
\mu_m\!\left.\frac{\partial B(\nu_n)}{\partial\tau_{\nu_n}}\right|_{\tau_{\rm max}},
\end{equation}
or by a symmetric lower boundary condition (for accretion disks)
\begin{equation}
\label{rtebc4}
\mu_m \left.\frac{dj_{nm}}{d\tau_n}\right|_{\tau_{\rm max}}\!\!\! = 0.
\end{equation}

All frequency points in Eqs. (\ref{rtef}) -- (\ref{rtebc4}) are independent,
so that they can by solved for one frequency at a time. We skip the  frequency
index $n$ and discretize in depth, described by index $d$.
One introduces a column vector ${\bf j}_d \equiv (j_{d,1}, j_{d,2},\ldots,j_{d,N\!A})$,
and writes Eqs. (\ref{rtef}) -- (\ref{rtebc4})  as a linear matrix equation
\begin{equation}
\label{rtefm} 
-{\bf A}_d {\bf j}_{d-1} +{\bf B}_d {\bf j}_{d} -{\bf C}_d {\bf j}_{d+1} = {\bf L}_d,
\end{equation}
where ${\bf A}_d$, ${\bf B}_d$,  and ${\bf C}_d$, are $N\!A \times N\!A$ matrices;
${\bf A}$ and ${\bf C}$ are diagonal, while ${\bf B}$ is full.
The system is solved by the standard Gauss-Jordan elimination, equivalent to
\index{Gauss-Jordan elimination}
Eqs. (\ref{elim1}) - (\ref{elim3}).
In terms of $j$, the mean intensity and the Eddington factor are given by
\index{Eddington factor}
\begin{equation}
\label{rteff}
J_n = \sum_{m=1}^{N\!A} w_m j_{nm}, \quad {\rm and}\quad
f_n = \sum_{m=1}^{N\!A} w_m \mu_m^2 j_{nm} \Big/ J_n.
\end{equation}
The program offers several  variants of the Feautrier scheme:
\index{Feautrier method!offered variants}
scheme \\ [-8pt]

$\bullet\ $ ordinary second--order Feautrier (1964) scheme;

$\bullet\ $  improved scheme by Rybicki \& Hummer (1991);

$\bullet\ $  spline collocation scheme  (Mihalas \& Hummer (1974);

$\bullet\ $  Auer (1976) fourth-order Hermitian scheme.\\ [2pt]
\noindent
By the nature of the Feautrier scheme, all these methods solve
the transfer equation for one frequency, but all angle points, at a
time. This involves solving the transfer equation for vectors of specific
intensities for all angles (with number $N\!A$, with a need of inverting 
a number of $N\!A \times N\!A$ matrices. Since the typical value of
$N\!A$ is quite low ($N\!A=3$ by default, which corresponds to 6
actual discretized angles) inverting such matrices does not present
any problem or any appreciable time consumption. The basic advantage
of the Feautrier scheme is that it treats scattering directly, without any
need to iterate.

However, if the number of angles is large (for as comparison
purposes, or some very specific applications), or if an atmospheric
structure exhibits very sharp variations with depth (e.g., sharp ionization
fronts), it is advantageous to use another offered scheme:

\smallskip
\noindent
$\bullet\ $  Discontinuous Finite Element (DFE) scheme by Castor, 
Dykema, \& Klein (1992).
It solves the linear transfer equation (\ref{rte0})
directly for the specific intensity, and therefore if scattering
is present, which is essentially always, the scattering part of the source
function has to be treated iteratively. To this end, a simple ALI-based procedure
is used; its setup is described through the corresponding input parameters -- see
Paper~III, \S\,\refnonstrte\  and \S\,\refnsttwortetwo.

\subsubsection{Global formal solution}
\label{formal_glob}

The main part of the global formal solution is a
simultaneous solution of the radiative transfer equation
and the set of kinetic equilibrium equations for all explicit levels.
Notice that in this step the levels that form a group that is
linearized as a single level are now treated separately, so that the
populations of the individual levels in the group are updated.

This is a typical NLTE line formation problem. The main point is
that the solution does not have to be perfect; it is only supposed to
provide a somewhat more consistent values of level populations
and radiation intensities before entering the next global iteration step.
Therefore, in the past, one employed several ordinary Lambda iterations,
that is, an iterative solution that alternates between solving the
transfer equation for the current values of level populations, and solving
the kinetic equilibrium equations with the current values of radiation intensities.

Later, this procedure was upgraded to treat the coupled problem more efficiently 
using the ALI technique together with preconditioning,  developed
by Rybicki \& Hummer (1991, 1992); for a description see Appendix B1,
and for more details refer to
Hubeny \& Mihalas (2014; \S\,14.5). This scheme offers an interesting possibility
to solve the so-called restricted NLTE problem (line formation with fixed 
atmospheric structure), without a linearization, provided that one allows
 for enough iterations of the global formal solution step. We recall that there
 are several computer programs designed specifically for this problem
 -- the ``Kitt Peak code'' of Auer (1973), PANDORA (Avrett \& Loeser 1982),
 MULTI (Carlsson 1986), DETAIL/SURFACE (Butler \& Giddings 1978)
 and others.

In parallel with, or on top of, this procedure, one can perform other
``formal'' solutions, essentially updating one state parameter by solving
the appropriate equation, while keeping other state parameters fixed. 
These include:\\
-- updating electron density by solving the charge conservation equation;\\
-- updating pressure by solving the hydrostatic equilibrium equation;\\
-- updating temperature, by solving the radiative (or radiative + convective) 
equilibrium equation.

A particularly important set of procedures is devised for models with
convection, where in the global formal solution step one has to iteratively
improve the temperature and other state parameters to smooth the solution 
that follows  directly from the previous iteration step. In many cases, not 
using such procedures would have disastrous consequences for the
convergence properties, or even lead to a violent divergence of the 
iteration scheme. These procedures will be described in Appendix B2.

% -------------------------------------------------------------------------------

\subsection{Discretization parameters}
\label{discr}

As stated above, the program is fully data-oriented. Both the genuinely discrete
quantities, (number of explicit atoms, ions, levels, transitions, etc,), as well as
discretized quantities (number of depth points, frequency points, etc.),  are
either set up directly (e.g. the number of depth points), but typically they are
being computed by the code based on the actual input. 
Therefore, they are not known a priori.

Since {\sc tlusty} can be used for a wide range of applications, these numbers 
can be vastly different for various cases. As described in the next chapter, 
the code can be compiled differently for different applications 
in such a way that it does not require an unreasonable amount of core
memory, and still reflects the needs based on the selected setup.

We list most of  these important numbers below, using the names trey are
referred to in the {\sc tlusty} source code.
\begin{description}
\item[NATOM] -- number of explicit atoms. Each explicit atom is composed of
one or several ionization stages, called explicit ions. The highest ionization stage
has to be considered as a one-level ion.
\item[NION] -- number of explicit ions. The highest ionization stage is {\em not}
counted in the number of explicit ions.
\item[NLEVEL] -- number of explicit levels, defined such as these are the energy
levels for which the kinetic equilibrium equation is being solved, and whose 
populations can therefore depart from their LTE values.
The one-level highest ions of the explicit species are now counted into the total 
number of levels. The individual superlevels and merged levels are counted as  
one level each.
\index{Superlevels}
\item[NLVEXP] -- number of explicit levels whose populations are linearized. This
number is equal to NLEVEL if all levels are treated individually, but is lower than
NLEVEL if one introduces level groups. In many cases, NLVEXP is significantly
lower that NLEVEL; for instance in the example in Paper~III, \S\,\refexampostwo, 
NLEVEL=1127 while NLVEXP=222 (there, we show them as MLEVEL and MLVEXP
in an output from the code {\sc pretlus} where they represent the actual values
of these parameters for the given model).
\item[NTRANS] -- number of transitions, both  bound-bound and bound-free,
between explicit energy levels. All transitions that are somehow taken into
account are counted into this number even if they are set in detailed radiative
balance or are dipole-forbidden, since in both these cases, the collisional rates
still have non-zero values.
\item[ND] -- number of discretized depth points
\item[NFREQ] -- the total number of discretized frequency points. Since the
selection of frequencies is very important for the overall accuracy of the resulting
model, the selection of frequency points is set by the program based a number 
of input parameters -- see Paper~III, \S\,\refnonstfreq. 
When computing metal line--blanketed models
with a large number of superlevels and superlines, the program first sets the
frequency points independently for each line and superline, and computes the
\index{Superlines}
corresponding cross section in these frequencies. Since there is typically a large
number of line overlaps, the program then removes some unnecessary points
originating from line overlaps.
\item[NFREQP] -- the number of auxiliary frequency points in the original setup. 
This number
is typically larger than MFREQ, and the reason that is kept separate is that
there are only a few arrays in the code that have to have dimension NFREQP
or larger, while there is a number of arrays, many of them multidimensional,
for the final set of frequencies with number NFREQ,
\item[NFREQC] -- number of frequency points in the continua. This set contains
the frequencies that do not specifically belong to any line, although the opacity
of some wide lines (e.g. Lyman $\alpha$) may still contribute to the total opacity
in such frequencies.
\item[NFREX] -- number of frequency points in which the mean intensity is
linearized. When using the hybrid CL/ALI method, this number is significantly
smaller than NFREQ.
\item[NFREQL] -- the maximum number of frequency points per line.
\item[NTOT$\equiv$NN] -- the dimension of the state vector ${\bf\psi}$, 
that is, the number of the state parameters. It is given by 
NFREX+NLVEXP+NC, where NC is the number of structural 
parameters other than the mean intensities and level populations that are
linearized. As explained in \S\,\ref{cl}, NC can attain values between 0
(when the atmospheric structure is held fixed), to 6 (for accretion disk
with convection and with including fictitious massive particle density;
the physical quantities then being $T$, $N$, $n_{\rm e}$, $n_m$, $z$, and
$\nabla$). A typical value for stellar atmospheres is 3, for $T$, $N$, and
$n_{\rm e}$.
\index{Eddington factor}
\item[NMU] -- number of angle points for the formal solution of the transfer
equation and determination of the Eddington factor. Its default value is 3,
but can be changed by an appropriate input parameter -- see Paper~III, 
\S\,\refnonstdiscr.
\end{description}
There are several secondary numbers that specify the sizes of
some multi-dimensional arrays. (The lengths of 1-dimensional arrays are
not important; they may oversized to satisfy the requirements following
from any physical setup without causing memory problems). They include
the following:
\begin{description}
\item[NLINES(IJ)] - number of lines that contribute to the opacity at 
frequency point IJ (that is, the total number of line overlaps at frequency IJ)
\item[NBF] -- the number of bound-free transitions
\item[NFIT] -- the maximum number of fit points for the input of the
photoionization cross sections when using the Opacity Project data.
\item[NCDW] -- number of levels with pseudocontinua
\item[NMER] -- number of merged levels
\item[NVOIGT] -- number of lines with a Voigt profile
\end{description}
There are also several arrays that are set only for the treatment of
superline cross sections, that is, when computing metal line-blanketed
\index{Superlines}
models. These arrays may be rather big and take a lot of memory. It
it not advised to set them to the maximum values for all kinds of models.
The dimensions of these files are the following:
\begin{description}
\item[NDODDF] -- number of depth points for storing the superline cross sections.
\index{Superlevels}
\item[NKULEV] -- the maximum number of internal energy levels for an
ion treated with superlevels. The data for these levels are read from Kurucz
files, e.g. {\tt gf2601.gam} for Fe II.
\item[NLINE] -- number of internal (genuine) lines for an ion; data for them are 
read from Kurucz files, e.g.  {\tt gf2601.lin} for Fe II. Based on these data, the
superline cross sections are constructed.
\index{Superlines}
\item[NCFE] -- the total number of internal frequency points
used for computing and storing the cross section or the individual lines, from
which the superline cross section is constructed.
\end{description}

%-------------------------------------------------------------------------------

\section{Initial LTE-gray model}
\label{gray}

\subsection {Stellar atmospheres}
\label{gray_atmos}

The procedure to construct the initial LTE-gray model is very similar to
that described by Kurucz (1970).

One first sets up a scale in the Rosseland optical depth, typically logarithmically
equidistant between $\tau_1$ and $\tau_D$, which are input parameters of
the model construction; typically chosen $\tau_1 \approx 10^{-7}$ and 
$\tau_D \approx 10^2$. Temperature is a known function of the Rosseland optical
depth (e.g. Hubeny \& Mihalas 2014, \$\,17.2 and 17.7):
\begin{equation}
\label{tgray}
T^4(\tau) = (3/4) T_{\rm eff}^4 [\tau + q(\tau)]+ (\pi/\sigma_R)H^{\rm ext}
\end{equation}
where $q(\tau)$ is the Hopf function, and $H^{\rm ext} = \int_0^\infty H_\nu^{\rm ext} d\nu$
is the frequency-integrated external irradiation flux.

The hydrostatic equilibrium equation is written as
\begin{equation}
\frac{d\ln P}{d\ln\tau} = \frac{g\tau}{\kappa P},
\end{equation}
because $\tau$ and $P$ span many orders of magnitude, so it is advantageous to 
integrate the equation for logarithms. Here $\kappa$ is the Rosseland mean opacity.

The total pressure $P= P_{\rm gas} + P_{\rm rad}$ (we neglect the turbulent
pressure here). The radiation pressure is expressed
as follows: Its gradient can be approximated as 
\begin{equation}
\label{hetau}
dP_{\rm rad}/d\tau \approx (4\pi/c) \int_0^\infty (dK_\nu/d\tau_\nu)\, d\nu
= (4\pi/c) \int_0^\infty H_\nu d\nu
= (\sigma_R/c) T_{\rm eff}^4,
\end{equation}
where the first equality follows from Eq. (\ref{radpres}), and the last one from the
definition of the effective temperature. The radiation pressure is thus given by
\begin{equation}
P_{\rm rad}(\tau) = (\sigma_R/c) T_{\rm eff}^4 \tau + P_{\rm rad}^0,
\end{equation}
where $P_{\rm rad}^0$ is the radiation pressure at the surface. Using the
Eddington approximation [$K_\nu=(1/3) J_\nu$, and $H_\nu^0=(1/\sqrt{3}) J_\nu^0$],
it is approximated by
\begin{equation}
P_{\rm rad}^0 = (4\pi/c)\int_0^\infty K_\nu d\nu 
\approx (\sigma_R/c)(1/\sqrt{3})\, T_{\rm eff}^4.
\end{equation}

One then proceeds by solving Eq.  (\ref{hetau}) from the top of the atmosphere
to the bottom. At the first depth point, $\tau_1$, one makes a first estimate of
the Rosseland mean opacity, $\kappa_1$, and assuming it is constant from this
point upward, and using the boundary condition $P_{\rm gas}(0)=0$, one obtains
the first estimate of the total pressure,
\begin{equation}
\label{pr1}
P_1 = (g/\kappa_1) \tau_1 + P^0_{\rm rad}.
\end{equation}
Having an estimate for the total pressure,
one uses the following procedure which is valid for every depth point $d$:
\begin{itemize}
\item from current total pressure $P_d$, at depth point $d$, one first extracts the
gas pressure, $P_{\rm gas} = P -  (\sigma_R/c) T_{\rm eff}^4 \tau_d - P_{\rm rad}^0$;
\item from the known temperature $T(\tau_d)$, given by Eq. (\ref{tgray}), we compute the
total particle number density $N=P_{\rm gas}/(kT)$;
\item with known $T$ and $N$, one determines the electron density $n_{\rm e}$
by solving the set of Saha equations and the charge equilibrium equation;
\item with known $T$ and $n_{\rm e}$, one computes LTE level populations of
all explicit levels of all explicit ions;
\item using these level populations, one computes monochromatic opacities for
all selected frequency points, and consequently the new value of the Rosseland
mean opacity $\kappa$ from its definition,
\begin{equation}
\frac{1}{\kappa} = \frac{\int_0^\infty (1/\chi_\nu)(dB_\nu/dT)\, d\nu}
{(dB/dT) }.
\end{equation}
\end{itemize}
We will refer to this procedure as $P\!\rightarrow\!\kappa$.
With the new value of $\kappa$, one returns to Eq. (\ref{pr1}), evaluates an
improved estimate of $P_1$, and repeats the procedure $P\!\rightarrow\!\kappa$
until convergence. Once this is done, one proceeds to the next depth point.

For the next three depth points, $d=2,\ldots,4$,
one obtains the first estimate (a predictor step)
of the total pressure by:
\begin{equation}
\ln P_d^{\rm pred} = \ln P_{d-1} + \Delta\ln P_{d-1} ,
\end{equation}
which is followed by a $P\!\rightarrow\!\kappa$ procedure, and with the new $\kappa$
one goes to the corrector step,
\begin{equation}
\ln P_d = (\ln P_{d}^{\rm pred} + 2 \ln P_{d-1} + \Delta\ln P_{d} +  \Delta\ln P_{d-1})/3,
\end{equation}
where
\begin{equation}
\Delta\ln P_d = \frac{g\tau_d}{\kappa_d P_d} (\ln\tau_d- \ln\tau_{d-1}).
\end{equation}
For the subsequent depth points, one uses the Hamming's predictor-corrector
scheme (see Kurucz 1970), where the predictor step is
\begin{equation}
\ln P_d = (3\ln P_{d-4} + 8 \ln P_{d-1}- 4\Delta\ln P_{d-2} + 8 \Delta\ln P_{d-3})/3,
\end{equation}
and the corrector step
\begin{eqnarray}
\ln P_d = (126\ln P_{d-1} -14 \ln P_{d-3}+ 9 \ln P_{d-4} 
+42 \Delta\ln P_{d} \nonumber \\
+ 108 \Delta\ln P_{d-1}-54\Delta\ln P_{d-2} + 24 \Delta\ln P_{d-3})/121.
\end{eqnarray}
After completing the above procedure for all depths, one constructs the
column mass scale, which will subsequently be used as the basic depth
scale, as
\begin{equation}
m_d = (P_d - P_{\rm rad}^0)/g.
\end{equation}

When convection is taken into account, one first computes the radiative gradient
of temperature,
\begin{equation}
\nabla_{{\rm rad},d} = \frac{(T_d - T_{d-1})} {(P_d - P_{d-1})}\frac {(P_d + P_{d-1})}{(T_d + T_{d-1})},
\end{equation}
and compares to the adiabatic gradient, $\nabla_{\rm add}$. 
If $\nabla_{\rm rad} > \nabla_{\rm add}$, the criterion for stability against 
convection is violated, we must determine the true 
gradient $\nabla$, where $\nabla_\mathrm{ad}\leq\nabla\leq\nabla_\mathrm{rad}$, 
that gives the correct total, radiative plus convective, flux.
If the instability occurs deep enough for the
diffusion approximation to be valid, then $(F_\mathrm{rad}/F)=
(\nabla/\nabla_\mathrm{ad})$, and the energy balance equation reads (see Hubeny \&
Mihalas 2014, \S\,17.4),
\begin{equation}
\label{eq:17.4c-10}
\mathcal{A}\big(\nabla-\nabla_\mathrm{el}\big)^{3/2}=\nabla_\mathrm{rad}-\nabla,
\end{equation}
where $\nabla_{\rm el}$ is the gradient of convective elements, and
\begin{equation}
\label{eq:17.4c-11}
\mathcal{A} = (\nabla_\mathrm{rad}/\sigma_\mathrm{R}T_\mathrm{eff}^4) 
(gQH_P/32)^{1/2}(\rho c_P T) (\ell/H_P)^2 .
\end{equation}
We see that $\mathcal{A}$ depends only on local variables. Adding
$\big(\nabla-\nabla_\mathrm{el}\big)+\big(\nabla_\mathrm{el}
-\nabla_\mathrm{ad}\big)$ to both sides of (\ref{eq:17.4c-10}), and using
the expression $\nabla_{\rm el} - \nabla_{\rm ad}= B \sqrt{\nabla-\nabla_{\rm el}}$,
where $B$ is given by Eq. (\ref{convb}), to eliminate
$\big(\nabla_\mathrm{el}-\nabla_\mathrm{ad}\big)$, we obtain a cubic equation
for $x\equiv\big(\nabla-\nabla_\mathrm{el}\big)^{1/2}$, namely
\begin{equation}
\mathcal{A}\big(\nabla-\nabla_\mathrm{el}\big)^{3/2}+\big(\nabla-\nabla_\mathrm{el}\big)+
B\big(\nabla-\nabla_\mathrm{el})^{1/2}=\big(\nabla_\mathrm{rad}-\nabla_\mathrm{ad}\big).
\end{equation}
or
\begin{equation}
\mathcal{A}x^3+x^2+Bx= \big(\nabla_\mathrm{rad}-\nabla_\mathrm{ad}\big),
\end{equation}
which can be solved numerically for the root $x_0$. We thus obtain the true
gradient $\nabla=\nabla_\mathrm{ad}+\mathcal{B}x_0+x_0^2$, and can proceed with
the integration, now regarding $T$ as a function of $P$ and $\nabla$.

%-----------------------------------------------------------------------------------------

\subsection{Accretion disks}
\label{gray_disk}

The adopted procedure closely follows that described in detail by Hubeny (1990).
An evaluation of the LTE-gray model proceeds in three, possibly four,  basic steps:
\medskip

\noindent {\em 1. Initialization of $m$, $\rho$, $P$, and $z$}.

Unlike the case of stellar atmospheres, one first sets up a column mass scale, 
based on an
empirically chosen $m_1$, the column mass at the first depth point, and $m_0$, the
column mass at the central plane. The latter is determined as described in \S\,\ref{disks}.
The individual mass-depth points are set logarithmically equidistant between $m_1$ and
$m_0$. One then calculates an initial estimate of the density $\rho$ and the vertical distance
$z$ corresponding to the column masses $m_d$, using the following procedure:

One introduces the characteristic gas pressure and radiation pressure scale heights 
$H_g$ and $H_r$ as
\begin{eqnarray}
H_g = (2 c_g^2/Q)^{1/2} \\
H_r = (\sigma_R/c) T_{\rm eff}^4 \kappa/Q,
\end{eqnarray}
where $Q$ is the gravity acceleration parameter defined by Eq. (\ref{qdef}),
$\kappa$ is the Rosseland mean opacity, and
$c_g$ is the isothermal sound speed associated to the gas pressure,
$c_g = (P_g/\rho)^{1/2}$. It is generally given by
\begin{equation}
c_g^2 = \frac{k}{\mu m_H} \frac{N}{N-n_{\rm e}} \, T,
\end{equation}
where $\mu$ is the mean molecular weight, and $m_H$ is the hydrogen atom mass.
The factor $N/(N-n_{\rm e})$ accounts
for a varying degree of ionization of the material; for a pure-hydrogen gas it
attains values between 1 (for completely neutral gas) to $1/2$ (for completely ionized gas).

The sound speed is initially taken as depth-independent, corresponding 
to the effective temperature,
The initial estimate  of $\kappa$ for the first depth
point is $\kappa_1 = \sigma_{\rm e}/(\mu m_H)$. One introduces two dimensionless
parameters,
\begin{equation}
r\equiv H_r/H_g,\quad y\equiv H/H_g,
\end{equation}
where $H$ is a combined scale height, given by the solution of the following
transcendental equation
\begin{equation}
y=\frac{\sqrt{\pi}}{2}\left(\frac{y}{y-r}\right)^{1/2}\!\!
\Big\{1-{\rm erfc}\Big[\sqrt{y(y-r)}\Big]\Big\} + {\rm erfc}(y-r) \exp[-r(y-r)],
\end{equation}
which is solved by the Newton-Raphson method, with the initial estimate\\
$y_0=r+(1/r)$ for $r>1$, and $y_0=\sqrt{\pi}/2$ for $r\leq 0$. The geometrical distance
from the central plane expressed in units of gas pressure scale height, $x\equiv z/H_g$,
is given by (Hubeny 1990),
\begin{eqnarray}
x=r+{\rm inverfc}\left\{\frac{m}{m_0}\frac{2y}{\sqrt{\pi}}\exp[r(y-r)] \right\}\,\, {\rm for}\,\,\,
x\leq y,\\
x=\sqrt{\frac{y}{y-r}} {\rm inverfc}\left\{\sqrt{\frac{4y(y-r)}{\pi}} \frac{m-m_y}{m_0}
+ {\rm erfc}\Big[ \sqrt{y(y-r)} \Big] \right\} \,\, {\rm for}\,\,\, x > y,
\end{eqnarray}
where
\begin{equation}
m_y=m_0 \frac{\sqrt{\pi}}{2y} \exp[r(y-r)]\, {\rm erfc}(y-r),
\end{equation}
and ${\rm inverfc}(x)$ is the inverse complementary error function, which can be
evaluated by a suitable fitting formula (see Hubeny 1990).

The first estimate of density is given by
\begin{eqnarray}
\rho(x) = \rho_0 \exp\left[-x^2(1-r/h)\right] \quad {\rm fort}\,\,\, x \leq y,\\
\rho(x) = \rho_0 \exp[-(x-r)^2] \exp[-r(y-r)]\quad {\rm for}\,\,\, x > y,
\end{eqnarray}
where $\rho_0$ is the density at the central plane, $\rho_0=m_0/(yH_g)$.
The initial estimate of the gas pressure is then obtained using the sound speed for the
characteristic temperature, $P_g(m) = c_g^2 \rho(m)$. Finally, the initial estimate
of the vertical distance from the central plane is obtained by solving the $z$-$m$ relation,
$dz = -dm/\rho$.

\medskip
\noindent {\em 2. Initial estimate of the temperature.}

The procedure differs depending on whether or not the Compton scattering
is taken into account.

Without Compton scattering:
The temperature as a function of the Rosseland optical depth
is given by a modification of the $T(\tau)$ relation for stellar
atmospheres (Hubeny 1990), which in a simplified form that assumes the Eddington 
approximation reads
\begin{equation}
\label{tdisk}
T^4 = \frac{3}{4} T_{\rm eff}^4 \left[ \tau \left(1-\frac{\tau}{\tau_{\rm tot}}\right) +\frac{1}{\sqrt{3}}
+ \frac{1}{3\epsilon\tau_{\rm tot}}\frac{w}{\bar w}\right],
\end{equation}
where $\tau$ is the Rosseland optical depth, $\tau_{\rm tot}$ is the $\tau$ at the central plane,
and $\epsilon=\kappa_B/\kappa$, that is the ratio of the Planck-mean to the Rosseland mean
opacity. In a strict gray model, $\epsilon=1$. If the total optical thickness of the disk
is large, $\tau_{\rm tot} \gg 1$, the last term in Eq. (\ref{tdisk}) is small. 

With Compton scattering:
The formalism is taken from Hubeny et al. (2001; their Eqs. (42-44). 
The local temperature is given by
the solution of a fourth-order algebraic equation
\begin{equation}
\label{tdiskc}
\bar\epsilon \left(\frac{T}{T_{\rm eff}}\right)^4 =
\frac{3}{4}\left[ \tau \left(1-\frac{\tau}{2\tau_{\rm tot}}\right) +\frac{1}{\sqrt{3}} \right]
(\bar\epsilon - 2.867\times 10^{-11} T)
+ \frac{1}{4\tau_{\rm tot}}\frac{w}{\bar w},
\end{equation}
where 
\begin{equation}
\bar\epsilon = \frac{\kappa_B}{n_{\rm e} \sigma_{\rm T}} \equiv
\frac{\int_0^\infty \kappa_\nu B_\nu d\nu}{\int_0^\infty B_\nu d\nu}
\frac{1}{n_{\rm e} \sigma_{\rm T}}.
\end{equation}
Equation (\ref{tdiskc}) is solved by a Newton-Raphson method.
\smallskip

For each depth, starting with $d=1$, the following iteration loop is performed: \\
$(a)$ first the increment of the Rosseland mean opacity is estimated (taken equal to the
increment at the previous depth $d-1$; the initial values at $d=1$ are given by the input
values); then the optical depth corresponding to depth $d$ is determined. \\
$(b)$ The temperature from Eq. (\ref{tdisk}) or (\ref{tdiskc}) is calculated. \\
$(c)$ Given the current values of $T$
and $P_g$, one performs a $P_g\!\rightarrow\!\kappa$ procedure (that is, a part of the
$P\!\rightarrow\!\kappa$ without its first step because we started already with the gas
pressure). \\
$(d)$ With the new value of $\kappa$ one computes an updated  optical depth $\tau$
and returns to step $(b)$. \\
The loop $(b)-(d)$  is repeated several times until the relative
changes of $T_d$ are sufficiently small.

\medskip
\noindent
{\em 3. Refinement of pressure, density, and vertical distance.}

The values of the structural parameters are improved by a simultaneous solution of the
hydrostatic equilibrium equation and the $z$-$m$ relation. It turned out that a numerically
more stable form of the hydrostatic equation is obtained by differentiating its original
form, $dP/dm=Qz$ once more over $m$ and using the exact expression $dz/dm=-1/\rho$,
to obtain a second-order equation for $P$,
\begin{equation}
\frac{d^2P}{dm^2} = -\frac{Q}{\rho} = -\frac{Q}{c_s^2 P},
\end{equation}
where the total sound speed $c_S = (P^{\rm old}/\rho^{\rm old})^{1/2}$ is taken as a 
known function of depth. The upper boundary condition is derived assuming that the temperature 
is constant for $m<m_1$, and integrating the hydrostatic equilibrium equation form depth $m_1$
upward. One obtains (Hubeny 1990)
\begin{equation}
\label{he1disk}
m_1 = H_g \rho(z_1) f\left(\frac{z-H_r}{H_g}\right),
\end{equation}
where
\begin{equation}
\label{he1f}
f(x)\equiv (\sqrt{\pi}/2) \exp(x^2)\, {\rm erfc}(x).
\end{equation}
Here, $H_g$, $H_r$ and $c_S$ are evaluated using the current values 
of the state parameters at $d=1$.

\medskip
\noindent {\em 4. Changing the structure in the convection zone.}

The procedure for changing temperature, and consequently the other state parameters,
in the regions where the material is unstable against convection, is exactly the same
as in the case of stellar atmospheres.

\section{Conclusion}

This document, which forms Part II of the three-paper series of a detailed user's
guide for {\sc tlusty} and {\sc synspec}, contains an overview of physical assumptions,
basic structural equations, and the description of the numerical methods to solve them.
This paper thus provides a theoretical background for the next paper, which will
provide a practical guide for working with {\sc tlusty}. It will
cover computational issues, namely a description of the input data and output
files, a selection of appropriate options to fine-tune physical and numerical setup of the
model construction, and basic troubleshooting.

% --------------------------------------------------------------------------

\section*{Acknowledgements}

We are grateful to Peter Nemeth, Yeisson Ossorio, and Klaus Werner for their
very careful reading of the manuscript and making many useful comments.
I.H. gratefully acknowledges the support from the Alexander von Humboldt Foundation,
and wishes to thank especially to Klaus Werner for his hospitality at the Institute of Astronomy 
and Astrophysics of the University of T\"ubingen, where a part of the work on this paper was done. 

% --------------------------------------------------------------------------

\section*{Appendix A: Details of the formal solution of the transfer equation}
\addcontentsline{toc}{section}{Appendix A: Details of the formal solution of the transfer equation} 

As mentioned above, {\sc tlusty} allows for several possibilities to perform
a formal solution of the transfer equation. We recall that by the term {\em formal
solution} we understand a solution with the thermal source function fully
specified. If the source function contains a scattering term, which is essentially
always the case, this term is {\em not} assumed as given; instead it is treated
self-consistently. The total number of depth points is $N\!D$; depth index $d$ 
attains values between 1 (representing the uppermost point) and $d=N\!D$,
representing the deepest point in the atmosphere. The angle points are
labelled $i=1,\ldots, N\!A$, with $N\!A$ being the total number of angle
points; their order is arbitrary.

We consider here the case where the transfer equation does
not contain any coupling of the individual frequency points, and therefore
it can be solved frequency by frequency. We thus skip an indication of the
frequency dependence of the corresponding quantities, as well as the indices
of discretized frequency points.

\subsection*{A1. Feautrier scheme}
\addcontentsline{toc}{subsection}{A1. Feautrier scheme}

This is the standard method to solve the transfer equation with the
known thermal source function. Here we also assume that the source
function does not depend on angle.

The angle-dependent transfer equation is discretized as follows:\\ [2pt]
\noindent $\bullet$ for inner depth points, 
$d=2,\dots,N\!D-1$,\ \ 
\begin{equation}
\label{feaain}
\frac{\mu_i^2 j_{d-1,i}}{\Delta\tau_{d-{1/2}}\Delta\tau_{d}}
-\frac{\mu_i^2 j_{di}}{\Delta\tau_{d}}\!
         \left(\!\frac{1}{\Delta\tau_{d-{1/2}}}
                +\frac{1}{\Delta\tau_{d+{1/2}}}\!\right)\!
+\!\frac{\mu_i^2 j_{d+1,i}}{\Delta\tau_{d+{1/2}}\Delta\tau_{d}}
= j_{di} - S_{d},
\end{equation}
where
\begin{equation}
\label{feaadt}
\Delta\tau_{d}\equiv 
\frac{1}{2}(\Delta\tau_{d-1/2}+\Delta\tau_{d+1/2}),
\end{equation}
The total source function is given by
\begin{equation}
\label{feaasf}
S_{d}=\frac{\sigma_d}{\chi_d}\sum_{j=1}^{N\!A}w_{dj} j_{dj}+\frac{\eta_{d}}{\chi_d}.
\end{equation}

\noindent $\bullet$ upper boundary condition, in the second-order form, is given by
\begin{equation}
\label{feaaub}
\mu_l \frac{j_{2i}-j_{1i}}{\Delta\tau_{3/2}} = 
j_{1i}- I_i^{\rm ext} +\frac{\Delta\tau_{3/2}}{2\mu_i} (j_{1i}-S_{1l})
\end{equation}
\noindent $\bullet$ For the lower boundary, $d=N\!D$, there are two possible forms, one for
stellar atmospheres, with diffusion approximation, and one for accretion disks, with
symmetry boundary condition, $dj/d\tau|_{N\!D} =0$. Both are considered in the second-order
form.

-- For stellar atmospheres 
\begin{equation}
\label{feaalb}
\mu_i\frac{j_{di}-j_{d-1,i}}{\Delta\tau_{d-1/2}}=I_i^+-j_{di}-
\frac{\Delta\tau_{d-1/2}}{2\mu_i}(j_{di}-S_{d}),
\end{equation}
where
\begin{equation}
\label{feaaip}
I^+_i = B_{N\!D} + \mu_i \frac{B_{N\!D}-B_{N\!D-1}}{\Delta\tau_{N\!D-1/2}}.
\end{equation}

-- For accretion disks
\begin{equation}
\label{feaalbd}
\mu_i\frac{j_{di}-j_{d-1,i}}{\Delta\tau_{d-1/2}}=-
\frac{\Delta\tau_{d-1/2}}{2\mu_i}(j_{di}-S_{d}),
\end{equation}

One introduces a column vector 
$ {\bf j}_d\equiv \{ j_{d1},j_{d2},\ldots,j_{d,N\!A}\}^T$, and the set
of equations (\ref{feaain}), (\ref{feaasf}) - (\ref{feaaip}) is expressed as
a block tridiagonal system of the form
\begin{equation}
\label{feautri}
-{\bf A}_{d}\,{\bf j}_{d-1}+{\bf B}_{d}\,{\bf j}_d-{\bf C}_{d}\,{\bf j}_{d+1}=
{\bf R}_d\quad (d=1,\dots,N\!D).
\end{equation}

${\bf A}_d$, ${\bf B}_d$, and ${\bf C}_d$ are $(N\!A\times N\!A)$ matrices.
${\bf A}_d$ and ${\bf C}_d$ are diagonal, containing the
finite--difference terms in (\ref{feaain}) that couple the angle--frequency
components of ${\bf j}_{di}$ at depth point $d$ to those at depth
points $d-1$ and $d+1$ respectively.
${\bf R}_d$ is a column vector of length $N\!A$, containing the thermal 
(i.e.~non--scattering) source terms in (\ref{feaasf}).
 
At an interior point $(d=2,\ \dots,\ N\!D-1)$, $(i,j=1,\ \dots,\ N\!A)$, 
\begin{equation}
\label{eq:12.2-26}
({\bf A}_d)_{ij}= \mu_i^2/(\Delta\tau_{d-{1/2}}\,\Delta\tau_{d})\,\delta_{ij},
\end{equation}
\begin{equation}
\label{eq:12.2-27}
({\bf C}_d)_{ij}=\mu_i^2/(\Delta\tau_{d+{1/2}}\,\Delta\tau_{d})\,\delta_{ij},
\end{equation}
\begin{equation}
\label{eq:12.2-28}
({\bf B}_d)_{ij}=\delta_{ij}+({\bf A}_d)_{ij}+({\bf C}_d)_{ij}- 
(\sigma_{d}/\chi_d) w_{j},
\end{equation}
\begin{equation}
({\bf R}_d)_i = \eta_d/\chi_d.
\end{equation}
Here $\delta_{ij}$ is the Kronecker symbol. 

At the upper boundary, $d=1$, one has $({\bf A}_1)_{ij}\equiv 0$, and
\begin{equation}
\label{bubatm}
({\bf B}_d)_{ij} = [1 + (2\mu_i/\Delta\tau_{d+1/2})
+2(\mu_i/\Delta\tau_{d+1/2})^2]\,\delta_{ij} 
- (\sigma_{d}/\chi_d) w_{j} ,
\end{equation}
\begin{equation}
\label{cubatm}
({\bf C}_d)_{ij} = 2(\mu_i/\Delta\tau_{d+1/2})^2\,\delta_{ij},
\end{equation}
and
\begin{equation}
\label{rubatm}
{\bf R}_{di} = \eta_{d}/\chi_d + (2\mu_i/\Delta\tau_{d+1/2}) I_i^{\rm ext}.
\end{equation}

For the lower boundary, $d=N\!D$,  one has $({\bf C}_{N\!D})_{ij} = 0$, and

-- for stellar atmospheres
\begin{equation}
\label{bndatm}
({\bf B}_{d})_{ij}=[1 + (2\mu_i/\Delta\tau_{d-1/2}) 
+2(\mu_i/\Delta\tau_{d-1/2})^2]\delta_{ij}
 - (\sigma_{d}/\chi_{d})\, w_j, 
\end{equation}
\begin{equation}
\label{andatm}
({\bf A}_{d})_{ij} = 2(\mu_i/\Delta\tau_{d-1/2})^2\,\delta_{ij},
\end{equation}
\begin{equation}
\label{rndatm}
({\bf R}_{d})_{i} = \sigma_{d}/\chi_{d}-(2\mu_i/\Delta\tau_{d-1/2})I^+_{d,i}.
\end{equation}

-- for accretion disks
\begin{equation}
\label{bnddis}
({\bf B}_{d})_{ij}=[1 + 2(\mu_i/\Delta\tau_{d-1/2})^2]\,\delta_{ij}
 - (\sigma_{d}/\chi_{d})\, w_j, 
\end{equation}
\begin{equation}
\label{anddis}
({\bf A}_{d})_{ij} = 2(\mu_i/\Delta\tau_{d-1/2})^2\,\delta_{ij},
\end{equation}
and
\begin{equation}
\label{rnddis}
({\bf R}_{d})_{i} = \sigma_{d}/\chi_{d},
\end{equation}

Equation (\ref{feautri}) is solved by a standard Gauss-Jordan elimination 
\index{Gauss-Jordan elimination}
that consists of a forward elimination, analogously to that employed for
complete linearization, Eqs, (\ref{elim1}) - (\ref{elim3}),
\begin{equation}
\label{elimj1}
{\bf D}_1 = {\bf B}_1^{-1} {\bf C}_1,\quad {\rm and}\quad
{\bf D}_d = ({\bf B}_d - {\bf A}_d {\bf D}_{d-1})^{-1} {\bf C}_d,\quad d=2,\ldots,N\!D,
\end{equation}
and
\begin{equation}
\label{elimj2}
{\bf Z}_1 = {\bf B}_1^{-1} {\bf L}_1,\quad {\rm and}\quad
{\bf Z}_d = ({\bf B}_d - {\bf A}_d {\bf D}_{d-1})^{-1} 
({\bf L}_d + {\bf A}_d {\bf Z}_{d-1}) ,\quad d=2,\ldots,N.
\end{equation}
followed by a back-substitution
\begin{equation}
\label{elimj3}
{\bf \delta\psi}_{\!N\!D} = {\bf Z}_{N\!D},\quad {\rm and}\quad
{\bf \delta\psi}_d = {\bf D}_d  {\bf \delta\psi}_{d+1} +  {\bf Z}_{d} ,
\quad d=N\!D-1,\ldots,1.
\end{equation}

\subsubsection*{Scalar Feautrier scheme}
 
Once the solution for ${\bf j}$ is obtained by the procedure 
described by Eqs. (\ref{elimj1}) - (\ref{elimj3}),
the mean intensity and the Eddington factors are computed using
(\ref{rteff}). This procedure may be sufficient. However, for the overall consistency,
the transfer equation needs to be solved again, now for the mean intensity using
the  newly determined Eddington factors. The reason for this is that such an equation 
is being considered as one of the structural equations. Although its solution for
the mean intensity is mathematically equivalent to the corresponding integral of 
the solution of the angle-dependent transfer equation, these solutions are
slightly different numerically due to rounding errors, and due to inaccuracies
in numerical integration over angles.  One should therefore enter the next linearization
step with a current solution of the same equation that is linearized; otherwise
the global iteration process would stop converging at small but non-zero relative changes.

The discretized combined moment equation is written as follows:

\noindent $\bullet$ For inner depth points, $d=2,\ldots,N\!D-1$, one has
\begin{eqnarray}
\label{feabip}
\frac{f_{d-1}J_{d-1}}{\Delta\tau_{d-1/2}\Delta\tau_{d}} -
\frac{f_{d}J_d}{\Delta\tau_{d}}\left(\frac{1}{\Delta\tau_{d-1/2}}+
               \frac{1}{\Delta\tau_{d+1/2}}\right)
 +\frac{f_{d+1}J_{d+1}}{\Delta\tau_{d+1/2}\Delta\tau_{d}} =
J_{d} - S_d.
\end{eqnarray}
The right--hand side of equation (\ref{feabip}) can be written as
\begin{equation}
\label{feabrh}
J_{d} - S_d = \epsilon_{d} J_{d} -\frac{\eta_{d}}{ \chi_{d}},
\end{equation}
with
\begin{equation}
\label{feabep}
\epsilon_{d}  \equiv 1 - \sigma_{d}/\chi_{d} 
= \kappa_{d}/\chi_{d}.
\end{equation}

\noindent $\bullet$ The upper boundary condition ($d=1$), in the second--order form
that follows from integrating equation (\ref{feaaub}) over angles, is
\begin{equation}
\label{feablu}
\frac{f_{d+1}J_{d+1}-f_{d}J_{d}} { \Delta\tau_{d+1/2}} =
g J_{d} - H^\mathrm{ext} + \frac {\Delta\tau_{d+1/2}}{2} 
\left(\epsilon_{d} J_{d} - \frac{\eta_{d}}{\chi_{d}} \right).
\end{equation}

\noindent $\bullet$ The lower boundary condition ($d=N\!D$) is again
different for semi-infinite atmospheres and for accretion
disks.

-- For stellar atmospheres, assuming the diffusion approximation,
\begin{equation}
\label{feablb}
\frac{f_{s}J_{s}-f_{d-1}J_{d-1} } {\Delta\tau_{d-1/2}}=
\frac{1}{2}({\cal B}_{d}-J_{d})  
+\frac{{\cal B}_{d}-{\cal B}_{d-1}}{3\,\Delta\tau_{d-1/2}}
-\frac{\Delta\tau_{d-1/2}}{2}
\left(\epsilon_{d}J_{d} - \frac{\eta_{d}}{\chi_{d}} \right),
\end{equation}
where ${\cal B}$ denotes the Planck function.

-- For accretion disks, one also employs a second-order form in which one
uses the symmetry condition $(dj/d\tau)_{N\!D}=0$,
%\
\begin{equation}
\label{feabdis}
\frac{2}{\Delta\tau_{d-1/2}}\frac{f_d J_d - f_{d-1}J_{d-1}}{\Delta\tau_{d-1/2}}
+\epsilon_d = \frac{\eta_d}{\chi_d}.
\end{equation}

Equations (\ref{feabip}), (\ref{feablu}), and (\ref{feablb}) also form a tridiagonal system
\begin{equation}
\label{feabtr}
-A_d J_{d-1} + B_d J_d -C_d J_{d+1} = R_d,
\end{equation}
where now $A_d$, $B_d$, and $C_d$ are scalars (real numbers).
They are given by
\begin{equation}
A_d =
\left\{ \begin{array}{ll}
0, & {\rm for\ \ } d=1, \\ [2pt]
f_{d-1}/[\Delta\tau_{d-1/2}\Delta\tau_{d}], & {\rm for\ \ } d\geq 2,
\end{array} \right.
\end{equation}
\begin{equation}
B_d =
\left\{ \begin{array}{ll}
(f_d/\Delta\tau_{d+1/2}) +g+\epsilon\Delta\tau_{d+1/2}/2, 
& d=1, \\ [2pt]
(f_d/\Delta\tau_d)\left[\Delta\tau_{d-1/2}^{-1}+\Delta\tau_{d+1/2}^{-1}\right]+\epsilon,
& d=2,\ldots,N\!D-1, \\ [2pt]
(f_d/\Delta\tau_{d-1/2}) +(1/2) +\epsilon\Delta\tau_{d-1/2}/2,
& d=N\!D, 
\end{array} \right.
\end{equation}
\begin{equation}
C_d =
\left\{ \begin{array}{ll}
f_{d+1}/[\Delta\tau_{d+1/2}\Delta\tau_{d}], & 
d<N\!D,\\ [2pt]
0 &  
d=N\!D, 
\end{array} \right.
\end{equation}
and
\begin{equation}
R_d =
\left\{ \begin{array}{ll}
H^{\rm ext} + (\Delta\tau_{d+1/2}/2) (\eta_d/\chi_d), & d=1, \\ [2pt]
(\eta_d/\chi_d), & d=2,\ldots,N\!D-1, \\ [2pt]
X^+ -\epsilon\Delta\tau_{d-1/2}/2,
& d=N\!D, 
\end{array} \right.
\end{equation}
where 
\begin{equation}
X^+={\cal B}_{N\!D}/2 + ({\cal B}_{N\!D}-{\cal B}_{N\!D-1})/(3\Delta\tau_{N\!D-1/2})
\end{equation}
Equation (\ref{feabtr}) is solved  analogously as described above.

% -----------------------------

\addcontentsline{toc}{subsection}{A2. Fourth-order Hermitian scheme}
\subsection*{A2. Fourth-order Hermitian scheme}
This efficient and very accurate modification of the standard Feautrier
scheme was suggested by Auer (1976). The finite difference scheme
originally expressed by Eq. (\ref{feaain}) is replaced by
\begin{equation}
\label{eq:12.4-2}
-A_d\,j_{d-1}+B_d\,j_d-C_d\,j_{d+1}+\alpha_d\,j_{d-1}^{\prime\prime} 
+\beta_d\,j_{d}^{\prime\prime}+\gamma_d\,j_{d+1}^{\prime\prime} = 0,
\end{equation}
where $j_d^{\prime\prime}\equiv \mu^2\, d^2 j/d\tau^2|_d= j_d -S_d$.
Expanding $j_{d\pm 1}$ and $j^{\prime\prime}_{d\pm 1}$ to fourth
order Taylor series, one obtains after some algebra (see Auer 1967; or
Hubeny \& Mihalas 2014, \$\,12.4) a block-tridiagonal system analogous
to the standard Feautrier scheme, Eq. (\ref{feautri}), where now
\begin{eqnarray}
({\bf A}_d)_{ij} &=& (a_{di}-\alpha_{di}) \delta_{ij} + \alpha_{di}\, (\sigma_{d-1}/\chi_{d-1})\, w_j, \\ [2pt]
({\bf C}_d)_{ij} &=& (c_{di}-\gamma_{di}) \delta_{ij} + \gamma_{di}(\sigma_{d+1}/\chi_{d+1})\, w_j, \\ [2pt]
({\bf B}_d)_{ij} &=& (b_{di}+\beta_{di}) \delta_{ij} - \beta_{di}(\sigma_d/\chi_d)\, w_j,
\end{eqnarray}
and
\begin{equation}
{\bf R}_{di} = \alpha_{di}(\eta_{d-1}/\chi_{d-1}) + \beta_{di} (\eta_d/\chi_d) +
\gamma_{di}(\eta_{d+1}/\chi_{d+1}), 
\end{equation}
where the individual auxiliary quantities are given by
\begin{eqnarray}
a_{di}&=&\mu_i^2/(\Delta\tau_{d}\Delta\tau_{d-1/2}), \\ [2pt]
c_{di}&=&\mu_i^2/(\Delta\tau_{d}\Delta\tau_{d+1/2}), \\ [2pt]
b_{di}&=&a_{di}+c_{di}, \\ [2pt]
\alpha_{di} &=& [1-a_{di}\Delta\tau_{d+1/2}^2/(2\mu_i^2)]/6, \\ [2pt]
\gamma_{di} &=& [1-c_{di}\Delta\tau_{d-1/2}^2/(2\mu_i^2)]/6, \\ [2pt]
\beta_{di}&=&1-\alpha_{di} - \gamma_{di},
\end{eqnarray}
Although one can construct a third-order form for the boundary condition,
the usual second-order form expressed by Eqs. (\ref{bubatm}) - (\ref{rnddis})
is satisfactory, and is being used in {\sc tlusty}.

% --------------------

\subsection*{A3. Improved  Rybicki-Hummer solution algorithm}
\addcontentsline{toc}{subsection}{A3. Improved  Rybicki-Hummer solution algorithm}

In this variant of the Feautrier scheme, the basic expressions
Eq. (\ref{feaain}) - (\ref{feaalbd}) remain unchanged, the only point which
is changed is the method of the solution of the resulting tri-diagonal system
(\ref{feautri}). It can obviously be used in conjunction with the 4-th
order Hermitian scheme as well.

This scheme has better numerical properties if $\Delta\tau \ll 1$, which may
easily happen near the surface. In this case, the terms proportional to
$\Delta\tau^{-2}$ are very large, which cause the other terms ("1" from
the Kronecker $\delta$, and the term corresponding to scattering source
function term), may be  lost because of a limited numerical representation,  
or be inaccurate due to numerical noise. In this case, one changes the original 
algorithm by introducing an auxiliary quantity 
\begin{equation}
H_d = -A_d + B_d - C_d,
\end{equation}
(with $A_1= C_{N\!D}=0$),
and
\begin{equation}
F_d=D_d^{-1} - 1.
\end{equation}
and the solution algorithm, originally described by Eqs. (\ref{elim1}) - (\ref{elim3}), 
proceeds now as follows
\begin{eqnarray}
F_d = C_d^{-1}\{ H_d + A_d \cdot[1-(1+F_{d-1})^{-1}]\}, \\
E_d = (1+F_d)^{-1} C_d^{-1} (R_d + A_d E_{d-1}),
\end{eqnarray}
The reverse sweep starts with $j_{N\!D} = E_{N\!D}$, followed by
for $d\!=\!N\!D-1,\ldots,\!1$,
\begin{equation}
j_d = (1+F_d)^{-1} j_{d+1} + E_d.
\end{equation}
This formalism applies both for the block-tridiagonal systems of equations
(\ref{feautri}), in which case $A$, $B$, $C$, $H$, $D$, $F$, and $E$
are matrixes, as well as by Eq. (\ref{feabtr}), in which case they are scalars.

%------------------------------

\subsection*{A4. Discontinuous Finite Elements}
\addcontentsline{toc}{subsection}{A4. Discontinuous Finite Elements}

If an atmospheric structure exhibits very sharp variations 
with depth, it is advantageous to use the Discontinuous Finite Element (DFE) 
scheme by Castor et al. (1992). It solves the linear transfer equation 
\begin{equation}
\label{rtestd}
\frac{dI_\nu}{d\widetilde{\tau}_\nu} = I_\nu-S_\nu,
\end{equation}
where $\widetilde\tau_\nu \equiv \tau_\nu/|\mu|$
is the optical depth along the line of propagation of radiation. Equation (\ref{rtestd})
is solved directly for the specific intensity, and therefore the scattering part of the source
function has to be treated iteratively. To this end, a simple ALI-based procedure
is used. It is outlined below. Here we describe the
method assuming that the total source function is fully specified.

The method is essentially an application of the Galerkin method. An idea is to
divide a medium into a set of cells, and to represent the source function
within a cell by a simple polynomial, in this case by a linear segment.
The crucial point is that the segments are assumed to have
step discontinuities at grid points.  The specific intensity at grid point
$d$ is thus characterized by two values $I_d^+$ and $I_d^-$ appropriate
for cells $(\widetilde{\tau}_d, \widetilde{\tau}_{d+1})$ and 
$(\widetilde{\tau}_{d-1}, \widetilde{\tau}_d)$, 
respectively. Notice that we are dealing with an intensity in a given direction; 
the superscripts ``$+$'' and ``$-$'' thus
do not denote intensities in opposite directions as it is usually used in
the radiative transfer theory.  The actual value of the specific intensity
$I(\widetilde{\tau}_d)$ is given as an appropriate linear combination of $I_d^+$ and
$I_d^-$.
We skip all details here; suffice to say that after some algebra one
obtains simple recurrence relations for $I_d^+$ and $I_d^-$, 
for $d=1,\ldots,N\!D-1$,
\begin{equation}
\label{dfem}
a_d I_{d+1}^- = 2 I_{d}^- + \Delta\widetilde{\tau}_{d+1/2} S_{d} + b_d S_{d+1} ,
\end{equation}
\begin{equation}
\label{dfep}
a_d I_{d}^+ = 2(\Delta\widetilde{\tau}_{d+1/2} + 1)\, I_{d}^- + b_d S_{d} - 
\Delta\widetilde{\tau}_{d+1/2} S_{d+1},
\end{equation}
where
\begin{equation}
\label{dfea}
a_d=\Delta\widetilde{\tau}_{d+1/2}^2 + 2\Delta\widetilde{\tau}_{d+1/2} + 2,
\end{equation}
\begin{equation}
\label{dfeb}
b_d=\Delta\widetilde{\tau}_{d+1/2} ( \Delta\widetilde{\tau}_{d+1/2} +1),
\end{equation}
and
\begin{equation}
\Delta\widetilde{\tau}_{d+1/2} = \widetilde{\tau}_{d+1}-\widetilde{\tau}_d,
\end{equation}
The boundary condition is $I_1^-=I^{\rm ext}$, where $I^{\rm ext}$ is the specific
intensity of external irradiation (for inward-directed rays, $\mu<0$).

For outward-directed rays ($\mu>0$), one can either use the same expressions as above,
renumbering the depth points such as $N\!D \rightarrow 1, N\!D-1 \rightarrow 2,
\ldots,  1\rightarrow N\!D$; or use the same numbering of depth points while setting
the recursion, for $d=N\!D-1,\ldots,1$, as
\begin{equation}
a_d I_{d}^- = 2 I_{d+1}^- + \Delta\widetilde{\tau}_{d+1/2} S_{d+1} + b_d S_{d} ,
\end{equation}
\begin{equation}
\label{dfepm}
a_d I_{d+1}^+ = 2(\Delta\widetilde{\tau}_{d+1/2} + 1)\, I_{d+1}^- + b_d S_{d+1} - 
\Delta\widetilde{\tau}_{d+1/2} S_{d},
\end{equation}
with $I_{d}^-=B_d+\mu (B_d-B_{d-1})/\Delta\widetilde{\tau}_{d-1/2}$, 
i.e., assuming the diffusion approximation, for $d=N\!D$.

Finally, the resulting specific intensity at $\widetilde{\tau}_d$ is given by a linear combination
of the ``discontinuous" intensities $I_d^-$ and $I_d^+$ as
\begin{equation}
\label{dfei}
I_d = \frac{I_d^- \Delta\widetilde{\tau}_{d+1/2}+ I_d^+ \Delta\widetilde{\tau}_{d-1/2}}
{\Delta\widetilde{\tau}_{d+1/2}+  \Delta\widetilde{\tau}_{d-1/2}}.
\end{equation}
At the boundary points, $d=1$ and $d=N\!D$, we set $I_d = I_d^-$.
As was shown by Castor et al. (1992), it is exactly the linear combination of the
discontinuous intensities expressed in Eq. (\ref{dfei}) which makes the method 
second-order accurate. 
Since one does not need to evaluate any exponentials, the method is 
also very fast.  

We stress again that the above scheme applies for a solution of the transfer
equation along a single line of sight; that is, for a single angle of propagation. The source
function is assumed to be given. Therefore, when scattering is not negligible, one
has to iterate on the source function. This is done using an application of the
Accelerated Lambda Iteration (ALI) method.

Here is an algorithm to use the ALI scheme in this context, assuming a diagonal
$\Lambda^\ast$ operator. For more details refer to Hubeny \& Mihalas (2014, \S\,13.5):
\begin{itemize}
\item[(i)]  For a given $S^{\rm old}$ (with an initial estimate $S^{\rm old}=B$ or
some other suitable value), perform a formal solution of the transfer equation,
one frequency and direction (given $\mu$) at a time. This yields
a new value of the specific intensity $I_{\mu}$ and also the values of the
$\Lambda_{\mu}^\ast$, angle-dependent approximate operator -- 
see below.
\item[(ii)] By integrating over directions using 
\begin{equation}
\label{jfs}
J^{\rm FS} = \frac{1}{2} \int_{-1}^1 d\mu\, \Lambda_{\mu}
[S^{\rm old}] 
\end{equation}
obtain new values of
the formal-solution mean intensity $J^{\rm FS}$. Here, the action of the $\Lambda$
operator simply means obtaining the specific intensity by solving the transfer equation 
using the old source function. Analogously, compute the angle-integrated 
$\bar\Lambda^\ast$ as
\begin{equation}
\bar\Lambda^\ast = \frac{1}{2} \int_{-1}^1 d\mu\, \Lambda_{\mu}^\ast,
\end{equation}
\item[(iii)] Evaluate a new iterate of the mean intensity as $J^{\rm new}=J^{\rm old} + \delta J$,
where
\begin{equation}
\label{jfsa}
\delta J = \frac{J^{\rm FS} - J^{\rm old}}{1-(1-\epsilon)\bar\Lambda^\ast}.
\end{equation}
\item[(iv)] If the mean intensity found in step (iii) differs from that used in step (i),
update the source function from (\ref{feaasf}) using the newly found mean intensity and 
repeat steps (i) to (iii) to convergence.
\end{itemize}

%-------------------------------------------------------------------------------

\subsection*{A5. Construction of the approximate $\Lambda^\ast$ operator}
\addcontentsline{toc}{subsection}{A5. Construction of the approximate $\Lambda^\ast$ operator}

As mentioned above, one has to construct the approximate $\Lambda^\ast$ operator
in the formal solution of the transfer equation.  This operator is then held fixed during
the subsequent step of the linearization procedure.  

As explained in Hubeny \& Mihalas (2014; \S\,13.3), the matrix elements of the $\Lambda$
operator can be evaluated by setting the source function to be the unit pulse function,
$S(\tau_{d})=\delta(\tau-\tau_d)$, so that
\begin{equation}
\Lambda_{dd^\prime} = \Lambda_{\tau_d}[\delta(\tau_{d^\prime}-\tau)],
\end{equation}
In practice, one does not have to solve the full transfer equation, but only to collect
coefficients that stand at $S_d$ in the expressions to evaluate $I_d$. 
The actual evaluation depends on the type of the formal solver of the transfer
equation

\subsubsection*{Using the Feautrier scheme}

The procedure, following Rybicki \& Hummer (1991), is as follows. Let $\bf{T}$ be an
$N \times N$ tridiagonal matrix and let its inverse be
$\Lambda\equiv{\bf T}^{-1}$. The equation for the inverse can be written as
$\bf{T}\cdot\Lambda = 1$, or, in component form
\begin{equation}
\label{eq:13.3-14}
-A_i\lambda_{i-1,j}+B_i\lambda_{ij}-C_{i}\lambda_{i+1,j} =\delta_{ij}.
\end{equation}
For any fixed value of $j$ this equation can be solved by one of the forms of
Gaussian elimination. In the usual implementation  the elimination proceeds from
$i=1$ to $i=N$, followed by back--substitution from $i = N$ to $i=1$,
\begin{equation}
\label{eq:13.3-15}
D_i = (B_i-A_iD_{i-1})^{-1}C_i,
\end{equation}
\begin{equation}
\label{eq:13.3-16}
Z_{ij} = (B_i-A_i D_{i-1})^{-1}(\delta_{ij}+A_i Z_{i-1,j}),
\end{equation}
and
\begin{equation}
\label{eq:13.3-17}
\lambda_{ij} = D_i\lambda_{i+1,j}+Z_{ij}.
\end{equation}
It is also possible to implement the method in reverse order,
\begin{equation}
\label{eq:13.3-18}
E_i = (B_i-C_i E_{i+1})^{-1}A_i,
\end{equation}
\begin{equation}
\label{eq:13.3-19}
W_{ij} = (B_i-C_i E_{i+1})^{-1}(\delta_{ij}+C_i W_{i+1,j}),
\end{equation}
and
\begin{equation}
\label{eq:13.3-20}
\lambda_{ij} = E_i\lambda_{i-1,j} + W_{ij}.
\end{equation}
The crucial idea of the method is to use parts of \textit{both} of these
implementations to find the diagonal elements $\lambda_{ii}$.

Since $\delta_{ij}=0$ for $i\neq j$, it follows from (\ref{eq:13.3-16})
and (\ref{eq:13.3-19}) that $Z_{ij}= 0$ for $i<j$, and $W_{ij}= 0$ for $i > j$.
Thus, from (\ref{eq:13.3-16}) and (\ref{eq:13.3-17}) we obtain, for special 
choices of $i$ and $j$,
\begin{equation}
\label{eq:13.3-21}
Z_{ii} = (B_i-A_i D_{i-1})^{-1},
\end{equation}
\begin{equation}
\label{eq:13.3-22}
\lambda_{ii} = D_{i}\lambda_{i+1,i}+Z_{ii},
\end{equation}
\begin{equation}
\label{eq:13.3-23}
\lambda_{i-1,i} = D_{i-1}\lambda_{ii},
\end{equation}
and, from (\ref{eq:13.3-19}) and (\ref{eq:13.3-20})
\begin{equation}
\label{eq:13.3-24}
W_{ii} = (B_i-C_i E_{i+1})^{-1},
\end{equation}
\begin{equation}
\label{eq:13.3-25}
\lambda_{ii} = E_{i}\lambda_{i-1,i} + W_{ii},
\end{equation}
and
\begin{equation}
\label{eq:13.3-26}
\lambda_{i+1,i} = E_{i+1}\lambda_{ii}.
\end{equation}
Using (\ref{eq:13.3-21}), (\ref{eq:13.3-22}), and (\ref{eq:13.3-26}) we
eliminate $Z_{ii}$ and $\lambda_{i+1,i}$ to obtain
\begin{equation}
\label{eq:13.3-27}
\lambda_{ii} = (1-D_i E_{i+1})^{-1}(B_i-A_i D_{i-1})^{-1}.
\end{equation}
The right hand side now depends only on the single--indexed quantities $A_i$ and
$B_i$, which are given, and $D_i$ and $E_i$ which can be found by two passes
through the depth grid, using the recursion relations (\ref{eq:13.3-15}) and
(\ref{eq:13.3-18}). Thus $\lambda_{ii}$ can be found in order $N$ operations.

An evaluation of the approximate operator requires only little extra work in the
formal solution. The
quantities $A_i$, $B_i$, $C_i$, and $D_i$ are common to both problems, and one
needs only to include the recursion relation (\ref{eq:13.3-18}) as part of the
back--substitution to find the auxiliary quantities $E_i$.

%----------------------

\subsubsection*{Using the Discontinuous Finite element method}

When using the DFE scheme for the formal solution of the transfer equation, 
one proceeds along the recurrence relations (\ref{dfem}) and (\ref{dfep}) to compute
\begin{eqnarray}
\label{lpm}
L_{d+1}^- &=& b_d/a_d,\\
L_{d}^+ &=& [2(\Delta\widetilde{\tau}_{d+1/2} + 1)\, L_{d}^- + b_d]/a_d
\end{eqnarray}
where $a_d$ and $b_d$ are given by (\ref{dfea}) and (\ref{dfeb}). The complete
diagonal element of the (angle-dependent)  elementary operator is obtained, 
in parallel with Eq. (\ref{dfei}), as
\begin{equation}
\label{dfelam}
\Lambda^\ast_d(\mu,\phi) \equiv \Lambda_{dd} = 
\frac{L_d^- \Delta\widetilde{\tau}_{d+1/2}+ L_d^+ \Delta\widetilde{\tau}_{d-1/2}}
{\Delta\widetilde{\tau}_{d+1/2}+  \Delta\widetilde{\tau}_{d-1/2}}.
\end{equation}
The values at the boundaries are $\Lambda_{dd}=0$ for $d=1$, and $\Lambda_{dd}=L_d^-$
for $d=N\!D$. The evaluation of the diagonal elements for outward-directed rays is
analogous,
\begin{eqnarray}
\label{lpmm}
L_{d}^- &=& b_d/a_d,\\
	L_{d+1}^+ &=& [2(\Delta\widetilde{\tau}_{d+1/2} + 1)\, L_{d+1}^- + b_d]/a_d
\end{eqnarray}

As stressed above, the solution of the transfer equation using the DFE method
is performed for one direction at a time, so $L$ and $\Lambda$ in Eqs. (\ref{lpm}) -
(\ref{dfelam}) are also specified
for given $\mu$ and $\phi$. The total approximate operator needed to evaluate the
new iterate of the source function or the mean intensity is given by
\begin{equation}
\bar\Lambda^\ast_d = \frac{1}{4\pi}\int_0^{2\pi}\!\!d\phi  \int_{-1}^1\! d\mu\, 
\Lambda^\ast_d(\mu,\phi).
\end{equation}
%
%--------------------------------------------------------------------------------------

\subsection*{A6. Compton scattering}
\addcontentsline{toc}{subsection}{A6. Compton scattering}

The basic complication in the formal solution is that one can no longer
solve the transfer equation frequency by frequency independently, as it
is done in the standard treatment, but the coupling of frequencies
expressed by Eq. (\ref{compt}) has to be taken into account. Although
one can devise an ALI-based method that would be efficient and universal,
this is not yet done in {\sc tlusty}, where one resorts to the direct scheme.

Such a scheme was developed by Hubeny et al. (2001); here we point
out only the basic features.
The angle-averaged  Compton scattering 
source function is given by
\begin{equation}
\label{comptsf}
S_\nu^{\rm Compt} = (1-x) J_\nu +(x-3\Theta)J_\nu^\prime 
+ \Theta J_\nu^{\prime\prime} + \frac{c^2}{2h\nu^3}J_\nu 2x 
(J_\nu^\prime - J_\nu),
\end{equation}
where 
\begin{equation}
x=\frac{h\nu}{m_{\rm e}c^2}, \quad \Theta=\frac{kT}{m_{\rm e}c^2}, 
\end{equation}
The second-order form of the radiative transfer equation, discretized in 
frequency, is written for the $i$-th frequency point as [see Hubeny 
et al. (2001), eq. (A51)],
\begin{equation}
\label{rtecom}
\frac{\partial^2(f_i J_i)}{\partial\tau_i^2} = J_i - \epsilon_i S_i^{\rm th}
- (\mathcal{A}_i J_{i-1} + \mathcal{B}_i J_i + \mathcal{C}_i J_{i+1})
+ J_i (\mathcal{U}_i J_{i-1} + \mathcal{E}_i J_i + \mathcal{V}_i J_{i+1}),
\end{equation}
where $\epsilon_i = \kappa_i^{\rm th}/(\kappa_i^{\rm th} + \sigma_i)$,
$\lambda_i=1-\epsilon_i$. 
Here, and in the following, we skip the depth index $d$.

Since Eq. (\ref{rtecom}) represents a discretized version of a partial
differential equation in depth and frequency, one need to invoke
specific initial conditions of the lowest and highest frequency.

We will first consider the internal frequency points, $\nu_i, 2 \leq i \leq N\!F-1$.
The coefficients  $\mathcal{A,B,C,E,U,V}$
are expressed by two possible ways. The original approach is that
developed in Hubeny et al. (2001), viz.
\begin{eqnarray}
\mathcal{A}_i &=& \lambda_i[(x_i -3\Theta)c_i^- + \Theta d_i^-],\\
\mathcal{B}_i &=& \lambda_i[(1-x_i) + (x_i-3\Theta)c_i^0 + \Theta d_i^0],  \\
\mathcal{C}_i &=& \lambda_i[(x_i -3\Theta)c_i^+ + \Theta d_i^+],\\
\mathcal{E}_i &=& \lambda_i[(2h\nu_i^3/c^2) 2x_i (c_i^0 - 1)],\\
\mathcal{U}_i &=& \lambda_i(2h\nu_i^3/c^2) 2x_i c_i^-,\\
\mathcal{V}_i &=& \lambda_i(2h\nu_i^3/c^2) 2x_i c_i^+.
\end{eqnarray}
There is another, preferable approach, based on the formalism of
Chang \& Cooper (1970), where,
\begin{eqnarray}
\mathcal{A}_i &=& \lambda_i[-\delta_{i-1} y_{i-1}+ \Theta d_i^-],\\
\mathcal{B}_i &=& \lambda_i[\delta_i y_{i+1} + (1-\delta_{i-1})y_{i-1}
                             + \Theta d_i^0 -\epsilon_i +1, \\
\mathcal{C}_i &=& \lambda_i[(1-\delta_{i}) y_{i+1}+ \Theta d_i^+],\\
\mathcal{E}_i &=& \mathcal{U}_i = \mathcal{V}_i  =0,
\end{eqnarray}
where 
\begin{eqnarray}
y_i &=& [(1-\delta_i) z_{i+1} + \delta_i z_i] \Delta_0, \\
\label{chcstim}
z_i &=& x_i[1+J_i c^2/(2h\nu_i^3)] - 3\Theta.
\end{eqnarray}
In Eq. (\ref{chcstim}) , one takes for $J_i$ the current, and thus known, 
specific intensity. This avoids a non-linearity of the transfer equation that would 
arise due to stimulated emission.

In both cases, the coefficients $c_i^0$, $c_i^-$, and $c_i^+$ come from discretizing
the first derivative terms in Eq. (\ref{comptsf}), 
\begin{eqnarray}
c_i^+ &=& (1-\delta_i)/\Delta_i, \quad c_i^- = -\delta_{i-1}/\Delta_i,\\
c_i^0 &=& -c_i^+ - c_i^-,
\end{eqnarray}
where 
\begin{eqnarray}
\Delta_{i-1/2} &=& \ln(\nu_i/\nu_{i-1}), \quad \Delta_{i+1/2} = \ln(\nu_{i+1}/\nu_{i}), \\
\Delta_i &=& \Delta_{i-1/2} + \Delta_{i+1/2}.
\end{eqnarray}
The coefficients $\delta_i$ are determined by solving a quadratic equation
\begin{equation}
\label{cqua}
a\,\delta_i^2 + b\,\delta_i +c =0,
\end{equation}
where the coefficients $a,b,c$ are given by, using the original approach
(Hubeny et al. 2011) as
\begin{eqnarray}
a &=& \Delta B_i\, \delta\beta_i, \\
b &=& \Delta B_i\,  (1+\beta_{i+1}-3/\bar\xi) + B_{i+1}\, \delta\beta_i, \\
c &=& B_{i+1}  (1+\beta_{i+1}-3/\bar\xi)- \Delta B_i/[\bar\xi \ln(\nu_{i+1}/\nu_i)],
\end{eqnarray}
where $\bar\xi = (h/kT) (\nu_i\nu_{i+1})^{1/2}$; 
or, using the Chang \& Cooper (1970) approach, as
\begin{eqnarray}
a &=& \Delta B_i\, \delta z_i, \\
b &=& \Delta B_i\,  z_{i+1} + B_{i+1}\, \delta z_i, \\
c &=& B_{i+1} z_{i+1} - \Theta\, \Delta B_i /\Delta_i
\end{eqnarray}
where, in both cases,
\begin{eqnarray}
z_i &=&x_i(1+\beta_i)-3\Theta, \quad \Delta z_i = z_i - z_{i+1},\\
\beta_i &=& [\exp(h\nu_i/kT) -1]^{-1},\\
B_i  &=& (2h\nu_i^3/c^2) \beta_i, \quad \Delta B_i = B_i - B_{i+1}.
\end{eqnarray}
In both cases, one picks the solution of (\ref{cqua}) which satisfies $0\leq\delta_i\leq 1/2$.

The analogous coefficients
$d_i$ come from the second derivative terms. 
\begin{eqnarray}
d_i^-&=& 2 (\Delta_{i-1/2} \Delta_i)^{-1}, \quad 
d_i^+= 2 (\Delta_{i+1/2} \Delta_i)^{-1}, \\
d_i^0 &=& -d_i^+ - d_i^- .
\end{eqnarray}

The non-linear terms in Eq.
(\ref{comptsf}), corresponding to the stimulated emission, are linearized by
using the ``old'' intensities, replacing $B_i$ by $B_i^\prime$, where
\begin{equation}
\mathcal{B}_i^\prime = \mathcal{B}_i + \mathcal{U}_i J_{i-1}^{\rm old} + \mathcal{E}_i J_i^{\rm old} + \mathcal{V}_i J_{i+1}^{\rm old}.
\end{equation}
In the case of the Chang \& Cooper approach, the stimulated emission terms were
already included, as is seen in Eq. (\ref{chcstim}),
so $\mathcal{B}_i^\prime = \mathcal{B}_i$.

The initial condition for the lowest frequency, $i=1$, which is assumed to be very low, is
simply
\begin{equation}
\mathcal{B}_i = 1-2x_i,
\end{equation}
and all the other coefficients $\mathcal{A}_i= \mathcal{C}_i=\mathcal{E}_i= \mathcal{U}_i= 
\mathcal{V}_i= 0$.

The initial condition for the highest frequency, $i=N\!F$, is more complicated. In the original
approach after Hubeny et al. (2001) one has
\begin{eqnarray}
\mathcal{A}_i &=& -\xi_0/\ln(\nu_i/\nu_{i-1}) + (1-\delta_{i-1}) \xi_1, \\
\mathcal{B}_i &=& \xi_0/\ln(\nu_i/\nu_{i-1}) + \delta_{i-1} \xi_1,
\end{eqnarray}
where
\begin{eqnarray}
\xi_0 &=& kT/[h(\nu_i\nu_{i-1})^{1/2}], \\
\xi_1 &=& 1-3\xi_0 + (1-\delta_{i-1}) e^{-h\nu_i/kT} + \delta_{i-1} e^{-h\nu_{i-1}/kT},
\end{eqnarray}
In the case of Chang \& Cooper approach, one has
\begin{eqnarray}
\mathcal{A}_i &=& - \Theta/\delta_{i-1}+ (1-\delta_{i-1}) \zeta_0, \\
\mathcal{B}_i &=& \Theta/\delta_{i-1}+ \delta_{i-1} \zeta_0,
\end{eqnarray}
where
\begin{eqnarray}
\zeta_0 &=& (1-\delta_{i-1}) \zeta_i + \delta_{i-1} \zeta_{i-1}, \\
\zeta_i &=& x_i \left(1+ e^{-h\nu_i/kT}\right)- 3\Theta.
\end{eqnarray}

The left-hand side of Eq. (\ref{rtecom}) is discretized as in the ordinary transfer
problem, so that the final algebraic equation for the mean intensities reads
\begin{equation}
\label{rtecom2}
-\alpha_{id}J_{i,d-1}-\gamma_{id}J_{i,d+1} + (\beta_{id}+1-\mathcal{B}_{id}^\prime)J_{id}
- \mathcal{A}_{id}J_{i-1,d} - \mathcal{C}_{id}J_{i+1,d} = \epsilon_{id} S_{id}^{\rm th}.
\end{equation}
One can use equation (\ref{rtecom2}) in two different ways, called in vague 
analogy with the hydrodynamical terminology as {\em explicit} or {\em implicit}.

\smallskip
(i) The {\bf implicit} way consists of solving Eq. (\ref{rtecom2}) with the full
coupling taken into account; that is, all the mean intensities, including
those in frequency points $i-1$ and $i+1$, are solved for. This approach is more
stable, but it involves a more time-consuming solution of the coupled problem.
It is solved by a standard scheme.
One introduces a column vector ${\bf J}_i \equiv (J_{i1}, J_{i2},\ldots,J_{iD})$,
where $D\equiv N\!D$ being the number of depth points,
and write Eq. (\ref{rtecom2})  as a linear matrix equation
\begin{equation}
\label{com_glob} 
-{\bf A}_i {\bf J}_{i-1} +{\bf B}_i {\bf J}_{i} -{\bf C}_i {\bf J}_{i+1} = {\bf L}_i,
\end{equation}
where the elements of the matrices are given by the corresponding coefficients
$\mathcal{A}_{id}$, $\mathcal{B}_{id}^\prime$, $\mathcal{C}_{id}$, $\alpha_{id}$, $\beta_{id}$,
and $\gamma_{id}$, and vector $({\bf L}_i)=\epsilon_{id}S_{id}^{\rm th}$. 
Matrices ${\bf B}$ are tridiagonal (because of a difference representation of
the second derivative with respect to depth), and matrices ${\bf A}$ and
${\bf C}$ are diagonal  (because the terms containing the frequency derivatives
are local).
Equation (\ref{com_glob}) is solved by a standard Gauss-Jordan elimination. 

(ii) The {\bf explicit} way consists of avoiding the frequency coupling 
by considering
the radiation intensity in the last three terms on the l.h.s. of Eq. (\ref{rtecom2}),
corresponding to frequency derivatives, being given by the ``old'' values of the
specific intensity. In practice, Eq. (\ref{rtecom2}) can be simplified to the
ordinary form,
\begin{equation}
\label{rtecom_ex}
-\alpha_{id}J_{i,d-1}-\gamma_{id}J_{i,d+1} + (\beta_{id}+1-\widetilde B_{id})J_{id}
= \epsilon_{id} S_{id}^{\rm th}.
\end{equation}
where
\begin{equation}
\widetilde B_{id} = \mathcal{B}_{id}^\prime + \mathcal{A}_{id}J_{i-1,d}^{\rm old} + 
\mathcal{C}_{id}J_{i+1,d}^{\rm old}.
\end{equation}
The numerical solution of this problem is analogous to the solution of the
standard problem without Compton scattering; the only difference being that
different coefficients enter in the linear equation. Obviously, the
explicit solution can be performed frequency by frequency.

It should be kept in mind that regardless whether the implicit or
explicit way is used, one still has to iterate to update (i) coefficient
$\mathcal{B}_{id}^\prime$ that describes the stimulated emission and which depends
on the radiation intensity, and (ii) the Eddington factor. If one uses an
explicit method, one has to introduce another nested iteration loop to
update coefficient $\widetilde B_{id}$ that accounts for the frequency derivatives,
and also depends on radiation intensities. 

To update the Eddington factor, one has to solve an angle-dependent transfer
equation for the specific intensity. To this end, we use the explicit form of the 
scattering term. As in the standard case, one takes the source function as 
angle-independent (that is, angle averaged), but unlike the standard case
it is taken as known, without an explicit integration of the specific intensity
over angle. Therefore, one  solves the transfer equation for one
frequency-angle point at a time, viz.
\begin{equation}
\label{rteang1}
\mu_j \frac{dI_{ij}}{d\tau_i} = I_{ij} - \epsilon_i S_I^{\rm th} - 
\widetilde B_i J_i^{\rm old},
\end{equation}
where $I_{ij}$ is the specific intensity at frequency point $i$ and
discretized angle point $j$; $\mu_j$ is the cosine of the polar angle.
One can employ equation (\ref{rteang1}) as is, when using the DFE method
for the formal solver,
or use the second-order (Feautrier) form,
\begin{equation}
\label{rteang2}
\mu_j^2\,\frac{d^2 j_{ij}}{d\tau_i^2} = j_{ij} - \epsilon_i S_I^{\rm th} - 
\widetilde B_i J_i^{\rm old},
\end{equation}
where $j_{ij} \equiv [I_i(\mu_j) + I_i(-\mu_j)]/2$ is the symmetrized (Feautrier)
intensity.

%--------------------------------------------------------------------------------------
\section*{Appendix B. Details of the global formal solution}
\addcontentsline{toc}{section}{Appendix B. Details of the global formal solution}

\subsection*{B1. Preconditioned kinetic equilibrium equations}
\addcontentsline{toc}{subsection}{B1. Preconditioned kinetic equilibrium equations}

As mentioned in \S\,\ref{formal_glob}, the crucial part of the global formal
solution, i.e., the bulk of calculations performed between the two successive
iterations of the linearization scheme, is the simultaneous solution of the
radiative transfer and kinetic equilibrium equations, keeping the current
atmospheric structure (temperature, density) fixed---the  so-called ``restricted 
NLTE problem". 

As mentioned above, one can use a simple Lambda iteration scheme, 
that is to perform an iterative solution that alternates between solving the
transfer equation for the current values of level populations, and solving
the kinetic equilibrium equations with the current values of radiation intensities.
However, there is a much better scheme that is based on the ALI scheme with preconditioning 
of the kinetic equilibrium, first suggested by Rybicki \& Hummer (1991, 1992).
We follow here a slightly modified formalism, presented in Hubeny \& Mihalas
(2014, \S 14.5). 

We use a variant of the scheme, called ``preconditioning within the same 
transition only'', in which the discretized kinetic equilibrium equations, 
with corresponding quadrature weights, denoted in the section 
as $\overline w_i$ to avoid confusion with occupation probabilities,  are written as
\begin{eqnarray}
\label{precond}
\nonumber
\hspace*{-3em}\sum_{l}n_l C_{lu}\!\!&
+ &\!\!\sum_{l}\sum_i  \overline w_i \frac{4\pi}{h\nu_i}\Big(n_l U_{lu}
+ n_l V_{lu}\Lambda_i[\eta_i^\mathrm{old}/\chi_i^\mathrm{old}]\\
\nonumber
\!\!&-&\!\!n_l V_{lu}\Lambda_i^\ast[n_u^\mathrm{old}U_{ul}/\chi_i^\mathrm{old}]
-n_l V_{lu}\Lambda_i^\ast[n_l^\mathrm{old} U_{lu}/\chi_i^\mathrm{old}]\\
\nonumber
\!\!&+&\!\!n_l^\mathrm{old}V_{lu}\Lambda_i^\ast[n_u U_{ul}/\chi_i^\mathrm{old}]
+n_l^\mathrm{old}V_{lu}\Lambda_i^\ast[n_l U_{lu}/\chi_i^\mathrm{old}]\Big) \\
\nonumber
\hspace*{-1em}=\sum_{l}n_u C_{ul}\!\!
&+&\!\!\sum_{l}\sum_i \overline w_i \frac{4\pi}{h\nu_i}\Big(n_u U_{ul}
+n_u V_{ul}\Lambda_i[\eta_i^\mathrm{old}/\chi_i^\mathrm{old}]\\
\nonumber
\!\!&-&\!\!n_u V_{ul}\Lambda_i^\ast[n_l^\mathrm{old}U_{lu}/\chi_i^\mathrm{old}]
-n_u V_{ul}\Lambda_i^\ast[n_u^\mathrm{old} U_{ul}/\chi_i^\mathrm{old}]\\
\!\!&+&\!\!n_u^\mathrm{old}V_{ul}\Lambda_\nu^\ast[n_l U_{lu}/\chi_i^\mathrm{old}]
+n_u^\mathrm{old}V_{ul}\Lambda_i^\ast[n_u U_{ul}/\chi_i^\mathrm{old}]\Big),
\end{eqnarray}
where for bound-bound transitions (generalizing the Rybicki \& Hummer
expressions to include the occupation probabilities $w_l$ and $w_u$),
\begin{eqnarray}
U_{ul}(\nu) &\equiv& (h\nu/4\pi) A_{ul} \phi_{lu}(\nu) w_l, \quad  u>l,\\
U_{lu}(\nu) &\equiv& 0, \quad u>l, \\
V_{lu}(\nu) &\equiv&  (h\nu/4\pi) B_{lu} \phi_{lu}(\nu) w_u,
\end{eqnarray}
and for bound-free transitions, and for $u>l$,
\begin{eqnarray}
U_{ul}(\nu) &\equiv& n_\mathrm{e} \Phi_{ul}(T) (2h\nu^3/c^2)\exp(-h\nu/kT)
       \sigma_{lu}(\nu) w_l,  \ \ \  \\
U_{lu}(\nu) &\equiv& 0,   \\
V_{ul}(\nu) &\equiv& n_\mathrm{e} \Phi_{ul}(T) \exp(-h\nu/kT)  
       \sigma_{lu}(\nu) w_l,   \\
V_{lu}(\nu) &\equiv& \sigma_{lu}(\nu) w_u,
\end{eqnarray}
Equations (\ref{precond}) form a linear set for the level populations
$n_u, u=2,\ldots,N\!L$.
The linearity is achieved by considering some level populations in
the bi-linear products as the ``old" populations. Similarly, the total opacity
$\chi_i$ that enters in the action of the $\Lambda^\ast$ operator, is evaluated
using the old populations. 

Recognizing that $\Lambda_i^\ast[\eta_i^{\rm old}/\chi_i^{\rm old}] = J_i^{\rm old}$,
and noting that for a diagonal $\Lambda^\ast$ operator the terms
$-n_l V_{lu}\Lambda_i^\ast[n_l^\mathrm{old} U_{lu}/\chi_i^\mathrm{old}]$ and
$n_l^\mathrm{old}V_{lu}\Lambda_i^\ast[n_l U_{lu}/\chi_i^\mathrm{old}]$
exactly cancel out (and analogously for the term with $n_u n_u^{\rm old}$),
we can rewrite Eq. (\ref{precond}) into a more traditional form,
\begin{equation}
\label{precond1}
\sum_{l\not= u}n_l (C_{lu} + R_{lu}^\prime) =
n_u \sum_{l\not= u}(C_{ul} + R_{ul}^\prime),
\end{equation}
where, for $l<u$
\begin{eqnarray}
R_{lu}^\prime &=& \sum_i \overline w_i \sigma_{lu}(\nu_i) J_i ^\prime w_u, \\
R_{ul}^\prime &=& \sum_i \overline w_i \sigma_{lu}(\nu_i) \left(J_i + \beta_i^\prime\right)\, G_{lu},
\end{eqnarray}
which are quite parallel to the original equations (\ref{rij}) and (\ref{rji}), replacing
$J$ by $J^\prime$ and $\beta=(2h\nu^3/c^2)$ by $\beta^\prime$, 
where
\begin{eqnarray}
J_i^\prime &=& J_i - \Lambda_i^\ast n_u^{\rm old} U_{ul}/\chi_i^{\rm old}, \\
\beta_i^\prime &=& (2h\nu_i^3/c^2) 
[1 - \Lambda_i^\ast n_l^{\rm old} \sigma_{lu}(\nu_i) w_u/\chi_i^{\rm old}].
\end{eqnarray}
One proceeds in exactly the same way as in the ordinary Lambda iteration.
The iteration loop consists in 
solving the transfer equation with the current populations to obtain a better
estimate of the mean intensities, and then solving 
the system (\ref{precond}) or (\ref{precond1}) with these mean intensities
to obtain the ``new'' populations $n_i$. Unlike the ordinary Lambda iteration,
the modified radiative rates contain not only the current mean intensities of
radiation, but also the approximate operator $\Lambda^\ast$ and
 the ``old'' populations, known from the previous iteration. The iteration process
 may be augmented by an application of the Ng acceleration.
 
 Usually, one does not perform many iterations because the aim of the procedure
 is only to provide an improved and more consistent values of mean intensities and level
 populations before entering the next step of the global iteration process. But if
 the procedure is used, for instance, to obtain an exact solution for the radiation
 field and level populations for a fixed atmospheric structure, then one should
 perform more iterations of the preconditioning scheme together with Ng acceleration.

The iteration scheme to obtain new populations and mean intensities of radiation,
being a classical Lambda iterations or a preconditioning scheme with ALI, is usually
supplemented by a simultaneous solution of the charge conservation equation. This
is done iteratively; one simply alternates between the solution of the (preconditioned)
kinetic equilibrium equations, e.g., Eq. (\ref{precond1}), for new populations
with a given electron density, and the solution of the charge conservation equation
(\ref{chce}) for a new electron density with given populations.

% -----------------------------------------

\subsection*{B2. Temperature correction in the convection zone}
\addcontentsline{toc}{subsection}{B2. Temperature correction in the convection zone}

If the convection is taken into account, 
the logarithmic gradient $\nabla$, in a discretized form,  is given
either by
\begin{equation}
\label{nabla1}
\nabla_d \equiv \nabla_{d-1/2} =
\frac{T_d - T_{d-1}}{P_d-P_{d-1}} \frac{P_d+P_{d-1}}{T_d + T_{d-1}}.
\end{equation}
or by
\begin{equation}
\label{nabla2}
\nabla_d  = \ln (T_d /T_{d-1}) / \ln (P_d /P_{d-1}).
\end{equation}
The energy balance equation is linearized as described in detail in
Appendix D1; see also
Hubeny \& Mihalas (2014, \S\, 18.2). Although the linearization
scheme may in principle converge without additional correction procedures,
in practice it is a very rare situation. The essential point is that a
linearization iteration may yield the actual values of temperature and other
state parameters such that, for instance, the actual logarithmic gradient in a
former convection zone may spuriously decrease below the adiabatic gradient
at certain depth points, so that these points would be declared as
convectively stable and the radiative flux would be demanded to be equal 
to the total flux, which would lead to a serious
destabilization of the overall scheme, likely ending in a fatal divergence.

It is therefore almost always necessary to perform certain correction
procedures to assure that the convection zone is not disturbed by
non-convective regions, and that the temperature and other state
parameters are smooth enough functions of depth before one enters the next
iteration of the overall linearization scheme.
{\sc tlusty} offers several such schemes:\\ [4pt]
\noindent $\bullet\ $ {\sf Definition of the convection zone}.

\noindent
After a completed iteration, the code examines the depth point in which
the actual gradient surpasses the adiabatic one. If such a point is solitary,
or if it occurs at much smaller column densities than the upper edge of
the previous convection zone, the point is declared as convectively stable,
and the traditional radiative equilibrium equation is solved for it in the
next iteration step. 
On the other hand, if there is/are depth points in which $\nabla < \nabla_{\rm ad}$
(so that they are seemingly convectively stable),
surrounded on both sides by points that are convectively unstable
$\nabla \geq \nabla_{\rm ad}$, these points are declared as convectively
unstable, and are considered to be part of the convection zone. Within such 
a newly defined convection zone, one or both of the following correction procedures
are performed:\\ [4pt]
\noindent $\bullet\ $ {\sf Standard correction procedure}.

\noindent
The idea of the correction is as follows. In view of Eq. (\ref{conv}), the
convective flux is given by
\begin{equation}
\label{conref1}
F_{\rm conv} = F_0 (\nabla-\nabla_{\rm el})^{3/2},
\end{equation}
where 
\begin{equation}
\label{conref2}
F_0 = (gQH_P/32)^{1/2}(\rho c_P T)(\ell/H_P)^2,
\end{equation}
After a completed iteration of the global linearization scheme, one takes
the current values of the state parameters and the radiation flux, and computes,
in the convection zone, the new convective flux corresponding 
to this radiation flux so that the total flux is perfectly conserved,
\begin{equation}
\label{conref3}
F_{\rm conv}^\ast =  F_{\rm tot}- F_{\rm rad},
\end{equation}
where $F_{\rm tot} = \sigma T_{\rm eff}^4$.
If $F_{\rm rad}$ is spuriously larger than $F_{\rm tot}$, then $F_{\rm rad}$
is set to $0.999 F_{\rm tot}$.
The new difference of temperature gradients 
corresponding to this convective flux is then
\begin{equation}
\label{conref4}
\nabla-\nabla_{\rm el} = (F_{\rm conv}^\ast/F_0)^{2/3},
\end{equation}
which is related to $\nabla-\nabla_{\rm ad}$ through
\begin{equation}
\label{conref5}
\nabla-\nabla_{\rm ad} = (\nabla-\nabla_{\rm el})+B \sqrt{\nabla-\nabla_{\rm el}}.
\end{equation}
where $B$ is given by Eq. (\ref{convb}). Both $B$ and $\nabla_{\rm ad}$
are computed through the current values of the state parameters. Equation
(\ref{conref5}) thus yields the new gradient $\nabla$ and, keeping pressure
fixed, the new temperature. With the new temperature, one recalculates the
thermodynamic variables, and iterates the process defined by
Eqs. (\ref{conref2}) - (\ref{conref5}) to convergence.

In solving Eq. (\ref{conref5}), one proceeds from the top of the convection zone
to the bottom, because the gradient $\nabla$ is given by Eq.
(\ref{nabla1}) or (\ref{nabla2}),
so in order to  evaluate $T_d$
one needs to know $T_{d-1}$ in the previous depth point.

This procedure works well if the convective flux is dominant, because it
should be kept in mind that the radiation flux is also imperfect. Therefore,
the temperature is corrected only at depths where the convective flux
is larger than some limiting value, $F_{\rm conv} > \gamma F_{\rm tot}$,
where $\gamma$ is by default taken as $\gamma=0.7$, or is set by 
input data -- see Paper~III, \S\,\refnonstconv.\\ [4pt]
\noindent $\bullet\ $ {\sf Refined correction procedure}

\noindent
The above procedure is improved by recognizing that the coefficient
$B$ is an explicit function of temperature, so $B$ can be expressed
as $B\equiv\beta T^3$. More importantly, the radiation flux is not kept
fixed, but is written as
\begin{equation}
\label{conref6}
F_{\rm rad} \equiv \alpha T^4 \nabla,
\end{equation}
so that instead of keeping $F_{\rm rad}$ fixed, one first computes
$\alpha$ from (\ref{conref6}) for the current values of $T$ and $\nabla$,
and  rewrites the combined equations (\ref{conref3}) --(\ref{conref5}) as a
non-linear equation for the temperature,
\begin{equation}
\label{conref7}
\nabla(T) = \nabla_{\rm ad} +
\left(\frac{F_{\rm tot} - \alpha T^4 \nabla(T)}{F_0}\right)^{2/3}
+ \beta T^3 \left(\frac{F_{\rm tot} - \alpha T^4 \nabla(T)}{F_0}\right)^{1/3},
\end{equation}
which is solved by the Newton-Raphson method, again going from the
\index{Newton-Raphson method}
top of the convection zone to the bottom. This procedure, if chosen
to be performed, is done after first completing the standard procedure described
above, and only after a certain number of global linearization iterations
(driven by input data -- see Paper~III, \S\,\refnonstconv\  and \S\,\refnsttwoconv).

Other, more sophisticated refinement procedures are in principle possible, 
but they were not yet implemented in {\sc tlusty}.

% -----------------------------------------

\subsection*{B3. Evaluation of the thermodynamic quantities}
\addcontentsline{toc}{subsection}{B3. Evaluation of the thermodynamic quantities}

The internal energy per unit volume is given by
\begin{equation}
E= \frac{3}{2}NkT + 3P_{\rm rad} + \sum_I N_I \left[ E_I 
+ \left(\frac{d\ln U_I}{d\ln T}\right) kT \right],
\end{equation}
where $N_I$, $U_I$, and $E_I$ are the total population, partition function,
and the ground state energy (measured from the ground state of the neutral
atom) of an ion $I$, respectively. 
The latter is thus given by the sum of ionization energies of all the lower ions.

Here, and in the following expressions, it is assumed that the 
relation between the radiation energy and pressure is given by
the equilibrium relation, $E_{\rm rad} = 3 P_{\rm rad}$, and 
$P_{\rm rad} = (a_R/3) T^4$, where $a_R$ is the radiation constant.
This approximation is made only for the purposes of describing convection.
The rationale for this approach is that in the situations when convection
is important (typically for cool models), the radiation pressure is usually
a small part of the total pressure.

The adiabatic gradient is given by
\begin{equation}
\nabla_{\rm ad} = \left(\frac{\partial\ln T}{\partial\ln P}\right)_{\!\!S} =
- \frac{P}{\rho c_P T} \left(\frac{\partial\ln \rho}{\partial\ln T}\right)_{\!\!P},
\end{equation}
where the specific heat is
\begin{eqnarray}
c_P &=& \left(\frac{\partial E}{\partial T}\right)_{\!P} - \frac{P}{\rho^2}
\left(\frac{\partial \rho}{\partial T}\right)_{\!P} \nonumber \\
&=&\left(\frac{\partial E}{\partial T}\right)_{\!P_{\rm gas}} -
\left(\frac{\partial E}{\partial P_{\rm gas}}\right)_{\!T}
\left(\frac{\partial P}{\partial T}\right)_{\!P_{\rm gas}} \nonumber \\
&-&
\frac{P}{\rho^2} \left[\left(\frac{\partial \rho}{\partial T}\right)_{\!P_{\rm gas}} -
\left(\frac{\partial \rho}{\partial P_{\rm gas}}\right)_{\!T} 
\left(\frac{\partial P}{\partial T}\right)_{\!P_{\rm gas}} \right],
\end{eqnarray}
and
\begin{equation}
\left(\frac{\partial\ln \rho}{\partial\ln T}\right)_{\!\!P} =\frac{T}{\rho}
\left[\left(\frac{\partial \rho}{\partial T}\right)_{\!P_{\rm gas}} -
\left(\frac{\partial \rho}{\partial P_{\rm gas}}\right)_{\!T} 
\left(\frac{\partial P}{\partial T}\right)_{\!P_{\rm gas}} \right].
\end{equation}
In view of the above approximations,
$(\partial P/\partial T)_{P_{\rm gas}} = 1$, and so
all the quantities are expressed in terms of four thermodynamic
derivatives, $(\partial E/\partial T)_{P_{\rm gas}}$,
$(\partial E/ \partial P_{\rm gas})_T$,
$(\partial\rho/\partial T)_{P_{\rm gas}}$,
$(\partial\rho/ \partial P_{\rm gas})_T$.
These derivatives are calculated as differences, e.g., 
$(\partial E/\partial T)_{P_{\rm gas}} = [E(T\!+\!\Delta T, P_{\rm gas})
- E(T, P_{\rm gas})]/\Delta T$, where $\Delta T = 0.001 T$.

The thermodynamic quantities may also be formulated through the entropy;
which is set by the keyword IFENTR. In this case
\begin{equation}
\nabla_{\rm ad} = - \left(\frac{\partial S}{\partial T}\right)_{\!\!P_{\rm gas}}\Bigg/
\left(\frac{\partial S}{\partial P_{\rm gas}}\right)_{\!\!T} \frac{P_{\rm gas}}{T},
\end{equation}
and
\begin{equation}
c_P = -\frac{P}{\rho T}\left(\frac{\partial\ln \rho}{\partial\ln T}\right)_{\!\!P}\bigg/
\nabla_{\rm ad}
\end{equation}
%

% -----------------------------------------

\subsection*{B4. Recalculation of vertical distance and density for
accretion disks}
\addcontentsline{toc}{subsection}{B4. Recalculation of vertical distance  
and density for accretion disks}

For computing the vertical structure of accretion disks, in which case the gravity 
acceleration depends on the vertical
distance from the central plane, it is important to recalculate the vertical distance
and density to be as consistent with the rest of the structural parameters as
possible.
{\sc tlusty} uses one of the two following procedures:

\noindent $\bullet\ $ {\sf Original procedure}.

\noindent
In order to update $z$, one can use
the $z$-$m$ relation (\ref{dm}),
written in a discretized form
\begin{equation}
\label{zdet}
z_d = z_{d+1} + (m_{d+1} - m_d)(1/\rho_d + 1/\rho_{d+1})/2, \quad z_{N\!D}=0.
\end{equation}
However, using this equation is not convenient in the case where $z$ determined
by the current linearization step exhibits an oscillatory or other unphysical behavior.
The reason is that Eq. (\ref{zdet}) was used in the linearization, so that any such behavior 
of $z$ was likely shared by a similar behavior of $\rho$, and thus
solving Eq. (\ref{zdet}) again does not necessarily help.

A better possibility is to employ a discretized hydrostatic equilibrium equation,
used as an equation for $z$, namely
\begin{equation}
Q (z_d+z_{d+1})/2 = (P_{d+1}-P_d)/(m_{d+1}-m_d),
\end{equation}
or 
\begin{equation}
\label{zhe1}
z_d = \frac{2}{Q} \frac{P_{d+1}-P_d}{m_{d+1}-m_d} - z_{d+1}; \quad z_{N\!D}=0,
\end{equation}
which follows from  equation (\ref{heqz}).
This equation can also be discretized as
\begin{equation}
\label{zhe2}
z_d = \frac{1}{2Q}\left(\frac{P_{d+1}-P_d}{m_{d+1}-m_d} + 
\frac{P_d - P_{d-1}}{m_d-m_{d-1}} \right).
\end{equation}
Here, $P=P^{\rm rad} + P^{\rm gas}$ is the total pressure; we do not consider 
an ill-defined turbulent pressure.

An iterative scheme to update the vertical distance and the density proceeds 
as follows:\\ [2pt]
(i) Compute the gradient of the radiation pressure as
\begin{equation}
\frac{P^{\rm rad}_d - P^{\rm rad}_{d-1}}{m_d-m_{d-1}} = \frac{4\pi}{c} \frac{1}{m_d-m_{d-1}} 
\sum_{i=1}^{N\!F}w_i (f_{di} J_{di} - f_{d-1,i} J_{d-1,i}),
\end{equation}
which is then being held fixed in the subsequent iteration scheme. \\ [2pt]
(ii) Compute the gas pressure for the current density and electron density as
\begin{equation}
P_d^{\rm gas} = kT_d N_d = kT_d [\rho_d/(\mu m_H) + n_{{\rm e},d}],
\end{equation}
(iii) Compute a new vertical distance $z$ using Eq. (\ref{zhe2}). If this procedure yields
an unphysical values of $z$ such as $z_d < z_{d+1}$ at certain depth $d$, one switches
to an evaluation of the new $x=z$ using Eq. (\ref{zhe1}). \\ [2pt]
(iv) Having determined a new $z$-scale, recalculate the mass density 
for $d < N\!D$ by using Eq. (\ref{zdet}), namely
\begin{equation}
\label{rhodet}
\rho_d = \frac{\Delta m_{d+1/2} \rho_{d+1}}{(z_d-z_{d+1}) \rho_{d+1} - \Delta m_{d+1/2} },
\end{equation}
where $\Delta m_{d+1/2} = (m_{d+1}-m_d)/2$. Before applying Eq. (\ref{rhodet}), one
computes the ratio $\xi_d = n_{{\rm e},d}/\rho_d$. After computing new $\rho_d$ for
all $d< N\!D$ from Eq. (\ref{rhodet}), one computes new electron density as
$n_{{\rm e},d} = \xi_d \rho_d$, which is based on a reasonable assumption that 
the ionization structure is not changed significantly by updating the density. \\ [2pt]
(v) Steps (ii) - (iv) are repeated until a relative change in $z$ is sufficiently small.\\

\noindent $\bullet\ $ {\sf Modified procedure}.

\noindent
A better overall procedure is to solve simultaneously six governing equations 
for the six unknowns that form a vector $\psi_d$, viz.
\begin{equation}
\psi_d \equiv \{ P_d, P^{\rm gas}_d, \rho_d, N_d, n_{{\rm e},d}, z_d \},
\end{equation}
where, at depth point $d$,  $P_d$ is the total pressure, $N_d$ is the total particle 
number density. and the other symbols have their usual meaning. The six governing
equations are
\begin{eqnarray}
\label{he6a}
(P_d - P_{d-1})/\Delta m_{d-1/2} - Q(z_d + z_{d-1}) &=& 0, \\
P_d - P^{\rm gas}_d - P^{\rm rad}_d &=& 0, \\
\rho_d - (\mu/m_H) (N_d - n_{{\rm e},d}) &=& 0, \\
P^{\rm gas}_d - N_d k T_d &=& 0, \\
n_{{\rm e},d} - \zeta_d N_d &=& 0, \\
z_d - z_{d+1}  - \Delta m_{d+1/2}\left(1/\rho_{d} + 1/\rho_{d+1}\right) &=& 0.
\label{he6b}
\end{eqnarray}
The first equation apply for $d>1$. For $d=1$ it is replaced by
(for details refer to Appendix D.2),
\begin{equation}
\label{he6a1}
\rho_1 H_g f(x_1) - m_1 = 0,
\end{equation}
and the last equation for $d=N\!D$ is replaced by
\begin{equation}
\label{he6b1}
z_{N\!D} = 0.
\end{equation}
The non-linear set of equations (\ref{he6a}) - (\ref{he6b1}) is solved by
linearization for the six components of vector $\psi$. The resulting set
of linearized equations can be written as
\begin{equation}
\label{he6tri}
-A_d \delta\psi_{d-1} + B_d \delta\psi_d - C_d \delta\psi_{d+1}=L_d,
\end{equation}
where $A_d$, $B_d$, and $C_d$ are $6\times 6$ matrices. The system
(\ref{he6tri}) is solved by a standard Gauss-Jordan forward-backward
elimination scheme described by Eqs. (\ref{elim1}) - (\ref{elim3}).
The temperature
$T_d$ and the radiation pressure $P^{\rm rad}_d$ are held fixed. The
ratio $\zeta_d = n_{{\rm e},d}/N_d$ is recomputed after each internal
iteration step and is held constant during the next one.  

From the physical point of view, the above procedure is equivalent to the
original one, as it solves the hydrostatic equilibrium equation together
with the $z-m$ relation (the rest of the six governing equations are essentially
the definitions of the individual quantities), but the iteration scheme to
obtain the solution is more efficient in the presert procedure.

%--------------------------------------------------------------------------------------

% --------------------------------------------------------------------------

\addcontentsline{toc}{section}{Appendix C. Linearization of the structural equations} 
\section*{Appendix C. Linearization of the structural equations}

\subsubsection*{Notation}

The order of the individual quantities in the state vector ${\bf \psi}$,
and the order of equations within the global operator ${\bf P}$ are arbitrary. Here
we chose, in keeping with the original approach introduced in Auer \& Mihalas (1969),
the state vector in the form (\ref{cl1}), and the corresponding order of
equations: $N\!F$ transfer equations for explicit (linearized) frequency points, 
hydrostatic equilibrium equation,
radiative equilibrium equation, charge conservation equation, and $N\!L$ kinetic
equilibrium equations supplemented by the corresponding particle conservation
(abundance definition) equations. To simplify the notation, the index
corresponding to hydrostatic equilibrium (or total particle density $N$) is
denoted $N\!H$; the index corresponding to radiative equilibrium (or temperature
$T$) as $N\!R$, and that corresponding to charge conservation (and the electron
density $n_\mathrm{e}$) as $N\!P$. In the present convention
\begin{equation}
\label{eq:18.2-5aa}
N\!H=N\!F+1,\quad N\!R=N\!F+2, \quad  N\!P=N\!F+3.
\end{equation}
in the case of accretion disks, there is another state parameter and corresponding
equation, namely the geometrical distance from the midplane, $z$. The corresponding
index is denoted $N\!Z$, and its usual value is $N\!Z = N\!F+4$.

However, {\sc tlusty} can accept any reasonable values for these parameters,
not just those specified by Eq. (\ref{eq:18.2-5aa}).

\subsection*{C1. Stellar atmospheres}
\addcontentsline{toc}{subsection}{C1. Stellar atmospheres}

\subsubsection*{Linearized transfer equation}

The discretized transfer equation was already considered in Appendix A, and is 
described by Eqs. (\ref{feabip}) - (\ref{feabdis}).
Let $i, i=1,\ldots,N\!F$, be a row corresponding to the transfer equation, and
let $P_i(\bf\psi)=0$ be a formal expression of the transfer equation. Then the
individual matrix element $ij$ represents the partial derivative of the $i$-th 
transfer equation with respect to the $j$-th mean intensity,\\

$\bullet$ 
For the upper boundary condition, $d=1$, and $j=1,\ldots,N\!F$
\begin{eqnarray}
(B_{1})_{ij} &\equiv& \frac{\partial P_{1,i}}{\partial J_{1,j}} =
\left[ \frac{f_{1,i}}{\Delta\tau_{3/2,i}}+g_i
+ \frac{\tau_{3/2,i}}{2} \epsilon_{1,i} \right]\delta_{ij} , \\ 
(C_{1})_{ij} &\equiv& -\frac{\partial P_{1,i}}{\partial J_{2,j}} =
\frac{f_{2,i}}{\Delta\tau_{3/2,i}}\,\delta_{ij} .
\end{eqnarray}
The other columns corresponding to the components $\psi_{d,k}$, for
$k > N\!F$, i.e. corresponding to the temperatures and the number densities
(total, electron, and level populations)
\begin{eqnarray}
(B_{1})_{ik} &\equiv& \frac{\partial P_{1,i} }{ \partial\psi_{1,k}} =
\left[ -\frac{f_{1,i}J_{1,i}-f_{2,i}J_{2,i}}{\Delta\tau_{3/2,i}^2}
+ \frac{1}{2}\left(\epsilon_{1,i}J_{1,i}-\frac{\eta_{1,i}}{\chi_{1,i}}\right) \right]
 \frac{\partial\Delta\tau_{3/2,i}}{\partial\psi_{1,k}} \nonumber \\
 &+& \frac{\Delta\tau_{3/2,i}}{2}\left[
 \frac{\partial\epsilon_{1,i}}{\partial\psi_{1,k}} J_{1,i}
- \frac{\eta_{1,i}}{\chi_{1,i}} \left(
\frac{1}{\eta_{1,i}} \frac{\partial\eta_{1,i}}{\partial\psi_{1,k}} -
\frac{1}{\chi_{1,i}} \frac{\partial\chi_{1,i}}{\partial\psi_{1,k}}  \right)\right] , \\
(C_{1})_{ik} &\equiv& -\frac{\partial P_{1,i}}{\partial\psi_{2,k}} =
\left( \frac{f_{1,i}J_{1,i}-f_{2,i}J_{2,i}}{\Delta\tau_{3/2,i}^2}\right)
\frac{\partial\Delta\tau_{3/2,i}}{\partial\psi_{2,k}},
\end{eqnarray}
where, generally,
\begin{equation}
 \frac{\partial\Delta\tau_{d-1/2,i}}{\partial\psi_{d,k}} =
 \frac{\Delta\tau_{d-1/2,i}}{\omega_d+\omega_{d-1}}
 \frac{\partial\omega_{d-1,i}}{\partial\psi_{d-1,k}},
 \end{equation}
 and $\omega_{di} = \chi_{di}/\rho_d$.
The components of the right-hand side vector ${\bf L}_1$ are given by
\begin{equation}
\label{eq:18.2-12}
L_{1i} \equiv -P_{1,i} = -\frac{f_{1,i}J_{1,i}-f_{2,i}J_{2,i}}{\Delta\tau_{3/2,i}}
- g_{i}J_{1,i} + H_i^\mathrm{ext} - \frac{\Delta\tau_{3/2,i}}{2}
\left(\epsilon_{1,i}J_{1,i}-\frac{\eta_{1,i}}{\chi_{1,i}}\right).
\end{equation}
$\bullet$ 
For the inner points, $d=2,\ldots,N\!D-1$, one has for $j=1,\ldots,N\!F$,
\begin{eqnarray}
\hspace*{-4em}(A_d)_{ij}&=&
\frac{f_{d-1,i}}{\Delta\tau_{d-1/2,i}\Delta\tau_{d,i}}\,\delta_{ij}, \\
\hspace*{-4em}(B_d)_{ij}&=&
\left[\frac{f_{d,i}}{\Delta\tau_{d,i}}\left(\frac{1}{\Delta\tau_{d-1/2,i}}+
\frac{1}{\Delta\tau_{d+1/2,i}}\right)+\epsilon_{d,i} \right]\,\delta_{ij}, \\
\hspace*{-4em}(C_d)_{ij}&=&
\frac{f_{d+1,i}}{\Delta\tau_{d+1/2,i}\Delta\tau_{d,i}}\delta_{ij}, 
\end{eqnarray}
and for $k>N\!F$,
\begin{eqnarray}
(A_{d})_{ik} &=& a_{di} \frac{\partial\omega_{d-i,i}}{\partial\psi_{d-1,k}}, \\
(C_{d})_{ik} &=& c_{di} \frac{\partial\omega_{d+1,i}}{\partial\psi_{d+1,k}}, \\
(B_{d})_{ik} &=& -(a_{di}+ c_{di}) \frac{\partial\omega_{d,i}}{ \partial\psi_{d,k}}
+  \frac{\partial\epsilon_{1,i}}{\partial\psi_{1,k}} \, J_{di} \nonumber \\
&-& \frac{\eta_{d,i}}{\chi_{d,i}} \left(
\frac{1}{\eta_{d,i}} \frac{\partial\eta_{d,i}}{\partial\psi_{d,k}} -
\frac{1}{\chi_{d,i}} \frac{\partial\chi_{d,i}}{\partial\psi_{d,k}}  \right),
\end{eqnarray}
where
\begin{eqnarray}
\alpha_{di} &=& \frac{f_{di} J_{di} - f_{d-1,i} J_{d-1,i}}{\Delta\tau_{d-1/2,i}\Delta\tau_{di}},\\\
\gamma_{di} &=& \frac{f_{di} J_{di} - f_{d+1,i} J_{d+1,i}}{\Delta\tau_{d+1/2,i}\Delta\tau_{di}},\\\
\beta_{di} &=& \alpha_{di} + \gamma_{di}, \\
a_{di} &=& \big[ \alpha_{di}+(1/2) \beta_{di}(\Delta\tau_{d-1/2,i}/\Delta\tau_{di})\big]
\big/ \big(\omega_{d-1,i} + \omega_{di} \big),\\
c_{di} &=& \big[ \gamma_{di}+(1/2) \beta_{di}(\Delta\tau_{d+1/2,i}/\Delta\tau_{di})\big]
\big/ \big(\omega_{d+1,i} + \omega_{di} \big),
\end{eqnarray}
and the right--hand--side vector
\begin{equation}
\label{eq:18.2-19}
L_{di} = -\beta_{di} - \epsilon_{di}J_{di} + \eta_{di}/\chi_{di}.
\end{equation}
$\bullet$ 
For the lower boundary condition, $d=N\!D$, we have, 
for $j\leq N\!F,\, k>N\!F$,
\begin{eqnarray}
(B_{d})_{ij}  &=&
\label{bndre0}
\left[ \frac{f_{di}}{\Delta\tau_{d-1/2,i}}+\frac{1}{2}
+ \frac{\tau_{d-1/2,i}}{2}\, \epsilon_{di} \right]\delta_{ij} , \\ 
(A_{d})_{ij}  &=&
\frac{f_{d-1,i}}{\Delta\tau_{d-1/2,i}}\,\delta_{ij} , \\
(B_{d})_{ik}  &=&
\label{bndre}
\left[ -\frac{f_{di}J_{di}-f_{d-1,i}J_{d-1,i}}{\Delta\tau_{d-1/2,i}^2}+b_i
+ \frac{1}{2}\left(\epsilon_{di}J_{di}-\frac{\eta_{di}}{\chi_{di}}\right) \right]
 \frac{\partial\Delta\tau_{d-1/2,i}}{\partial\psi_{d,k}} \nonumber \\
 &+& \frac{\Delta\tau_{d-1/2,i}}{2}\left[
 \frac{\partial\epsilon_{di}}{\partial\psi_{d,k}} J_{di}
- \frac{\eta_{di}}{\chi_{di}} \left(
\frac{1}{\eta_{di}} \frac{\partial\eta_{di}}{\partial\psi_{d,k}} -
\frac{1}{\chi_{di}} \frac{\partial\chi_{di}}{\partial\psi_{d,k}}  \right)\right]  \nonumber \\
&-& \left(\frac{1}{2}+\frac{1}{3\Delta\tau_{d-1/2,i}}\right)
\left(\frac{dB_i}{dT}\right)_{\!\!d} \delta_{k,N\!R},\\
(A_{d})_{ik}  &=&
\left( \frac{f_{di}J_{di}-f_{d-1,i}J_{d-1,i}}{\Delta\tau_{d-1/2,i}^2}-b_i\right)
\frac{\partial\Delta\tau_{d-1/2,i}}{\partial\psi_{d-1,k}} \nonumber \\
&-&\frac{1}{3\Delta\tau_{d-1/2,i}} 
\left(\frac{dB_i}{dT}\right)_{\!\!d-1} \!\delta_{k,N\!R},\
\label{andre}
\end{eqnarray}
where 
\begin{equation}
b_i \equiv \frac{1}{3} \frac{B_{di}-B_{d-1,i}}{\Delta\tau_{d-1/2,i}^2}.
\end{equation}
The last terms in Eqs.(\ref{bndre}) and (\ref{andre}), which only apply for 
$\psi_k = T$, i.e., $k=N\!R$, arise from the derivatives of the Planck function 
with respect to temperature. Finally
\begin{eqnarray}
\label{eq:18.2-27}
L_{d,i}  &=& -\frac{f_{di}J_{di}-f_{d-1,i}J_{d-1,i}}{\Delta\tau_{d-1/2,i}}
- \frac{1}{2}(J_{di}-B_{di})   
+ \frac{1}{3} \frac{B_{di}-B_{d-1,i}}{\Delta\tau_{d-1/2,i}}\nonumber \\
&-& \frac{\Delta\tau_{d-1/2,i}}{2}
\left(\epsilon_{di}J_{di}-\frac{\eta_{di}}{\chi_{di}}\right).
\end{eqnarray}

\subsubsection*{Linearized hydrostatic equilibrium equation}

The discretized form of  Eq. (\ref{he1}) reads  
\begin{equation}
\label{eq:18.1b-8}
N_dkT_d - N_{d-1}kT_{d-1} + \frac{4\pi}{ c}\sum_{i=1}^{N\!F}w_i(f_{di}J_{di}-
f_{d-1,i}J_{d-1,i}) = g(m_d - m_{d-1}).
\end{equation}
The upper boundary condition is derived from Eq. (\ref{he1})
assuming that the radiation force remains constant from the boundary surface
upward:
\begin{equation}
\label{eq:18.1b-9}
N_1kT_1 = m_1 \big[ g- (4\pi/c)\sum_{i=1}^{N\!F} w_i (\chi_{1i}/\rho_1) g_i J_{1i} \big].
\end{equation}
For accretion disks, Eqs. (\ref{eq:18.1b-8})  and (\ref{eq:18.1b-9}) remain unchanged, 
the only difference is in the expression for the gravity acceleration $g$ as
\begin{equation}
g \equiv g(z) = Q(z_d+z_{d-1})/2.
\end{equation}
The upper boundary condition is  given by Eqs. (\ref{he1disk}) and (\ref{he1f}).

The components of matrices ${\bf A}$, ${\bf B}$, and vector ${\bf L}$ corresponding to
the hydrostatic equilibrium equation, the row $N\!H =N\!F+1$, 
follow from linearizing Eqs. (\ref{eq:18.1b-8}) and (\ref{eq:18.1b-9}), are given by
\begin{eqnarray}
(B_1)_{N\!H,i} &=& (4\pi/c)\,w_i\omega_{1,i}g_i, \quad i\le N\!F,\\
(B_1)_{N\!H,N\!H} &=& kT_1,\\
(B_1)_{N\!H,N\!R} &=&
kN_1+\frac{4\pi}{c}\sum_{j=1}^{N\!F}w_jg_jJ_{1,j}\frac{\partial\omega_{1,j}}{\partial T_{1}},\\
(B_1)_{N\!H,n} &=& \frac{4\pi}{c}\sum_{j=1}^{N\!F}w_j g_jJ_{1,j}
\frac{\partial\omega_{1,j}}{\partial\psi_{1,n}},\,n\!>\!N\!R, \\
(L_1)_{N\!H}&=&gm_1 -N_1 kT_1-\frac{4\pi}{c}\sum_{j=1}^{N\!F}w_j\omega_{1j}g_{i}J_{1i},
\end{eqnarray}
and, for $d>1$,
\begin{eqnarray}
(A_d)_{N\!H,i}&=&(4\pi/c) w_i f_{d-1,i} ,\quad i\le N\!F,\\
(B_d)_{N\!H,i}&= &(4\pi/c) w_i f_{di} , ,\quad i\le N\!F,\\
(A_d)_{N\!H,N\!H}&=&kT_{d-1},\\
(B_d)_{N\!H,N\!H}&=&kT_{d},\\
(A_d)_{N\!H,N\!R}&=&kN_{d-1},\\
(B_d)_{N\!H,N\!R}&=&kN_{d},
\end{eqnarray}
\begin{eqnarray}
\label{eq:18.2-41}
(L_d)_{N\!H}&=&g(m_d - m_{d-1}) -N_d kT_d +N_{d-1}kT_{d-1} \nonumber\\
          &-&\frac{4\pi}{c}\sum_{j=1}^{N\!F}w_j(f_{di}J_{di}-f_{d-1,i}J_{d-1,i}) . 
\end{eqnarray}

\subsubsection*{Linearized radiative equilibrium equation}

By discretizing the radiative equilibrium equation (\ref{re}) one obtains
\begin{equation}
\label{eq:18.1b-10}
\alpha_d \sum_{i=1}^{N\!F}w_i (\kappa_{di}J_{di} - \eta_{di}) +
\beta_d \left[
\sum_{i=1}^{N\!F} w_i \frac{f_{di}J_{di}-f_{d-1,i}J_{d-1,i}}{\Delta\tau_{d-1/2,i}} -
\frac{\sigma_R}{4\pi} T_\mathrm{eff}^4  \right] = 0.
\end{equation}
For accretion disks, Eq. (\ref{eq:18.1b-10}) is modified as expressed in 
Eq. (\ref{re_con_disk}), 
\begin{eqnarray}
\label{eq:18.1b-10d}
\alpha_d \left[\sum_{i=1}^{N\!F}w_i (\kappa_{di}J_{di} - \eta_{di}) + E_{\rm diss,d}\right]
\ \ \ \ \ \ \ \ \ \ \ \ \ \ \ \ \ \ \ \ \ \ \ \ \ \ \ \   \nonumber \\
+ \beta_d \left[
\sum_{i=1}^{N\!F} w_i \frac{f_{di}J_{di}-f_{d-1,i}J_{d-1,i}}{\Delta\tau_{d-1/2,i}} -
\frac{\sigma_R}{4\pi} T_\mathrm{eff}^4(1-\theta_d)  \right] = 0,
\end{eqnarray}
where $E_{\rm diss}$ and $\theta$ are given by Eqs. (\ref{ediss}) and (\ref{theta}),
respectively.

Linearization of Eq. (\ref{eq:18.1b-10}), for $d > 1$ and $k\geq N\!R$,  yields
\begin{eqnarray}
(A_d)_{N\!R,i} &=& \beta_d\, w_i \frac{f_{d-1,i}}{\Delta\tau_{d-1/2,j}} , \quad i\le N\!F,\\
(A_d)_{N\!R,k} &=& \beta_d \sum_{j=1}^{N\!F} w_j \frac{f_{dj}J_{dj} - f_{d-1,j} J_{d-1,j}}
{\Delta\tau_{d-1/2,j}^2}
\frac{\partial\Delta\tau_{d-1/2,j}}{\partial\psi_{d-1,k}}, \\
(B_d)_{N\!R,i}  &=& \alpha_d w_i \kappa_{di}
                        + \beta_d w_i \frac{f_{di}}{\Delta\tau_{d-1/2,j}}, \quad i\le N\!F,\\
(B_d)_{N\!R,k} &=& \alpha_d \sum_{j=1}^{N\!F} w_j 
                               \left( \frac{\partial\kappa_{dj}}{\partial\psi_{dk }} J_{dj} -
                               \frac{\partial\eta_{dj}}{\partial\psi_{dk }} \right) \ \nonumber \\
                        &-& \beta_d \sum_{j=1}^{N\!F} w_j \frac{f_{dj}J_{dj} - f_{d-1,j} J_{d-1,j}}
                               {\Delta\tau_{d-1/2,j}^2}
                               \frac{\partial\Delta\tau_{d-1/2,j}}{\partial\psi_{d,k}} , 
\end{eqnarray}
and
\begin{eqnarray}
\label{eq:18.2-49a}
(L_d)_{N\!R} &=& -\alpha_d \sum_{j=1}^{N\!F} w_j(\kappa_{dj}J_{dj}-\eta_{dj}) \nonumber \\
                    &+& \beta_d \left[\frac{\sigma_R}{4\pi}\,T_\mathrm{eff}^4 - \sum_{j=1}^{N\!F} w_j
\frac{f_{dj}J_{dj} - f_{d-1,j} J_{d-1,j}}{\Delta\tau_{d-1/2,j}} \right].
\end{eqnarray}
The upper boundary condition for $d=1$ is
\begin{eqnarray}
(B_d)_{N\!R,i}  &=& \alpha_d w_i \kappa_{di} 
                        + \beta_d w_i g_i,  \quad  i\le N\!F, \\
(L_d)_{N\!R} &=& -\alpha_d \sum_{j=1}^{N\!F} w_j(\kappa_{dj}J_{dj}-\eta_{dj}) \nonumber \\
                     &+& \beta_d \left[\frac{\sigma_R}{4\pi}\,T_\mathrm{eff}^4 - \sum_{j=1}^{N\!F} w_j
                             w_j(g_j J_{1j} - H_j^{\rm ext}) \right].
\end{eqnarray}

\subsubsection*{Linearized charge conservation equations}

The equation is local and simple, so the linearization is straightforward
\begin{eqnarray}
(B_d)_{N\!P,N\!P} &=& -1,\\
(B_d)_{N\!P,i}    &=& Z_i,\quad i>N\!P,
\end{eqnarray}
where $Z_i$ is the charge of the ion to which level $i$ belongs, and
\begin{equation}
\label{eq:18.2-51}
(L_d)_{N\!P} = n_\mathrm{e} - \sum_{i=1}^{N\!L}n_iZ_i.
\end{equation}
All other elements of the $N\!P$-th row of matrix ${\bf B}$, and all elements
of the $N\!P$-th row of matrices ${\bf A}$ and ${\bf C}$ are zero.%\\

\subsubsection*{Linearized kinetic equilibrium equations}

The radiative rates are written as quadrature sums,
\begin{eqnarray}
R_{lu} &=& \frac{4\pi}{ h} \sum_{i=1}^{N\!F} \frac{1}{\nu_i}\, w_i \sigma_{lu}(\nu_i)J_{di}, \\
R_{lu} &=& \frac{4\pi}{h}   \sum_{i=1}^{N\!F} \frac{1}{\nu_i}\, w_i  \sigma_{lu}(\nu_i) \left(
\frac{2h\nu_i^3}{c^2} + J_{di}\! \right) .
\end{eqnarray}
where the summations formally extend over all frequency points. It should
be kept in mind that the appropriate cross-section differ from zero only
in limited ranges of frequencies.

The kinetic equilibrium equations (\ref{ke1}) are local, therefore
$A_{ij} = C_{ij} = 0$ for $i > N\!P$ and all $j$. Because the form of matrix
elements is the same for all depth points, we drop the depth index $d$.
The matrix elements are as follows
\begin{eqnarray}
B_{m,i} &=& \sum_{j=1}^{N\!L} \frac{\partial\mathcal{A}_{mj}}{\partial J_i}\, n_j,\quad i\le N\!F,\\
B_{m,NR} &=& \sum_{j=1}^{N\!L} \frac{\partial\mathcal{A}_{mj}}{\partial T}\, n_j,\\
B_{m,NP} &=& \sum_{j=1}^{N\!L} \frac{\partial\mathcal{A}_{mj}}{\partial n_\mathrm{e}}\, n_j,\\
B_{mj} &=& \mathcal{A}_{mj}, \quad j > N\!P ,\\
L_m &=& b_m - \sum_{j=1}^{N\!L} \mathcal{A}_{mj} n_j.
\end{eqnarray}
For each atomic species $I$, the rate equation for a characteristic level, say $k$,
is replaced by the particle conservation equation, (\ref{ade}). The
corresponding matrix elements are
\begin{eqnarray}
B_{ki} &=&1,\\
B_{k,N\!H} &=& -\alpha_I, \\
B_{k,N\!P} &=& \alpha_I,
\end{eqnarray}
where $i$ labels all levels considered for species $I$,
and $\alpha_I = A_I/\sum_J A_J$ is the fractional abundance of species $I$.%\\

\subsubsection*{Linearized equation for convection}

When convection is taken into account, it is advantageous to consider the
logarithmic gradient of temperature $\nabla$ as one of the state parameters,
and include it in the state vector ${\bf \psi}$,
\begin{equation}
\label{eq:18.2c-1}
{\bf \psi}_d = \{J_{d1},\ldots,J_{d,N\!F},N,T,n_{\mathrm{e},d},n_{d1},\ldots,n_{d,N\!L},\nabla_d\},
\end{equation}
where we adopt the convention that the gradient $\nabla_d$ is the one
corresponding to depth $d-1/2$, i.e.,
\begin{equation}
\label{eq:18.1c-9}
\nabla_{d-1/2} = (\ln T_d - \ln T_{d-1})/ (\ln P_d - \ln P_{d-1}).
\end{equation}
The gradient may also be evaluated as
\begin{equation}
\label{eq:18.1c-10}
\nabla_{d-1/2} = \frac{T_d - T_{d-1}}{ T_d + T_{d-1} } \cdot
                              \frac{P_d + P_{d-1}}{ P_d - P_{d-1} }.
\end{equation}
In the following, we use equation (\ref{eq:18.1c-10}) to represent the
logarithmic temperature gradient. The corresponding adiabatic gradient also
has to be evaluated at the mid-point $d-\frac{1}{2}$, 
$\nabla_\mathrm{ad}= \nabla_\mathrm{ad}(T_{d-1/2},P_{d-1/2})$,
with $T_{d-1/2} = (T_d +T_{d-1})/2$, and a similar equation for
$P_{d-1/2}$. The mid-point values may also be evaluated as geometric
means, i.e.  $T_{d-1/2} = (T_d T_{d-1})^{1/2}$, but in the following
we use the arithmetic mean.
The convective flux is a quantity that also corresponds to the grid mid-points,
hence it should be written as
$F_\mathrm{conv} = F_{\mathrm{conv},d-{1/2}}$.

At the depth points where the Schwarzschild criterion is satisfied, the
radiative equilibrium equation has to be modified to radiative+convective 
equilibrium equation which reads
\begin{eqnarray}
\label{eq:18.1c-11}
\alpha_d \left[ \sum_i w_i(\kappa_i J_i -\eta_i) + \frac{\rho_d}{4\pi}
\frac{F^{\mathrm{conv}}_{d+1/2}({\bf\psi_d},{\bf\psi}_{d+1}) -
F^{\mathrm{conv}}_{d-1/2}({\bf\psi_d},{\bf\psi}_{d-1})}{\Delta m_d} \right] \nonumber \\
+ \beta_d \left[
\sum_{i=1}^{N\!F} w_i \frac{f_{di}J_{di}-f_{d-1,i}J_{d-1,i}}{\Delta\tau_{d-1/2,i}} -
\frac{\sigma_R}{4\pi} T_\mathrm{eff}^4  
+ \frac{F^{{\rm conv}}_{d-1/2}({\mathbf\psi_d},{\bf\psi}_{d-1})}{4\pi} \right] = 0,
\end{eqnarray}
where $\Delta m_d = (m_{d+1} - m_{d-1/2})/2$. Here we show explicitly a dependence 
of the convective flux of the state parameters, because for instance $F^{{\rm conv}}_{d-1/2}$
is evaluated at the midpoint, that is, for temperature $T=(T_d+T_{d-1})/2$, and analogously for
other quantities. For accretion disks, Eq. (\ref{eq:18.1c-11}) is modified analogously to
Eq. (\ref{eq:18.1b-10d}).

There are four modifications of matrices ${\bf A}$, ${\bf B}$, ${\bf C}$, and vector
${\bf L}$ when convection is taken into account:\\
-- a modification of the row $N\!R=N\!F+2$, corresponding to radiative
equilibrium, now being modified to radiative + convective equilibrium;\\
-- an addition of a column in all matrices corresponding to $\nabla$; using the
convention of equation (\ref{eq:18.2c-1}), it is the column $N\!N=N\!F+N\!L+4$;\\
-- an addition of a row $N\!N$, corresponding to $\nabla$;\\
-- a modification of vector ${\bf L}$, namely changing the $N\!R$ element, and
adding the $N\!N$ element.

$\bullet$ Because the convective flux depends only on $T$, $P$, and
$\nabla$, and writing $P=NkT$, the only new elements of the row $N\!R$ 
of matrices ${\bf A}$, ${\bf B}$, ${\bf C}$ are those corresponding to $N$, $T$, and $\nabla$. 
In the radiative zone, $\nabla < \nabla_\mathrm{ad}$, all new elements are zero.
In the convection zone, $\nabla \ge \nabla_\mathrm{ad}$, there are the following
additions to the matrix elements, denoted by superscript ``conv'', that follow
from linearizing equation (\ref{eq:18.1c-11}), where we introduce 
$H^{\rm conv}\equiv  F^\mathrm{conv}/4\pi$,
\begin{eqnarray}
A_{N\!R,N\!H}^\mathrm{conv} &=& \alpha_d 
   \left( \frac{\partial H^{\rm conv} }{ \partial P} \right)_{\! d-1/2}
   \frac{\rho_d}{ 2 \Delta m_d} kT_{d-1} \nonumber \\
  &+& \frac{\beta_d}{2} \left( \frac{\partial H^{\rm conv} }{ \partial P} \right)_{\! d-1/2}\!\!
   kT_{d-1}, \\
A_{N\!R,N\!R}^\mathrm{conv} &=& \alpha_d \left[
   \left( \frac{\partial H^{\rm conv} }{ \partial T} \right)_{\! d-1/2} \!\!\!+
   \left( \frac{\partial H^{\rm conv} }{ \partial P} \right)_{\! d-1/2} 
   \!\! kN_{d-1} \right]
   \frac{\rho_d}{ 2 \Delta m_d} \nonumber  \\
&+&\beta_d \left[
 \frac{1}{2}  \left( \frac{\partial H^{\rm conv} }{ \partial T} \right)_{\! d-1/2} +
  \frac{1}{2} \left( \frac{\partial H^{\rm conv} }{ \partial P} \right)_{\! d-1/2}
  \!\! kN_{d-1} \right], \\
A_{N\!R,N\!N}^\mathrm{conv} &=& \alpha_d 
  \frac{H^{\mathrm{conv}}_{d-1/2} }{ (\nabla_d -\nabla_{\mathrm{el},d-1/2})}
   \frac{3\rho_d}{ 4 \Delta m_d}, \\
B_{N\!R,N\!H}^\mathrm{conv} &=& \alpha_d \left[
   \left( \frac{\partial H^{\rm conv} }{ \partial P} \right)_{\! d+1/2}\!\!\!  -
   \left( \frac{\partial H^{\rm conv} }{ \partial P} \right)_{\! d-1/2} 
   \right] \frac{\rho_d}{ 2 \Delta m_d} kT_{d} \nonumber \\
   &+& \frac{\rho_d}{(N_d-n_{\mathrm{e},d})}
    \frac{H^{\mathrm{conv}}_{d+1/2}
   - H^{\mathrm{conv}}_{d-1/2}}{\Delta m_d} \nonumber \\
 &+& \frac{\beta_d}{2}  \left( \frac{\partial H^{\rm conv} }{ \partial P} \right)_{\! d-1/2} 
 kT_{d} ,\\
B_{N\!R,N\!R}^\mathrm{conv} &=& \alpha_d \left[
   \left( \frac{\partial H^{\rm conv} }{ \partial T} \right)_{\! d+1/2}\!\!\!   +
   \left( \frac{\partial H^{\rm conv} }{ \partial P} \right)_{\! d+1/2}  kN_d 
   \right. \nonumber \\ 
  &-& \left.\left( \frac{\partial H^\mathrm{conv} }{ \partial T} \right)_{\! d-1/2}\!\!\!   -
   \left( \frac{\partial H^\mathrm{conv} }{\partial P} \right)_{\! d-1/2}\!\!\! kN_d 
   \right] \frac{\rho_d}{ 2 \Delta m_d} \nonumber \\
&+& \beta_d \left[
 \frac{1}{2}  \left( \frac{\partial H^{\rm conv} }{ \partial T} \right)_{\! d-1/2} +
 \frac{1}{2}  \left( \frac{\partial H^{\rm conv} }{ \partial P} \right)_{\! d-1/2} 
 \!\! kN_{d} \right], \\
B_{N\!R,N\!P}^\mathrm{conv} &=& 
   -\alpha_d \frac{\rho_d}{(N_d-n_{\mathrm{e},d})}
    \frac{H^{\mathrm{conv}}_{d+1/2}
   - H^{\mathrm{conv}}_{d-1/2}}{\Delta m_d} ,\\
B_{N\!R,N\!N}^\mathrm{conv} &=& \alpha_d \left[
  \frac{H^{\mathrm{conv}}_{d+1/2} }{ \nabla_{d+1} -\nabla_{\mathrm{el},d+1/2}} -
  \frac{H^{\mathrm{conv}}_{d-1/2} }{ \nabla_{d} -\nabla_{\mathrm{el},d-1/2}}
   \right] \frac{3 \rho_d}{ 4 \Delta m_d}\nonumber  \\
 &+& \beta_d \frac{3}{2}
  \frac{H^{\mathrm{conv}}_{d-1/2} }{ 
  \nabla_d -\nabla_{\mathrm{el},d-1/2}}, \\
C_{N\!R,N\!H}^\mathrm{conv} &=& 
   -\alpha_d \left( \frac{\partial H^{\rm conv} }{ \partial P} \right)_{\! d+1/2}
   \frac{\rho_d}{2 \Delta m_d} kT_{d+1} ,\\
C_{N\!R,N\!R}^\mathrm{conv} &=& -\alpha_d \left[
   \left( \frac{\partial H^{\rm conv} }{ \partial T} \right)_{\! d+1/2}\!\!\!  +
   \left( \frac{\partial H^{\rm conv} }{ \partial P} \right)_{\! d+1/2} \!\!\! 
   kN_{d-1} \right]
   \frac{\rho_d}{2 \Delta m_d} , \ \ \  \\
\vspace*{-.8em}C_{N\!R,N\!N}^\mathrm{conv} &=&- \alpha_d 
  \frac{H^{\mathrm{conv}}_{d+1/2} }{ (\nabla_d -\nabla_{\mathrm{el},d+1/2})}
   \frac{3\rho_d}{ 4 \Delta m_d}, \\ 
L_{N\!R}^\mathrm{conv} &=& -\alpha_d \frac{\rho_d}{\Delta m_d} (H^{\mathrm{conv}}_{d+1/2}
   - H^{\mathrm{conv}}_{d-1/2}) - \beta_d H^{{\rm conv}}_{d-1/2},
\end{eqnarray}
where we used  $\partial T_{d-1/2} / \partial T_d = 1/2$.
The partial derivatives of the convective flux with respect to temperature
and pressure are obtained numerically,
\begin{equation}
\label{eq:18.2c-8}
 \frac{\partial H^{\rm conv} }{ \partial T} =
 \frac{H^{\rm conv}(T+\delta T,P,\nabla) -  H^{\rm conv}(T,P,\nabla)  }{\delta T},
\end{equation}
and analogously for the derivative with respect to pressure. Here, $\delta T$
is an arbitrary small quantity; it is typically chosen $\delta T = 0.01\, T$.
The derivative with respect to $\nabla$ is computed analytically.

$\bullet$ The additional row corresponding to $\nabla$, that is a
linearized equation (\ref{eq:18.1c-10}), is simple
\begin{eqnarray}
A_{N\!N,N\!H}^\mathrm{conv} &=& 
  -\frac{2P_{d}\nabla_d }{ P_d^2 - P_{d-1}^2 }\,  kT_{d-1}, \\
A_{N\!N,N\!R}^\mathrm{conv} &=& 
   \frac{2T_{d} \nabla_d}{ T_d^2 - T_{d-1}^2 } 
   -\frac{2P_{d}\nabla_d }{ P_d^2 - P_{d-1}^2 }\, kN_{d-1}, \\
B_{N\!N,N\!H}^\mathrm{conv} &=& 
  - \frac{2P_{d-1}\nabla_d }{ P_d^2 - P_{d-1}^2 }\, kT_d, \\
B_{N\!N,N\!R}^\mathrm{conv} &=& 
  \frac{2T_{d-1}\nabla_d }{ T_d^2 - T_{d-1}^2 }  
  - \frac{2P_{d-1} \nabla_d}{ P_d^2 - P_{d-1}^2 }\, kN_d, \\
B_{N\!N,N\!N}^\mathrm{conv} &=& -1,
\end{eqnarray}
and
\begin{equation}
\label{eq:18.2c-22}
L_{N\!N} = \nabla_{d} - \frac{T_d - T_{d-1}} {T_d + T_{d-1} } \cdot
                        \frac{P_d + P_{d-1}}{ P_d - P_{d-1} }.\\
\end{equation}

%---------------------------------------------------

\subsection*{C2. Accretion disks}
\addcontentsline{toc}{subsection}{C2. Accretion disks}

\subsubsection*{Radiative transfer equation}

It is the same as for stellar atmospheres; the only difference is in the lower
boundary condition, $d=N\!D$. 
Its discretized form follows from integrating Eq. (\ref{feabdis}) over angles.
The corresponding matrix elements are given by the same expressions
as in Eqs. ({\ref{bndre0}) - (\ref{andre}); the only difference is that in Eqs.
(\ref{bndre}) and (\ref{andre}) the specific terms for $k=N\!R$ are  not present;
formally it is expressed as the term $\delta_{k,N\!R}=0$ for all values of $k$.
The elements of the right-hand side vector are given by
\begin{equation}
L_{d,i} =  -\frac{f_{di}J_{di}-f_{d-1,i}J_{d-1,i}}{\Delta\tau_{d-1/2,i}}
- \frac{\Delta\tau_{d-1/2,i}}{2}
\left(\epsilon_{di}J_{di}-\frac{\eta_{di}}{\chi_{di}}\right).
\end{equation}

\subsubsection*{Vertical hydrostatic equilibrium equation}

All matrix elements $A_{N\!H,i}$ and $B_{N\!H,i}$ and vector $L_{N\!H}$ 
remain
unchanged, the only difference is the addition of elements corresponding to vertical
distance $z$,
\begin{eqnarray}
(A_d)_{N\!H,N\!Z} &=& Q(m_d-m_{d-1})/2, \\
(B_d)_{N\!H,N\!Z} &=& -Q(m_d-m_{d-1})/2
\end{eqnarray}

The upper boundary condition can either be considered to be the same as in the
case of stellar atmospheres, or is taken in a different form,
based on the formalism put forward in \S\,\ref{gray_disk}, specifically Eqs. 
(\ref{he1disk}) and (\ref{he1f}). However, the scale heights $H_g$ and $H_r$ are
not approximated, but instead are treated  using actual values of the state parameters,
and are thus linearized.  Specifically,
\begin{eqnarray}
H_g &=& \left(\frac{2N_1 k T_1}{\rho_1 Q}\right)^{1/2}, \\
H_r &=& \frac{4\pi}{cQ\rho_1} \sum_{i+1}^{N\!F} w_i\chi_i\! \left(g_i J_i- H_i^{\rm ext}\right)
\end{eqnarray}
The right-hand side is given by
\begin{equation}
(L_1)_{N\!H} = m_1 - H_g \rho_1 f(x_1),
\end{equation}
where
\begin{equation}
x_1 = \frac{z_1-H_r}{H_g}, \quad {\rm and}\quad f(x)=\frac{\sqrt\pi}{2}\exp(x^2){\rm erfc}(x).
\end{equation}
and the matrix elements are
\begin{eqnarray}
(B_1)_{N\!H,i} &=& -\frac{4\pi}{cQ} f_1^\prime w_i g_i \chi_{1i},\quad i=1,\ldots,N\!F, \\
(B_1)_{N\!H,N\!H} &=& \frac{1}{N_1}(\rho_1 H_g f_1 + H_r \rho_1 f_1^\prime), \\
(B_1)_{N\!H,N\!R} &=& \frac{1}{2T_1}[\rho_1 H_g f_1 + \rho_1 f_1^\prime(H_r -z_1)]\nonumber \\
&-& \frac{4\pi}{cQ} f_1^\prime 
\sum_{i=1}^{N\!F} w_i \frac{\partial\chi_{1i}}{\partial T_1} \left(g_i J_i- H_i^{\rm ext}\right), \\
(B_1)_{N\!H,k} &=& - \frac{4\pi}{cQ} f_1^\prime 
\sum_{i=1}^{N\!F} w_i\frac{\partial\chi_{1i}}{\partial\psi_{1k}} \left(g_i J_i- H_i^{\rm ext}\right), \\
(B_1)_{N\!H,N\!Z} &=& \rho_1 f_1^\prime,
\end{eqnarray}
where $k >N\!R$ and $k\not= N\!Z$, and
\begin{equation}
f_1 = f(x_1), \quad f_1^\prime = \left. \frac{df}{dx}\right|_1,
\end{equation}
and $f^\prime$ is evaluated as $f^\prime(x)=[f(x+\Delta x)-f(x)]/\Delta x$ with
$\Delta x= 0.01 x$.

\subsubsection*{Energy balance equation}

As follows from Eq. (\ref{eq:18.1b-10d}), the matrix elements are very similar to 
those for the stellar atmosphere case, the only difference is the replacement
$\sigma_R T_{\rm eff}^4$ by $\sigma_R T_{\rm eff}^4(1-\theta_d)$, which enters
only the right-hand side vector (namely, its differential equation part). There are several
additional terms in the integral equation part which we summarize below,
\begin{eqnarray}
(A_d)_{N\!R,N\!H}^{\rm add} &=& \alpha_d E_0 w_d\, \mu m_H, \\
(B_d)_{N\!R,N\!H}^{\rm add} &=& -\alpha_d E_0 w_d\, \mu m_H, \\
(L_d)_{N\!R}^{\rm add} &=& -\alpha_d E_0 w_d\, \rho_d,
\end{eqnarray}
where
\begin{equation}
E_0 = E_{\rm diss}/\rho w = \frac{9}{16\pi} \frac{GM}{R^3} \left(\frac{A}{B}\right)^2,
\end{equation}
as follows from Eq. (\ref{ediss}).

\subsubsection*{Equation for vertical distance}

Matrix elements for $d<N\!D$ are
\begin{eqnarray}
(B_d)_{N\!Z,N\!H} &=& \mu m_H (m_{d+1} -m_d)/2\rho_d^2, \\
(C_d)_{N\!Z,N\!H} &=& -\mu m_H (m_{d+1} -m_d)/2\rho_{d+1}^2, \\
(B_d)_{N\!Z,N\!P} &=& -\mu m_H (m_{d+1} -m_d)/2\rho_d^2, \\
(C_d)_{N\!Z,N\!P} &=& \mu m_H (m_{d+1} -m_d)/2\rho_{d+1}^2, \\
(B_d)_{N\!Z,N\!Z} &=& (C_d)_{N\!Z,N\!Z} =1, \\
(L_d)_{N\!Z} &=& z_{d+1} - z_d + 
\frac{m_{d+1} -m_d}{2} \left(\frac{1}{\rho_d}+ \frac{1}{\rho_{d+1}} \right),
\end{eqnarray}
and for $d=N\!D$ they are simply
\begin{equation}
(B_d)_{N\!Z,N\!Z} = 1, \quad (L_d)_{N\!Z} = 0.
\end{equation}
%
%---------------------------------------------------

\subsection*{C3. Compton scattering}
\addcontentsline{toc}{subsection}{C3. Compton scattering}

Within the present setup, Compton scattering can only be treated using the
traditional full complete linearization approach, because using the hybrid CL/ALI scheme
is very unstable, and the Rybicki scheme cannot be easily generalized to this case
while preserving its favorable numerical properties.

Here we summarize the additional terms in matrices ${\bf A}$, ${\bf B}$, ${\bf C}$,
and in the right-hand side vector ${\bf L}$, described above in Appendix C.1.
We denote these additional terms with superscript ``compt".

$\bullet$ Matrix elements corresponding to the transfer equation:
For the highest frequency, which in the linearization has index $i=1$,
\begin{eqnarray}
(B_d)_{ii}^{\rm compt} &=& \mathcal {B}_i,  \\
(B_d)_{i,i+1}^{\rm compt} &=& \mathcal {C}_i,  \\
(L_d)_{i}^{\rm compt} &=&  -\mathcal {B}_i J_i -\mathcal {C}_i J_{i+1}.
\end{eqnarray}
For other frequencies, one has
\begin{eqnarray}
(B_d)_{i,i-1}^{\rm compt} &=& \mathcal {A}_i,  \\
(B_d)_{ij}^{\rm compt} &=& \mathcal {B}_i,  \\
(B_d)_{i,i+1}^{\rm compt} &=& \mathcal {C}_i,  \\
(B_d)_{i,N\!R}^{\rm compt} &=& [ (-3 c_i^- + d_i^-) J_{i-1} 
+ (-3 c_i^0 +d_i^0) J_{i} \nonumber \\
&+& (-3 c_i^+ +d_i^+) J_{i+1}] (n_{\rm e}\sigma_{\rm e}/\chi_i)(\Theta/T)  \\
(B_d)_{i,N\!P}^{\rm compt} &=& S^{\rm compt}/n_{\rm e},  \\
(L_d)_{i}^{\rm compt} &=& -S_{\rm compt}.
\end{eqnarray}
with
\begin{equation}
S^{\rm compt} = \mathcal {A}_i J_{i-1} +\mathcal {B}_i J_i +\mathcal {C}_i J_{i+1},
\end{equation}

$\bullet$ Matrix elements corresponding to the radiative equilibrium equation:
\begin{eqnarray}
(B_d)_{N\!R,i}^{\rm compt} &=& -\alpha_d \chi_{di} w_i
   (\mathcal{A}_{i+1} +\mathcal{B}_{i} +\mathcal{C}_{i-1}) \\
(B_d)_{N\!R,N\!R}^{\rm compt} &=& -\alpha_d  \chi_{di} w_i  [ (-3 c_i^- + d_i^-) J_{i-1} 
+ (-3 c_i^0 +d_i^0) J_{i} \nonumber \\
&+& (-3 c_i^+ +d_i^+) J_{i+1}] (n_{\rm e}\sigma_{\rm e}/\chi_i)(\Theta/T)  \\
(B_d)_{N\!R,N\!P}^{\rm compt} &=& -\alpha_d  \chi_{di} w_i S^{\rm compt}/n_{\rm e}, \\
(L_d)_{N\!R}^{\rm compt} &=& \alpha_d  \chi_{di} w_i S_{\rm compt}.
\end{eqnarray}
It turns out that in some cases the overall linearization scheme is more stable if
one does not linearize the Compton terms in the radiative equilibrium, so the
above additional matrix elements $(B_d)_{N\!R,k},k=1,\ldots,N\!P$, are set to zero
if needed.

%----------------------------------------------------------

\section*{Appendix D. Details of the implementation of the hybrid CL/ALI method}
\addcontentsline{toc}{section}{Appendix D. Details of the implementation of the hybrid 
CL/ALI method}

As shown in \S\,\ref{hybrid}, the method does not linearize the mean intensity of radiation
in all frequency points, but only in selected few frequencies, call ``explicit" frequencies. 
For the remaining frequency points, called ``ALI frequencies", the mean intensity is 
expressed as
\begin{equation}
\label{eq:18.4-14}
J_{dj} = \Lambda_{d,j}^\ast (\eta_{dj}/\kappa_{dj}) + \Delta J_{dj}.
\end{equation}
The mean intensity in the ALI frequencies are thus removed form the state vector.
The parameter $N\!F$ now refers to the number of explicit frequencies.

Although {\sc tlusty} allows for both a diagonal as well as tri-diagonal operator,
here we consider a diagonal $\Lambda^\ast$ for simplicity. 
In any case, numerical experience showed that
although an application of the tridiagonal operator speeds up the convergence of
the iterations, a similar or even faster convergence is obtained with a
diagonal operator applied together with the Ng acceleration.

In the expressions for the elements of the Jacobian, the mean intensity in the
ALI frequency point $i$ is then viewed as a function of $T$, $n_{\rm e}$, and the set
of atomic level populations $n_i, i=1,\ldots,N\!L$.
Denoting  $x\equiv \{T, n_\mathrm{e}, n_i \}$, the corresponding derivatives are given by
\begin{equation}
\label{ddef}
\frac{\partial J_{di}}{\partial x_d} \equiv
D_{dj}^x = \Lambda_{dj}^\ast \frac{\eta_{dj}}{\kappa_{dj}} \left(
\frac{1}{\eta_{dj}} \frac{\partial\eta_{dj}}{\partial x} -
\frac{1}{\kappa_{dj}} \frac{\partial\kappa_{dj}}{\partial x} \right),
\end{equation}
where the term $\Delta J_{id}$ is known from the previous iteration and is therefore 
independent of $x$, and consequently does not appear in linearized equations.

The elements of matrices ${\bf A}$,  ${\bf B}$, and ${\bf C}$ are then modified as
follows. All the expressions for the first $N\!F$ rows are valid, provided that
$N\!F$ now refers to the number of explicit (linearized) frequency points. Analogously, the
expression for the first $N\!F$ columns are also unchanged. Other elements will contain
additional terms that correspond to the contribution of the ALI frequencies.  We will
summarize the additional terms below.

We use the following notation. As specified above, $x$ may mean any of the
state vector parameters $T$, $n_{\rm e}$, and $n_i, i=1,N\!L$. The corresponding
index is denoted as $N\!X$; i.e. $N\!X=N\!R$ for $x=T$, $N\!X=N\!P$ for $x=n_{\rm e}$,
etc. The total number of the ALI frequency points is denoted by $N\!F\!A$.

For the row corresponding to the hydrostatic equilibrium equation one has the
following additional terms, denoted with superscript ``ALI" ,
\begin{eqnarray}
(B_1)^{\rm ALI}_{N\!H,N\!X} &=& (4\pi/c)\sum_{j=1}^{N\!F\!A} w_j g_j 
\left[ \omega_{1j} D_{1j}^x + (\partial\omega_{1j}/\partial x_1) J_{1j} \right], \\
(L_1)^{\rm ALI}_{N\!H} &=& -(4\pi/c)\sum_{j=1}^{N\!F\!A} w_j \omega_{1j} g_j J_{1j}, 
\end{eqnarray}
for the upper boundary condition, and, for $d>1$,
\begin{eqnarray}
(A_d)^{\rm ALI}_{N\!H,N\!X} &=& (4\pi/c)\sum_{j=1}^{N\!F\!A} w_j f_{d-1,j} D_{d-1,j}^x \\
(B_d)^{\rm ALI}_{N\!H,N\!X} &=& (4\pi/c)\sum_{j=1}^{N\!F\!A} w_j f_{dj} D_{dj}^x, \\
(L_d)^{\rm ALI}_{N\!H} &=& -(4\pi/c)\sum_{j=1}^{N\!F\!A} w_j (f_{dj}J_{dj}-f_{d-1,j}J_{d-1,j}).
\end{eqnarray}

For the row corresponding to the radiative equilibrium, one has the following
additions. 
\begin{eqnarray}
(A_d)^{\rm ALI}_{N\!R,N\!X} &=& \beta_d\sum_{j=1}^{N\!F\!A} 
w_j f_{d-1,j} D_{d-1,j}^x/\Delta\tau_{d-1/2,j}, \\
(B_d)^{\rm ALI}_{N\!R,N\!X} &=& \alpha_d \sum_{j=1}^{N\!F\!A} w_j \left[
   \kappa_{dj} D_{dj}^x + \frac{\partial\kappa_{dj}}{\partial x_d} J_{dj} - 
\frac{\partial\eta_{dj}}{\partial x_d} \right] \nonumber \\
                        &+& \beta_d \sum_{j=1}^{N\!F\!A}  
                         w_j f_{dj} D_{dj}^x/\Delta\tau_{d-1/2,j}, \\
(L_d)^{\rm ALI}_{N\!R} &=& -\alpha_d \sum_{j=1}^{N\!F\!A} w_j 
(\kappa_{dj}J_{dj}-\eta_{dj}) \nonumber \\
&-&  \beta_d \sum_{j=1}^{N\!F\!A} w_j 
\frac{f_{dj}J_{dj}-f_{d-1,j}J_{d-1,j}}{\Delta\tau_{d-1/2,j}}
\end{eqnarray}

The set of kinetic equilibrium equations is written in the form
\begin{equation}
\label{eq:18.4-22}
\bfn - {\mbox\boldmath{\mathcal{A}}}^{-1} \cdot \bfb = 0,
\end{equation} 
from which one obtains 
\begin{equation}
\label{eq:18.4-23}
(\partial\bfn/\partial x) = - {\mbox\boldmath{\mathcal{A}}}^{-1} \cdot {\bf V}_x.
\end{equation}
Vector ${\bf V}$ represents the column of the Jacobi matrix corresponding to
quantity $x$; $x$ stands for any quantity of vector ${\bf\psi}$, including linearized
mean intensities.
\begin{equation}
\label{eq:18.4-24}
{\bf V}_x = (\partial  {\mbox\boldmath{\mathcal{A}}} /\partial x)\cdot \bfn -
 (\partial{\bf b}/\partial x).
\end{equation}
Let the transition $l \leftrightarrow u$ be represented by an arbitrary
combination of the ``explicit'' and ``ALI'' frequency points; either subset
is allowed to be empty. Generally, the contribution from this transition
comes only to the two following components of vector ${\bf V}_x$, namely
\begin{eqnarray}
 ({\bf V}_x)_l &=& \frac{\partial(R_{lu} + C_{lu})}{ \partial x}\, n_l - 
                      \frac{\partial(R_{ul} + C_{ul})}{ \partial x}\, n_u , \nonumber \\
\label{eq:18.4-25}
 ({\bf V}_x)_u &=& - ({\bf V}_x)_l .
\end{eqnarray}
The radiative rate is written in a discretized form as
\index{Radiative rates}
\begin{equation}
\label{eq:18.4-26}
R_{lu} = (4\pi/ h) \left[ \sum_{i=1}^{N\!F} w_i \sigma_{lu}(\nu_i) J_i/\nu_i +
                                       \sum_{j=1}^{N\!F\!A} w_j \sigma_{lu}(\nu_j) J_j/\nu_j, \right],
\end{equation}
and similarly for the downward rate. The contribution to ${\bf V}_x$ from the 
collisional rates and from the ``explicit'' frequency points is the same as in the
case of the standard CL, whereas the ALI contribution is given by
\begin{equation}
\label{eq:18.4-27}
(V_x)_l^\mathrm{ALI} = [n_l -n_u G_{lu}(\nu)](4\pi/h)\sum_{j=1}^{N\!F\!A}
w_j \sigma_{lu}(\nu_j) D_j^x/\nu_j,
\end{equation}
where $D_j^x$ is given by equation (\ref{ddef}). The important differences
from the standard complete linearization is that because the derivatives
$D_j^x$ are generally nonzero for $x$ being the individual level populations,
the Jacobian contains contributions from the populations. This of course
expresses the fact stated already that within the ALI formalism the kinetic
equilibrium equations are nonlinear in the level populations.

The charge conservation equation, and the additional terms for convection are
independent of the radiation field, and therefore are the same as in the traditional
complete linearization.

% --------------------------------------------------------------------------

\section*{Appendix E. Implementation of the Rybicki sche{-}me}
\addcontentsline{toc}{section}{Appendix E. Implementation of the Rybicki scheme}

The overall structure of the global Jacobian is reversed from the original
complete linearization, in the sense that the role of frequencies and depths is 
reversed. 
Although one may use the idea of the hybrid CL/ALI method, in the sense that only
some frequency points are taken as explicit, it is not very practical because it
slows down the convergence rate of the linearization iteration. Therefore, all 
the frequency points are taken as explicit.
For convenience, we repeat here the basic equations of both schemes.

For complete linearization
\begin{equation}
\label{cl4a}
-{\bf A}_d {\bf \delta\psi}_{d-1} +{\bf B}_d {\bf \delta\psi}_{d} 
-{\bf C}_d {\bf \delta\psi}_{d+1} = {\bf L}_d,
\end{equation}
where 
\begin{equation}
\label{cl1aa}
{\bf \psi}_d = \{ J_1, \ldots, J_{N\!F}, N, T, n_{\rm e}, n_1, 
\ldots, n_{N\!L},[n_m],[\nabla],[z]\}.
\end{equation}
Here, $A$, $B$, and $C$ are $N\!N \times N\!N$ matrices, and $L$ is
a residual error vector.

For the Rybicki scheme,
\begin{equation}
\label{ryb1aa}
{\bf U}_i \delta{\bf J}_i + {\bf R}_i \delta{\bf T} = {\bf E} ,
\end{equation}
with
\begin{equation}
\delta {\bf J}_i \equiv \{\delta J_{1i}, \delta J_{2i}, \ldots,\delta J_{N\!D,i} \},\quad
i=1,\ldots,N\!F,
\end{equation}
and 
\begin{equation}
\delta {\bf T} \equiv \{\delta T_1, \delta T_2, \ldots,\delta T_{N\!D} \}.
\end{equation}
Here, the only state parameter besides the mean intensities is the
temperature. 
Therefore, one has to view all the other state parameters
as functions of $T$.
Consequently, instead of employing the partial derivatives
$\partial\chi/\partial T$, $\partial\chi/\partial n_{\rm e}$, $\partial\chi/\partial n_i$, etc.,
one has to consider a total derivative $D\chi/DT$, and analogously for the derivatives of
the emission and scattering coefficients,

The components corresponding to the radiative transfer equation are the following.
The tridiagonal matrices ${\bf U}$ that contain derivatives with respect to the
mean intensities are exactly analogous to the corresponding elements of matrices
${\bf A}$, ${\bf B}$, and ${\bf C}$ of complete linearization, 
\begin{equation}
U_{dd,i} = (B_d)_{ii}, \quad U_{d,d-1,i} = (A_d)_{ii}, 
\quad U_{d,d+1,i} = (C_d)_{ii}.
\end{equation}
Components of matrices ${\bf R}$ are similar, but not identical, to the components
of the original matrices,
\begin{equation}
R_{dd,i} = (B_d)_{i,N\!R}, \quad R_{d,d-1,i} = (A_d)_{i,N\!R},
\end{equation}
where we replace the partial derivatives $\partial\chi/\partial T$, $\partial\eta/\partial T$,
$\partial\sigma/\partial T$ by the total derivatives $D\chi/DT$, $D\eta/DT$, and
$D\sigma/DT$, respectively. The right-hand side vector is given by
\begin{equation}
E_{d,i} = (L_d)_i.
\end{equation}

Similarly, for the radiative equilibrium equation,
\begin{eqnarray}
V_{dd,i} &=& (B_{d})_{N\!R,i}, \quad V_{d,d-1,i} \equiv (A_d)_{N\!R,i}, \\
\label{rybw}
W_{dd,i} &=& (B_{d})_{N\!R,N\!R}, \quad W_{d,d-1,i} = (A_d)_{N\!R,N\!R}, \\
F_d &=& (L_d)_{N\!R},
\end{eqnarray}
where again in Eq. (\ref{rybw}) one replaces $\partial\kappa/\partial T$ and  
$\partial\eta/\partial T$ by $D\kappa/DT$ and $D\eta/DT$. 

The total derivatives are computed numerically as, for instance for $\chi$,
\begin{equation}
\frac{D\chi_d}{DT_d} = \frac{\widetilde{\chi}_d - \chi_d}{\Delta T_d},
\end{equation}
where $\Delta T_d$ is set to $0.01 T_d$. We denote the new temperature $T_d +\Delta T_d$
as $\widetilde{T}_d$.
The corresponding $\widetilde{\chi}_d$
is computed by the following multi-step procedure: \\ [4pt]
(i) Update of the gas pressure, denoted here
as $\widetilde{P}$, by solving
the hydrostatic equilibrium equation,
\begin{eqnarray}
\widetilde{P}_1 &=& m_1 (g-g_1^{\rm rad}), \\
\label{rybpd}
\widetilde{P}_d &=& \widetilde{P}_{d-1} + (m_d- m_{d-1}) 
\left[ g- g_d^{\rm rad} (1 + 4\Delta T_d)\right], \quad d>1,
\end{eqnarray}
where
\begin{eqnarray}
g_1^{\rm rad}&=& (4\pi/c\rho_1)\sum_{i=1}^{N\!F} w_i \chi_{1i} g_i J_{1i}, \\
g_d^{\rm rad}&=& (4\pi/c)\sum_{i=1}^{N\!F} w_i 
\frac{ f_{di} J_{di}-f_{d-1,i}J_{d-1,i}}{m_d-m_{d-1}}, \quad d>1
\end{eqnarray}
is the current value of the radiation acceleration. The last term in Eq. (\ref{rybpd}) 
represents an approximate modification of the radiation pressure due to modified
temperature by expressing $P^{\rm rad} = \gamma P^{\rm rad, TE} \propto \gamma T^4$,
(with the superscript ``TE" denoting the thermodynamic equilibrium), and thus 
\begin{equation}
P^{\rm rad}(T+\Delta T) \approx P^{\rm rad}(T) + (dP^{\rm rad}/dT) \Delta T =
P^{\rm rad}(1 +4 \Delta T),
\end{equation}
which assumes that the non-equilibrium proportionality parameter $\gamma$
remains unchanged, and that the radiation pressure varies slowly with depth
so that the same thermodynamic relation can be used also for 
the radiation acceleration. \\ [4pt]
\noindent (ii) Compute new total particle density for the new gas pressure,
\begin{equation}
\widetilde{N}_d= \widetilde{P}_d/(k \widetilde{T}_d).
\end{equation}

\noindent (iii) For the new $\widetilde{N}_d$, compute the new electron number
density $\widetilde{n}_{\rm e}$ by solving a set of Saha equations and the charge
conservation equation. \\ [4pt]
(iv) Using the new $\widetilde{T}$ and $\widetilde{n}_{\rm e}$, and using current 
values of the radiative rates and collisional cross sections, solve the set of kinetic
equilibrium equations to obtain a new set of populations $\widetilde{n}_i, i=1,\ldots,
N\!L$.
Notice that in the case of LTE the kinetic equilibrium equations are replaced by
the set of Saha-Boltzmann expression for the level populations. \\ [4pt]
(v) Finally, the opacities and emissivities are evaluated by standard expressions
for the new values of $\widetilde{T}$, $\widetilde{n}_{\rm e}$, and $\widetilde{n}_i$,
\begin{equation}
\widetilde{\chi}_d = \chi_d(\widetilde{T}, \widetilde{n}_{\rm e}, \{\widetilde{n}_i\}),
\end{equation}
and analogously for $\eta$ and $\sigma$.

% --------------------------------------------------------------------------

\section*{Appendix F. Partial frequency redistribution}
\addcontentsline{toc}{section}{Appendix F. Partial frequency redistribution}

This process is not a main emphasis of {\sc tlusty},
but can be included in an approximate way.
In the present version of {\sc tlusty}, only the H I Lyman $\alpha$, the Mg I
resonance line at $\lambda$2852 \AA, and the Mg II resonance lines at
$\lambda\lambda$2796, 2803 \AA\ can be treated with 
partial redistribution, namely using the partial coherent scattering approximation.

Under the usual approximation of {\em complete frequency redistribution}
(CFR),
there is no correlation between the frequencies of an absorbed and an emitted
photon during the process of resonance scattering in a line, that is a radiative 
transition from level $i$ to level $j$, immediately followed by a radiative
transition from $j$ to $i$. 

A more general approach is called the {\em partial
frequency redistribution} (PFR, but often, and in the past, an acronym PRD
was used). In this case, there is a correlation between the absorbed and
emitted photons
described through the {\em redistribution function}. For a comprehensive
discussion refer to Hubeny \& Mihalas (2014; Chaps.10 and 15).
The most appropriate redistribution function for a resonance line is a linear
combination of complete redistribution, and coherent scattering on the
atom's frame, which in the observer's frame is described through the
redistribution function $R_{\rm II}$. The latter can further be approximated
by another linear combination of complete redistribution and coherent
scattering in the laboratory frame, called {\em partial coherent scattering
approximation} [Jefferies \& White (1960), Kneer (1975), 
Hubeny \&  Mihalas (2014, \S\,15.3)].

The presence of PFR effects changes in principle both the emission
coefficient in the line, as well as the radiative rates. As discussed for
instance in Hubeny \& Mihalas (2014, \S\,15.1), the latter effects are
usually very small, so in {\sc tlusty} only a modification of the emission
coefficient is taken into account. 
One makes another approximation
that the effects of partial redistribution are negligible in the line core,
which further simplifies the modifications needed to treat the PFR effects
in {\sc tlusty}.

For a line chosen to be with PFR, the emission coefficient for the ordinary
complete redistribution is given by, neglecting stimulated emission,
\begin{equation}
\eta_{ij}^{\rm CFR}(x) = n_j A_{ji} \phi_{ij}(x),
\end{equation}
where $x=(\nu-\nu_0)/\Delta\nu_D$ is the frequency difference from the
liner center expressed in units of Doppler width, $A_{ji}$ is the Einstein
coefficient for spontaneous emission from level $j$ to $i$, and $\phi(x)$
is the (normalized) absorption profile coefficient that expresses the
frequency dependence of the line absorption cross section, typically given by
a Voigt profile, or a special profile in the case of hydrogen.

The approximate treatment of PFR adopted in {\sc tlusty} consists of two
simple modifications of the standard CFR approach:
The modifications for the approximate approach adopted in {\sc tlusty} is:\\ [2pt]
(i) Replacing the emission coefficient with
\begin{equation}
\eta_{ij}^{\rm PFR}(x) = \eta_{ij}^{\rm CFR}(x) [1-\gamma_{ij} \bar s(x)] ,
\end{equation}
where $\bar s(x)$ is a step function, $\bar s(x) = 0$ for $x\leq x_{\rm div}$,
and $\bar s(x)= 1$ for $ > x_{\rm div}$; and $x_{\rm div}$ is the so-called
{\em division frequency}, which is an input parameter (XPDIV). $\gamma$
is the so-called {\em coherence fraction}, given approximately by, 
\begin{equation}
\label{coher}
\gamma_{ij} \approx A_{ji}/\Gamma_{ij},
\end{equation}
where $\Gamma_{ij}$ is the line broadening parameter [see Eq. (\ref{voigt})].
The simple form of Eq. (\ref{coher}) follows from neglecting various other
processes, such as inelastic collisions.\\ [2pt]
(ii) Adding a coherent scattering term to the original scattering terms, which is usually just
the electron (Thomson) scattering,
\begin{equation}
\sigma(\nu) = \sigma^{\rm orig}(\nu) + \kappa_{ij}(\nu)\gamma_{ij} \bar s(x).
\end{equation}
The above treatment of partial redistribution is admittedly very crude,
but it should be kept in mind that this approach is designed only to assess
possible modifications of the global atmospheric structure due to PFR in
some lines, which, in the vast majority of cases, is relatively small. 
To compute detailed line profiles with more accurate treatment
of PFR, one can always take an atmospheric structure computed by {\sc tlusty}
and use it in another codes specifically designed to produce line profiles
with a more sophisticated treatment of PFR.

% --------------------------------------------------------------------------

\section*{References}
\addcontentsline{toc}{section}{References} 

\def\reference{\par \leftskip20pt \parindent-20pt\parskip4pt}
\noindent
\reference Allende Prieto, C., Lambert, D., Hubeny, I., \& Lanz, T. 2003,
ApJS, 147, 363.
\reference Auer L.H. 1973, ApJ, 180, 469.
\reference Auer, L.H. 1976, JQSRT, 16, 931.
\reference Auer, L.H. 1984, in {\it Methods in Radiative Transfer}, ed. by
W. Kalkofen, Cambridge Univ. Press, Cambridge, p.101.
\reference Auer, L.H., \& Mihalas, D. 1969, ApJ 158, 641
\reference Avrett, E.H., \& Loeser, R. 1982, ASP Conf. Ser.  26, 489.
\reference Bergeron, P., Wesemael, F., \& Fontaine, G. 1992, ApJ 387, 288.
\reference Butler, K., \& Giddings, J. 1978, in Newsletter  on Analysis of Astron.
Spectra, 9 (Daresbury Lab., England).
\reference Carlsson, M. 1986, Uppsala Astron. Obs. Tech. Rep. 33.
\reference Castor, J.I., Dykema, P., \& Klein, R.I., 1992, ApJ 387, 561.
\reference Chang, J.S., \& Cooper, G., 1970, J. Comp. Phys., 6, 1.
\reference Daeppen, W., Anderson, L.S., \& Mihalas, D.  1987, ApJ, 319, 195.
\reference Feautrier, P. 1964, C. R. Acad. Sci. Paris., Ser. B, 258, 3189.
\reference Frank, J., King, A. \& Raine, D. 1992, {\it Accretion Power in 
Astrophysics}, 2nd ed., Cambridge Univ. Press, Cambridge.
\reference Grevesse, N., \& Sauval, A. 1998, Space Sci. Rev., 85, 161.
\reference Hubeny, I. 1988, Computer Physics Comm. 52, 103.
\reference Hubeny, I. 1990, ApJ, 351, 632.
\reference  Hubeny, I., Blaes, O., Agol, E., \& Krolik, J.H., 2001,
     ApJ, 559, 680.
\reference Hubeny, I., Burrows, A., \& Sudarsky, D. 2003, ApJ,   594, 1011.   
\reference Hubeny, I., \& Hubeny, V., 1998, ApJ,  505, 558.
\reference Hubeny, I., Hummer, D.G., \& Lanz, T. 1994, A\&A, 282, 151.
\reference Hubeny, I., \& Lanz, T. 1992, A\&A 262, 501.
\reference Hubeny, I., \& Lanz, T. 1995, ApJ, 439, 875.
\reference Hubeny, I., \& Lanz, T. 2017a,  {\it A Brief Introductory Guide to {\sc tlusty}
and {\sc synspec}}, (Paper~I).
\reference Hubeny, I., \& Lanz, T. 2017c,  {\sc tlusty} {\it User's Guide III: Operational Manual}
(Paper~III).
\reference Hubeny, I. \& Mihalas, D. 2014, {\it Theory of Stellar Atmospheres},
Princeton Univ. Press, Princeton.
\reference Hummer, D.G., \& Mihalas, D. 1988, ApJ  331, 794.
\reference Irwin, A.W. 1981, ApJS, 45, 621.
\reference Jefferies, J. \& White, O. 1960, ApJ, 132, 767.
\reference Kallman, T.R, 2000, {\it XSTAR: A Spectral Analysis Tool}, NASA
Goddard Space Flight Center.
\reference Kneer, F. 1975, ApJ, 200, 367.
\reference Kingdon, J., \& Ferland, G. 1996, ApJS, 106, 205.
\reference K\v r\'\i\v z, S., \& Hubeny, I., 1986, 
Bull. Astron. Inst. Czechosl., 37, 129.
\reference Krolik, J.H. 1999, {\it Active Galactic Nuclei}, Princeton Univ. Press, Princeton.
\reference Kurucz, R.L. 1970, SAO Spec. Rep. 309.
\reference Lanz, T., \& Hubeny, I. 2003, ApJS, 146, 417.
\reference Lanz, T., \& Hubeny, I. 2007, ApJS, 169, 83.
\reference Lemke, M., 1997, A\&AS, 122, 285.
\reference Lynden-Bell, D. \& Pringle, J.E., 1974, MNRAS, 168, 603.
\reference Mihalas, D., 1978, {\it Stellar Atmospheres}, 2nd ed., Freeman,
    San Francisco.
\reference Mihalas, D., \& Hummer, D.G. 1974, ApJS, 28. 343. 
\reference Novikov, I.D., \& Thorne K.S 1973, in {\it Black Holes}, ed. by.
C. de Witt \& B. de Witt, Gordon \& Breach, New York.
\reference Olson, G.,  \& Kunasz, P. 1987, JQSRT, 38, 325.
\reference Riffert, H., \& Harold, H. 1995, ApJ, 450, 508.
\reference Rybicki, G.B. 1969, JQSRT, 11, 589.
\reference Rybicki, G.B., \& Hummer, D.G., 1991, A\&A 245, 171.
\reference Rybicki, G.B., \& Hummer, D.G., 1992, A\&A 262, 209.
\reference Rybicki, G.B., \& Lightman, A.P. 1979, {\it Radiative Processes in
Astrophysics}, Willey \& Sons, New York.
\reference Seaton, M., editor, 1995, {\it The Opacity Project, Vol. I}, Inst. of Physics
Publishing, Bristol.
\reference Shakura, N. I., \& Sunyaev, R. A. 1973, A\&A  24, 337.
\reference Sharp, C.S., \& Burrows, A. 2007, ApJS, 168, 140.
\reference Sparks \& Fischel 1971, NASA Spec. Rep. 3066.
\reference Traving, G., Baschek, B., \& Holweger, H. 1966, Abhand. Hamburg.
C     Sternwarte. Band VIII, Nr. 1.
\reference Tremblay, P.-E., \& Bergeron, P. 2009, ApJ, 696,1755.
\reference Trujillo Bueno, J., \& Fabiani Bendicho, P. 1995, ApJ, 455, 646.
\reference Werner, K., 1984, A\&A, 139, 237.

\def\noreference{\par \leftskip0pt \parindent0pt\parskip4pt}

\noreference

%\newpage
%\addcontentsline{toc}{section}{index}
%\printindex

\end{document}